\documentclass[aps,prx,color,pdflatex,citeautoscript,superscriptaddress,twocolumn,longbibliography]{revtex4-1}
\usepackage[T1]{fontenc}
\usepackage{graphicx}
\usepackage{dcolumn}
\usepackage{comment}
\usepackage{amsmath,amssymb,bbold,bm}
\usepackage{hyperref}
\hypersetup{colorlinks=true,citecolor=blue,linkcolor=blue,urlcolor=blue,pdfstartview=FitH,pdfpagemode=PageWidth}
\usepackage{amsfonts}
\usepackage{listings}
\usepackage{braket}
\usepackage{float,latexsym}
\usepackage{multirow}
\usepackage{tikz}
\usepackage{color}
\definecolor{dkgreen}{rgb}{0,0.5,0}

\newcommand{\vect}[1]{\boldsymbol{#1}}

\begin{document}

\title{Fermionized parafermions and symmetry-enriched Majorana modes}
\author{Aaron Chew}
\affiliation{Department of Physics and Institute for Quantum Information and Matter, California Institute of Technology, Pasadena, CA 91125, USA}
\author{David F. Mross}
\affiliation{Department of Condensed Matter Physics, Weizmann Institute of Science, Rehovot, 76100, Israel}
\author{Jason Alicea}
\affiliation{Department of Physics and Institute for Quantum Information and Matter, California Institute of Technology, Pasadena, CA 91125, USA}
\affiliation{Walter Burke Institute for Theoretical Physics, California Institute of Technology, Pasadena, CA 91125, USA}

\date{\today}

\begin{abstract}
Parafermion zero modes are generalizations of Majorana modes that underlie comparatively rich non-Abelian-anyon properties. We introduce exact mappings that connect parafermion chains, which can emerge in two-dimensional fractionalized media, to \emph{strictly one-dimensional} fermionic systems.  In particular, we show that parafermion zero modes in the former setting translate into `symmetry-enriched Majorana modes' that intertwine with a bulk order parameter---yielding braiding and fusion properties that are impossible in standard Majorana platforms.  Fusion characteristics of symmetry-enriched Majorana modes are directly inherited from the associated parafermion setup and can be probed via two kinds of anomalous pumping cycles that we construct.  Most notably, our mappings relate $\mathbb{Z}_4$ parafermions to conventional electrons with time-reversal symmetry.  In this case, one of our pumping protocols entails fairly minimal experimental requirements: Cycling a weakly correlated wire between a trivial phase and time-reversal-invariant topological superconducting state produces an edge magnetization with quadrupled periodicity.  
Our work highlights new avenues for exploring `beyond-Majorana' physics in experimentally relevant one-dimensional electronic platforms, including proximitized ferromagnetic chains. 

\end{abstract}
\maketitle

\section{Introduction}
\label{Introduction}

Interacting quantum systems in two dimensions can host quasiparticle excitations whose properties are seemingly at odds with their microscopic origin.  In particular, ground states characterized by a subtle non-local entanglement structure---i.e., topological order---host `anyon' excitations that not only carry fractional quantum numbers, but additionally exhibit exchange statistics that is neither bosonic nor fermionic.  An especially interesting example is provided by `non-Abelian anyons', which display a number of fascinating properties.  
First, non-Abelian anyons carry exotic zero-energy degrees of freedom that generate a space of locally indistinguishable ground states.  Second, braiding the anyons rotates the system within this ground-state space---yielding the remarkable phenomenon of non-Abelian statistics.  And third, they exhibit nontrivial fusion rules, i.e., pairs of non-Abelian anyons can combine to form multiple quasiparticle types.  The above characteristics are also technologically relevant as they form the basis for inherently fault-tolerant topological quantum computation \cite{Kitaev:2003,Nayak:2008}.    
An experimentally relevant setting where such exotic excitations emerge is the Moore-Read fractional-quantum-Hall state \cite{Moore:1991}. There, charge-$e/4$ quasiparticles harbor Majorana zero modes that endow them with the braiding and fusion properties of `Ising' non-Abelian anyons.

One can alternatively harness non-Abelian-anyon physics through defects in simpler topological phases \cite{BarkeshliBonderson}.  Consider, for example, the Kitaev chain \cite{Kitaev:2001}, which describes a spinless one-dimensional (1D) $p$-wave superconductor.  Domain walls separating topological and trivial phases of the model harbor Majorana zero modes, and hence behave very similarly to non-Abelian anyons in the Moore-Read state.  The pursuit of Majorana modes in 1D superconducting devices has correspondingly become a vibrant (and oft-reviewed \cite{BeenakkerReview,AliceaReview,FlensbergReview,TewariReview,FranzReview,ChetanReview,SatoReview,AguadoReview,LutchynReview}) enterprise.  A more exotic example arises from `parafermion' chains \cite{Fendley:2012,AliceaFendleyReview}---1D systems with degrees of freedom that possess an intrinsic, unbreakable $\mathbb{Z}_N$ charge symmetry, analogous to the unbreakable $\mathbb{Z}_2$ parity symmetry of fermions.  Domain walls between topological and trivial phases for the chain bind \emph{parafermion zero modes}, which are $\mathbb{Z}_N$ Majorana generalizations that generate larger ground-state degeneracy, denser braid transformations, and richer fusion rules.  Because the chain is built from neither bosons nor fermions, realizing these non-Abelian defects is more challenging.  Nevertheless, numerous blueprints now exist for stabilizing parafermion zero modes at line defects within a two-dimensional, \emph{Abelian} topologically ordered host.  Possible host platforms include the toric code \cite{Bombin}, fractional Chern insulators \cite{Barkeshli:2012}, quantum-Hall bilayers \cite{Barkeshli:2014b}, quantum-Hall/superconductor hybrids \cite{Lindner:2012,Clarke:2013a,ChengBraiding,Vaezi:2013}, and more \cite{You:2012,Maghrebi}.

\begin{figure}
\includegraphics[width=0.8\columnwidth]{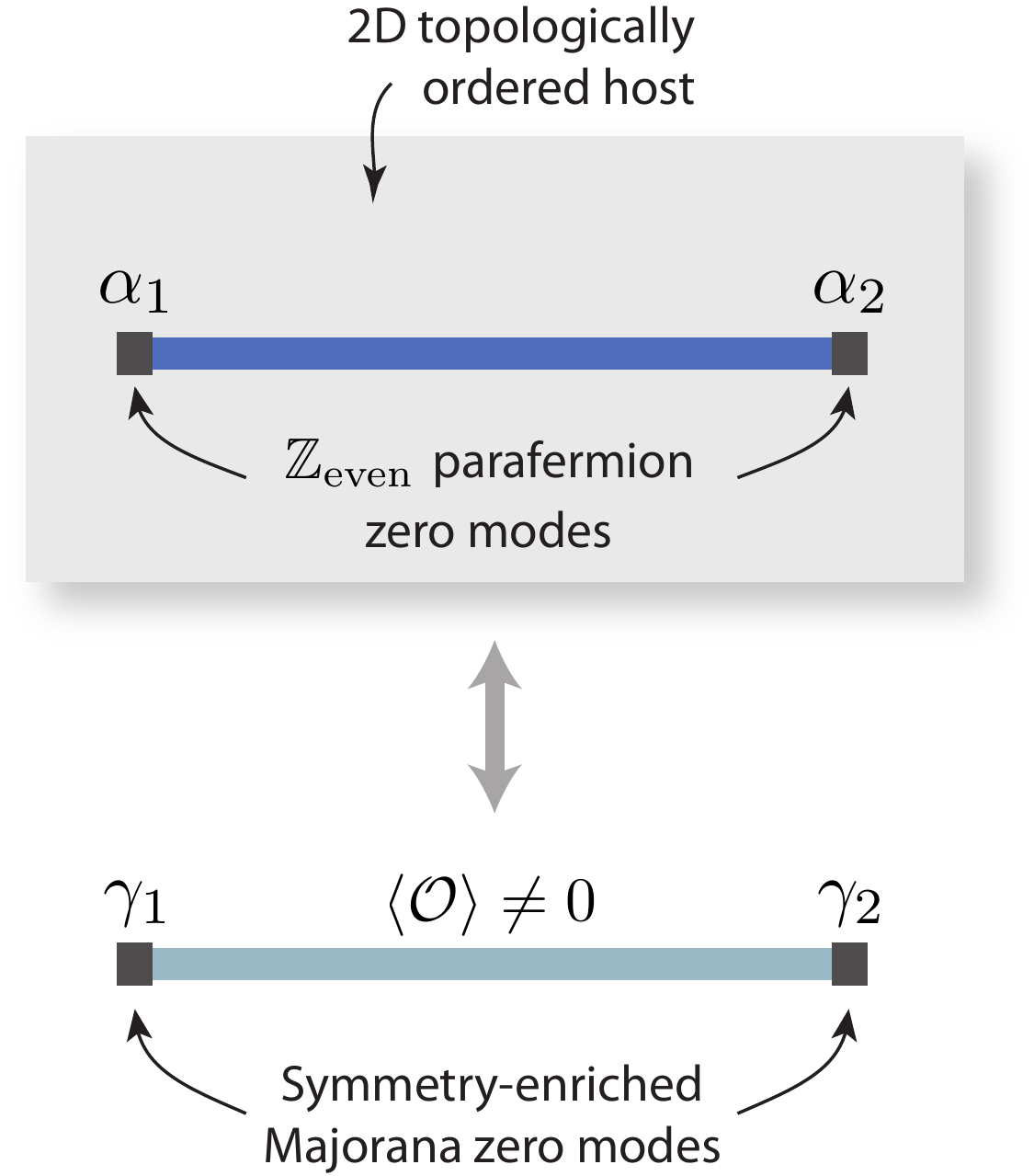}
\caption{Correspondence between non-Abelian defects in 2D topologically ordered phases and in strictly 1D fermionic systems.  Parafermion zero modes $\alpha_{1,2}$ translate into symmetry-enriched Majorana zero modes $\gamma_{1,2}$ intertwined with an order parameter $\mathcal{O}$.  We show that symmetry-enriched Majorana zero modes underlie physical properties not possible from conventional Majorana systems, including an enlarged set of braid transformations and anomalous pumping protocols that are closely related to nontrivial fusion rules in the associated parafermion platform. }
\label{Intro_fig}
\end{figure}

Here we rigorously establish a link between non-Abelian defects in such 2D topologically ordered phases and those that can arise in \emph{strictly 1D} fermion systems.  
To this end, we introduce exact, non-local mappings between arbitrary $\mathbb{Z}_{\rm even}$ parafermion chains and microscopic 1D fermionic models.  This machinery provides a `dictionary' connecting observables, phases, and any other quantity of interest between the two representations.  We in particular find that $\mathbb{Z}_{\rm even}$ parafermion zero modes translate into `symmetry-enriched Majorana zero modes' whose wavefunctions depend nontrivially on a spontaneously chosen order parameter for the fermions; see Fig.~\ref{Intro_fig}.  Although the degeneracy in the latter setting enjoys only partial topological protection, we demonstrate that symmetry-enriched Majorana modes give rise to phenomena that are not possible in conventional Majorana platforms.  

For one, braiding processes can alter the order-parameter configuration, thereby rotating the system within an enlarged subspace (though the braid matrices do \emph{not} match those arising from parafermions for reasons that we explain).  Moreover, we show that the richer fusion rules stemming from parafermion zero modes are directly manifested in the 1D fermion setting.  Imagine fusing two non-Abelian defects that bind parafermion zero modes.  One can define a pumping cycle that returns the Hamiltonian to its original form yet modifies the fusion channel for the defects.  The system thus exhibits an anomalous periodicity set by the number of available fusion channels.  Interestingly, one can realize pumping cycles with exactly the same periodicity by hybridizing symmetry-enriched Majorana modes in 1D fermion systems.  We study two implementations of such anomalous fermionic pumps.  One requires symmetry protection to maintain the same periodicity as in the parafermion realization, while the other relies only on locality and fermion-parity conservation.  

Useful insights can be obtained by specializing to the $\mathbb{Z}_4$ case, which we primarily focus on in this paper.  
In this limit the correspondences highlighted above can be anticipated from several angles.  First, each pair of $\mathbb{Z}_4$ parafermions contributes four states to the Hilbert space, just like two species of fermions.  Second, Ref.~\onlinecite{Pachos} used complementary analytical and numerical methods to infer 
that the eigenstates of certain $\mathbb{Z}_4$ parafermion chains can be described in terms of free fermions.  Third, Zhang and Kane \cite{ZhangKane} and Orth et al.~\cite{Orth} showed that proximitized edge states of a two-dimensional quantum-spin-Hall insulator can support zero modes reminiscent of $\mathbb{Z}_4$ parafermions (see also Refs.~\onlinecite{Peng,Hui,Vinkler}).  Finally, parafermion chains are related to bosonic clock models (for any $\mathbb{Z}_N$) \cite{FradkinKadanoff,Fendley:2012}---a relation that we will frequently exploit.  
In the $\mathbb{Z}_4$ limit one can decompose clock spins into two sets of Pauli matrices \cite{Kohmoto,Kohmoto2,Hutter} that can be fermionized by standard methods \footnote{For yet another take on fermionizing parafermions, see Ref.~\onlinecite{Calzona}.}.  We will later draw further connections to all of these works, particularly the results for quantum-spin-Hall systems.  

While the `fermionizability' of $\mathbb{Z}_4$ parafermion chains is thus natural, it is not clear a priori whether the associated 1D fermionic systems are at all physically relevant.  Importantly, in our fermionization scheme (which differs from the strategy noted above) $\mathbb{Z}_4$ parafermions map onto ordinary spinful electrons with familiar symmetries including time reversal and spin rotations.  Our dictionary thus indeed relates phases for parafermions to interesting, and in some cases already well-studied, 1D electronic states of matter.  The phase that supports symmetry-enriched Majorana modes (see again Fig.~\ref{Intro_fig}) corresponds to a topological superconductor accessed by spontaneously breaking time-reversal symmetry, which may already be realized in atomic-chain experiments \cite{Nadj-Perge,Ruby,Pawlak,Feldman,Jeon}.  As another noteworthy example, the parafermion chain supports a symmetry-protected topological phase that translates into a time-reversal invariant topological superconductor (TRITOPS) \cite{QiTRITOPS,ChungTRITOPS,ZhangTRITOPS,KeselmanTRITOPS,Haim1TRITOPS,Haim2TRITOPS,CamjayiTRITOPS} with a Kramers pair of Majorana zero modes at each end.  One of the anomalous pumping cycles we introduce involves modulating a fermionic wire between trivial and TRITOPS phases; the magnetization at the ends of the system exhibits quadrupled periodicity---reflecting the four fusion channels available in the corresponding parafermion platform.  We note that this pump is a strict-1D analogue of the $8\pi$-periodic Josephson effect identified for quantum-spin-Hall edges in Refs.~\onlinecite{ZhangKane,Orth}.  The experimental requirements for implementing the cycle are surprisingly minimal, thus providing a tantalizing opportunity for exploring certain aspects of parafermion physics using non-fractionalized 1D systems.

We organize the remainder of the paper as follows.  In Secs.~\ref{Mappings} through \ref{ExpImplications} we exclusively treat the $\mathbb{Z}_4$-parafermion case.  Section~\ref{Mappings} details our fermionization scheme, while Sec.~\ref{Phases} derives the correspondence between various phases in the clock, parafermion, and electronic representations.  We then turn to experimental implications in Sec.~\ref{ExpImplications}.  There we contrast the non-Abelian braiding properties arising from $\mathbb{Z}_4$ parafermion zero modes and symmetry-enriched Majorana modes, and analyze the anomalous pumping cycles.  Section~\ref{Z2Msec} generalizes these results to arbitrary $\mathbb{Z}_{\rm even}$ parafermions.  An executive summary appears in Sec.~\ref{Discussion} along with several future directions.  Finally, Appendices \ref{Inversion_Appendix} through \ref{HigherDictionaryAppendix} contain supplemental results and technical details.

\section{Operator Mappings}
\label{Mappings}

This section introduces non-local mappings that link bosonic $\mathbb{Z}_4$ clock operators, $\mathbb{Z}_4$ parafermions, and spinful fermions residing on a 1D lattice.  In what follows we primarily flesh out these mappings without recourse to specific Hamiltonians, which will instead be constructed and analyzed in Sec.~\ref{Phases}.  Sections~\ref{Z4clock} and \ref{Z4parafermion} below largely parallel the treatment of $\mathbb{Z}_3$ parafermions in Ref.~\onlinecite{Mong:2014b}.

\subsection{$\mathbb{Z}_4$ clock operators}
\label{Z4clock}

We first review the $\mathbb{Z}_4$ clock representation.  Each lattice site, labeled by integers $a$, contains a four-state `spin'.  The Hilbert space is spanned by unitary clock operators $\sigma_a$ and $\tau_a$ that satisfy
\begin{equation}
  \sigma_a^4 = \tau_a^4 = 1
\end{equation}
along with the commutation relation
\begin{equation}
  \sigma_a \tau_a = i \tau_a \sigma_a~.
  \label{clock_commutator}
\end{equation}
(Off site, the clock operators commute.)  The relations above imply that $\sigma_a$ and $\tau_a$ both exhibit eigenvalues $\pm1, \pm i$, with $\tau_a$ `winding' the eigenvalue of $\sigma_a$ and vice versa.  

We will be particularly interested in chains that exhibit a global $\mathbb{Z}_4$ symmetry, generated by
\begin{equation}
  Q = \prod_{a}\tau_a^\dagger~,
\end{equation}
as well as an antiunitary time-reversal symmetry $\mathcal{T}$ that satisfies $\mathcal{T}^2 = +1$.  The former acts according to
\begin{equation}
  Q \sigma_a Q^\dagger = i \sigma_a~,\qquad Q\tau_a Q^\dagger = \tau_a~.
  \label{Z_4}
\end{equation}
Note that if clock spins constitute physical degrees of freedom, $\mathbb{Z}_4$ symmetry can be broken either spontaneously or explicitly---a situation that we will later contrast with the cases where parafermions and fermions form the physical objects.  
Time reversal transforms clock operators as
\begin{equation}
  \mathcal{T} \sigma_a \mathcal{T} = \sigma_a^\dagger~,\qquad \mathcal{T} \tau_a \mathcal{T} = \tau_a~.
  \label{T}  
\end{equation}
We will also invoke a `charge conjugation' symmetry $\mathcal{C}$ that yields
\begin{align}
  \mathcal{C} \sigma_a \mathcal{C} = \sigma_a^\dagger~,\qquad\mathcal{C} \tau_a \mathcal{C} = \tau_a^\dagger~.
  \label{C}
\end{align}
Table~\ref{symmetry_table} summarizes these symmetry properties.  

\begin{figure}
\includegraphics[width=\columnwidth]{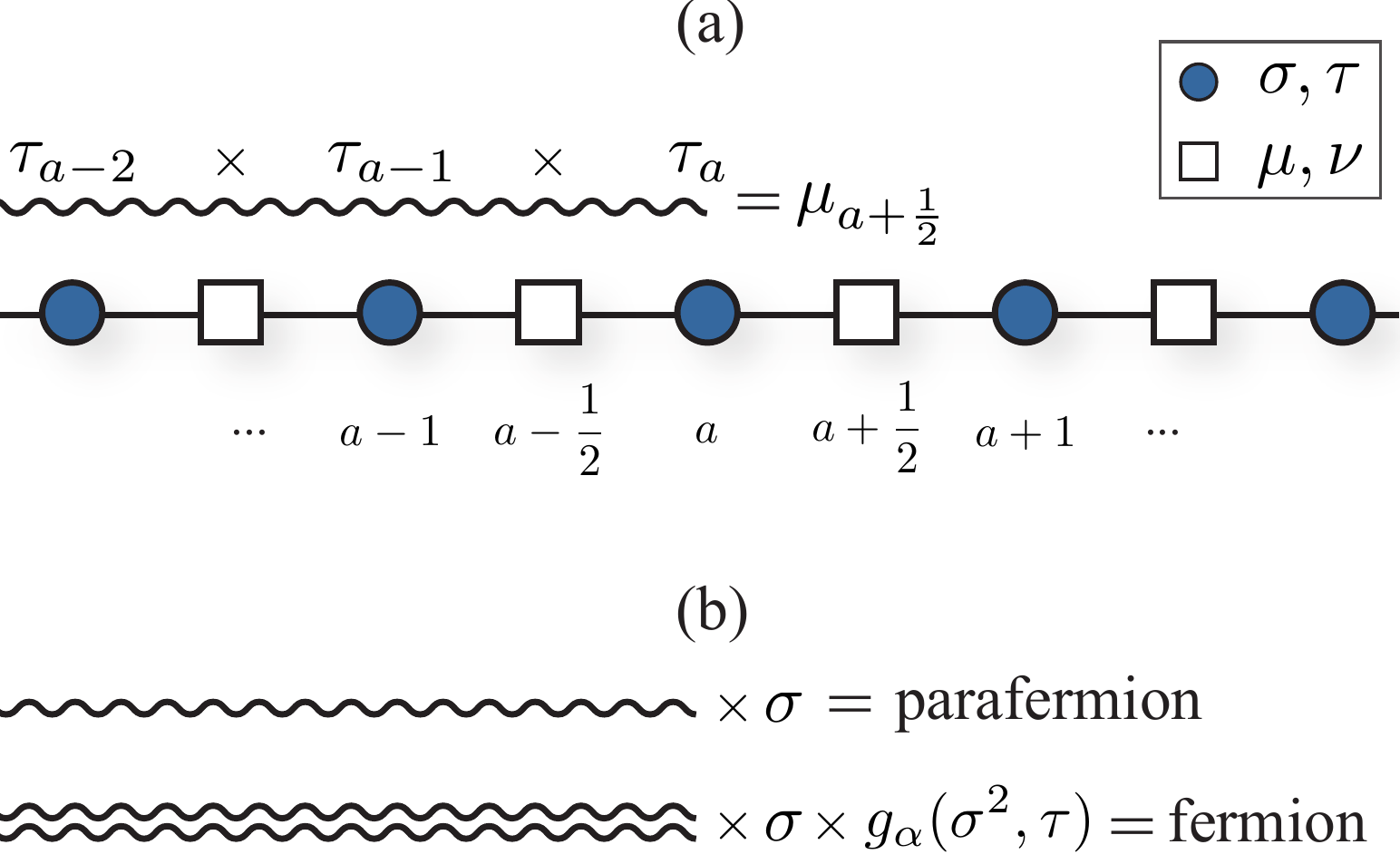}
\caption{(a) Chain of clock operators $\sigma_a,\tau_a$ together with their dual counterparts $\mu_{a+\frac{1}{2}},\nu_{a+\frac{1}{2}}$, which live on the dual lattice.  The dual operator $\mu_{a+\frac{1}{2}}$ corresponds to a non-local $\tau$ string (wavy line).  (b) Binding $\sigma$ and $\mu$ yields parafermion operators; attaching the double string $\mu^2$ to $\sigma \times g_\alpha(\sigma^2,\tau)$, where $g_\alpha(\sigma^2,\tau)$ is a local function of clock operators, gives fermions with spin $\alpha$.  See Secs.~\ref{Z4parafermion} and \ref{fermions} for precise expressions relating parafermions and fermions to clock variables.  }
\label{ops_fig}
\end{figure}

One can equivalently describe the system with dual operators
\begin{equation}
  \mu_{a+\frac{1}{2}} = \prod_{b < a+\frac{1}{2}} \tau_{b}~,\qquad \nu_{a+\frac{1}{2}} = \sigma_a^\dagger \sigma_{a+1}
  \label{duality}
\end{equation}
that reside on dual-lattice sites labeled by half-integers [see Fig.~\ref{ops_fig}(a)].  Similar to the original clock variables, the dual operators are unitary and satisfy
\begin{align}  
  \mu_{a+\frac{1}{2}}^4 = \nu_{a+\frac{1}{2}}^4 = 1~,\qquad\mu_{a+\frac{1}{2}}\nu_{a+\frac{1}{2}} = i \nu_{a+\frac{1}{2}}\mu_{a+\frac{1}{2}}~.
\end{align}
Their symmetry properties follow straightforwardly from Eqs.~\eqref{Z_4} through \eqref{C} and are also listed in Table~\ref{symmetry_table}.  

\begin{table}
\begin{center}
 \setlength\extrarowheight{2pt}
 \begin{tabular}{|c | c | c | c| c|} 
 \hline
  & $\mathbb{Z}_4$ & $\mathcal{C}$ & $\mathcal{T}$ \\ [0.05ex] 
 \hline\hline
 $\sigma \rightarrow$ & $i\sigma$ & $\sigma^\dagger$ & $\sigma^\dagger$ \\
 \hline
 $\tau \rightarrow$ & $\tau$ & $\tau^\dagger$ & $\tau$ \\
 \hline\hline
 $\mu \rightarrow$ & $\mu$ & $\mu^\dagger$ & $\mu$ \\
 \hline
 $\nu \rightarrow$ & $\nu$ & $\nu^\dagger$ & $\nu^\dagger$ \\
 \hline \hline
 $\alpha \rightarrow$ & $i \alpha$ & $\alpha^\dagger$ & $\alpha'^\dagger$ \\ 
 \hline
 $\alpha '\rightarrow$ & $i \alpha'$ & $\alpha'^\dagger$ & $\alpha^\dagger$ \\ 
 \hline\hline
 $f_{\uparrow}\rightarrow$ & $ie^{i\pi n_{\downarrow}}f_{\uparrow}$ & $e^{i\pi n_{\uparrow}}f_{\downarrow}$ & $ie^{i\pi n_{\uparrow}}f_{\downarrow}$ \\
 \hline
 $f_{\downarrow}\rightarrow$ & $-ie^{i\pi n_{\uparrow}}f_{\downarrow}$ & $e^{i\pi n_{\downarrow}}f_{\uparrow}$ & $ie^{i\pi n_{\downarrow}}f_{\uparrow}$ \\
 \hline  
\end{tabular}
\end{center}
\caption{Action of primitive symmetries on clock operators $\sigma,\tau$; dual clock operators $\mu,\nu$; two representations of parafermion operators $\alpha,\alpha'$; and spinful fermions $f_{\uparrow,\downarrow}$. Site labels are suppressed for brevity here and in other tables below.}  
\label{symmetry_table}
\end{table}

Suppose that $\mathbb{Z}_4$ symmetry is spontaneously broken, leading to $\langle \sigma_a\rangle \neq 0$.  Starting from such a broken-symmetry phase, the dual operator $\mu_{a+\frac{1}{2}}$ creates a domain-wall defect that winds all clock spins to the left of the dual site $a+\frac{1}{2}$.  Proliferation of these defects---i.e., $\langle \mu_{a+\frac{1}{2}} \rangle \neq 0$---destroys the order and restores $\mathbb{Z}_4$ symmetry.  In this sense $\sigma$ and $\mu$ respectively represent order and disorder operators.  Combining order and disorder operators generates $\mathbb{Z}_4$ parafermions \cite{FradkinKadanoff,Fendley:2012}, to which we turn next.

\subsection{$\mathbb{Z}_4$ parafermions}
\label{Z4parafermion}

We have some freedom for how to construct parafermions from order and disorder operators.  One choice binds $\sigma$ and $\mu$ to define lattice $\mathbb{Z}_4$ parafermions
\begin{equation}
  \alpha_{2a-1} = \sigma_a \mu_{a-\frac{1}{2}}~,\qquad
  \alpha_{2a} = e^{-i \frac{\pi}{4}} \sigma_a \mu_{a+\frac{1}{2}}~,
  \label{alpha}
\end{equation}
as sketched in Fig.~\ref{ops_fig}(b).
Like the clock variables, these unitary operators obey
\begin{equation}
  \alpha_a^4 = 1~.
\end{equation}
The $\tau$ string encoded in the disorder operators, however, yields the \emph{non-local} commutation relation
\begin{align}
  \alpha_a \alpha_{b>a} = i \alpha_{b}\alpha_a~.
  \label{alpha_commutator}
\end{align}
We could equally well bind $\sigma$ and $\mu^\dagger$ to define a non-locally related set of $\mathbb{Z}_4$ parafermion operators
\begin{equation}
  \alpha'_{2a-1} = \sigma_a \mu^\dagger_{a-\frac{1}{2}}~,\qquad
  \alpha'_{2a} = e^{i \frac{\pi}{4}} \sigma_a \mu^\dagger_{a+\frac{1}{2}}
\end{equation}
that similarly obey
\begin{equation}
  \alpha_a'^4 = 1~,\qquad \alpha'_a \alpha'_{b>a} = -i \alpha'_{b}\alpha'_a~.
  \label{alphap_commutator}
\end{equation}
While not independent, both representations are useful to consider since they transform into one another under 
time reversal $\mathcal{T}$.  Table~\ref{symmetry_table} lists their transformation properties, which are inherited from those of the clock operators and their duals.  Throughout this paper we mainly focus on the $\alpha_a$ representation for concreteness.  

Hereafter, we will define parafermions as physical degrees of freedom if the host system exhibits a $\mathbb{Z}_4$ symmetry (which sends $\alpha_a \rightarrow i \alpha_a$) that can never be broken \emph{explicitly} by any local perturbation.  
Consider, for example, $\mathbb{Z}_4$ parafermions germinated from extrinsic defects in a parent fractional-quantum-Hall medium.  The parafermion operator $\alpha_a^n$ adds nontrivial anyon charge to position $a$ provided $n \neq 0$ mod 4, while $(\alpha_a^\dagger)^n$ adds the opposite anyon charge.  Since the total anyon charge for the system must be trivial, all physical terms in the Hamiltonian must be invariant under $\mathbb{Z}_4$ symmetry.  

Next we discuss \emph{spontaneous} $\mathbb{Z}_4$ symmetry breaking, closely following Ref.~\onlinecite{Motruk:2013} (see also Refs.~\onlinecite{Bondesan:2013,Alexandradinata,Meidan}).  Due to the non-local commutation relation in Eq.~\eqref{alpha_commutator}, a parafermion system cannot spontaneously develop an expectation value $\langle \alpha_a \rangle \neq 0$ across the chain.  To see this, note that $\langle \alpha_a^\dagger \alpha_b \rangle = \pm i \langle \alpha_b \alpha_a^\dagger\rangle$; when $|a-b|\rightarrow \infty$, factorizing the left and right sides yields $\langle \alpha_a^\dagger\rangle \langle \alpha_b\rangle = \pm i \langle \alpha_b\rangle \langle \alpha_a^\dagger\rangle$, which admits only trivial solutions.  Since $[\alpha_a^2,\alpha_b^2] = 0$, however, no such obstruction exists for spontaneously developing an expectation value $\langle \alpha_a^2 \rangle \neq 0$.  The resulting `parafermion condensate' phase spontaneously breaks $\mathbb{Z}_4$ symmetry, but in a way that necessarily preserves $\mathbb{Z}_4^2$.  This is the maximal extent to which $\mathbb{Z}_4$ can be broken in a parafermion chain.  

Parafermions loosely exhibit a `self-dual structure' in that they arise from combinations of clock operators and their duals.  For a more precise statement consider the quantities
\begin{equation}
  e^{i\frac{\pi}{4}}\alpha_{2a-1}^\dagger \alpha_{2a} = \tau_a~,\quad e^{i \frac{\pi}{4}}\alpha_{2a}^\dagger\alpha_{2a+1} = \sigma_a^\dagger \sigma_{a+1}~.
\end{equation}
Duality swaps the role of the right-hand sides above, and hence implements a simple spatial translation of parafermion operators.

\subsection{Spinful fermions}
\label{fermions}

\begin{figure}
\includegraphics[width=0.8\columnwidth]{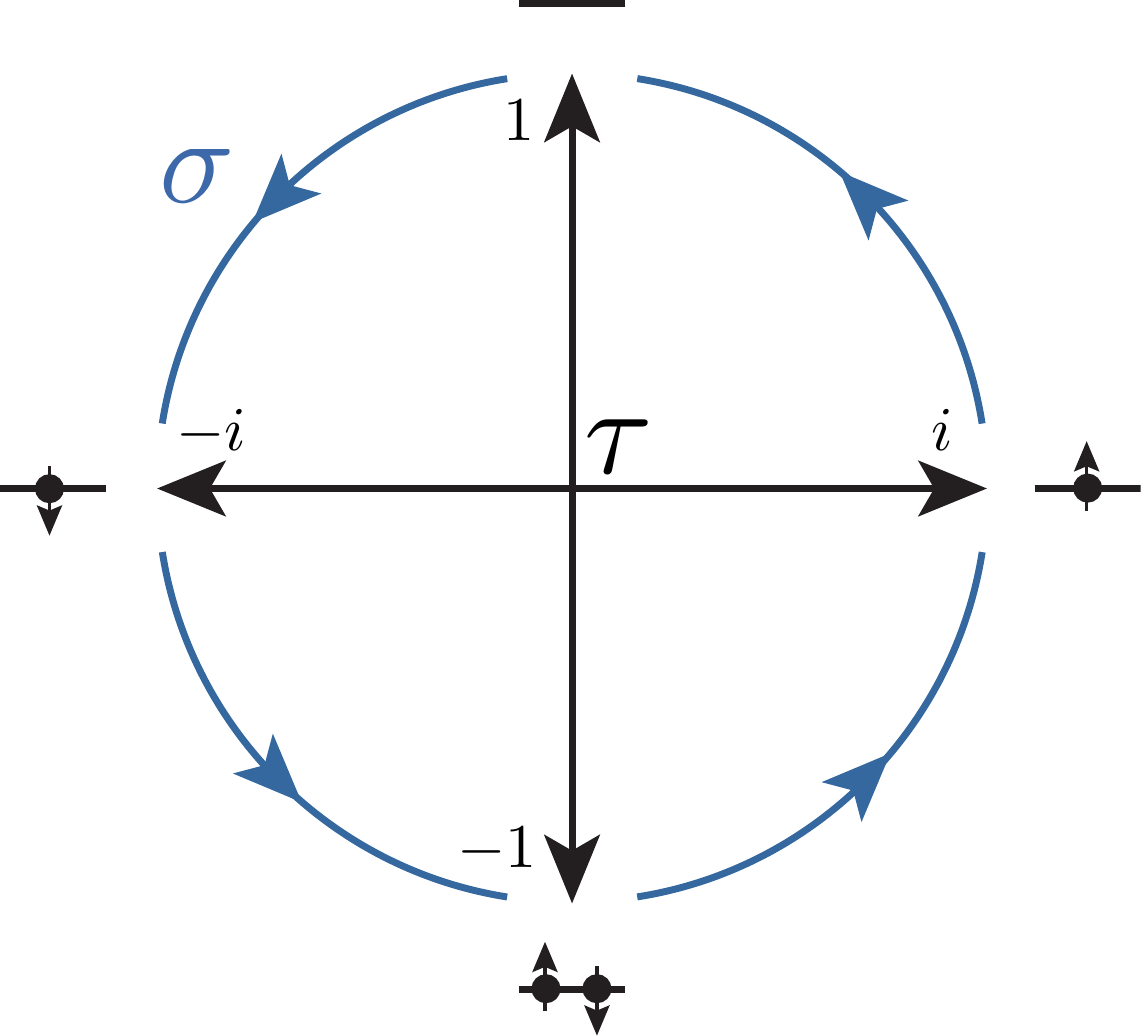}
\caption{Representation of $\mathbb{Z}_4$ clock-model operators in terms of spinful hard-core bosons.  Eigenstates of $\tau$ are encoded through boson number eigenstates, e.g., $\tau = +1$ is the boson vacuum while $\tau = -1$ corresponds to a state with both spins populated.  The operator $\sigma$ cycles through $\tau$ eigenstates and hence adds and removes bosons in a state-dependent fashion.  }
\label{clock_fig}
\end{figure}

In the previous subsection we saw that parafermionic commutation relations [Eq.~\eqref{alpha_commutator} or \eqref{alphap_commutator}] emerge upon combining the bosonic operator $\sigma$ with a string of $\tau$'s or $\tau^\dagger$'s.  `Doubling' the string as sketched in Fig.~\ref{ops_fig}(b)---i.e., attaching $\tau^2$'s to clock operators---instead naturally generates objects with fermionic statistics.  Since the doubled string is Hermitian, the freedom that led to multiple parafermion representations no longer exists here.  Recovering the full clock Hilbert space with four states per site, however, requires that the fermions carry an internal label that is profitably viewed as an electronic spin-1/2 degree of freedom.  

As a first step to formalizing this heuristic picture, we introduce spinful hard-core \emph{bosons} $b_{a,\uparrow}$ and $b_{a,\downarrow}$. 
Observe that one can decompose the $\tau_a$ clock operator via
\begin{equation}
  \tau_a = e^{i \frac{\pi}{2}(n_{a,\uparrow}-n_{a,\downarrow} + 2 n_{a,\uparrow}n_{a,\downarrow})},
  \label{tau_expansion}
\end{equation}
where $n_{a,\alpha} = b_{a,\alpha}^\dagger b_{a,\alpha}$ denote boson occupation numbers. In this representation $\tau_a = +1$ corresponds to the boson vacuum.  Starting from this state, adding a spin-down boson yields $\tau_a = -i$, further adding a spin-up boson yields $\tau_a = -1$, removing the spin-down boson gives $\tau_a = +i$, and finally removing the spin-up boson returns the $\tau_a = +1$ state.  
This sequence of $\tau_a$ windings is implemented by the conjugate clock operator $\sigma_a$ as Fig.~\ref{clock_fig} illustrates \footnote{This decomposition of $\sigma_a$ and $\tau_a$ in terms of hard-core bosons is not unique.  We could have instead expressed $\sigma_a$ in terms of boson densities and $\tau_a$ in terms of creation and annihilation operators that cycle $\sigma_a$ eigenvalues.  The latter parametrization is problematic, however, in that $\mathbb{Z}_4$-symmetric terms such as $-f(\tau_a + \tau_a^\dagger)$ become nonlocal upon fermionization (in contrast to our conventions, where such terms remain local).  }.  
To express $\sigma_a$ in terms of bosons it is convenient to introduce operators $P_\alpha(0) = 1-n_{a,\alpha}$ and $P_\alpha(1) = n_{a,\alpha}$ that respectively project onto the subspace with occupation numbers 0 and 1 for spin $\alpha$.  From Fig.~\ref{clock_fig} we see that 
\begin{align}
  \sigma_a &= b_{a,\downarrow}^\dagger P_\uparrow(0) P_\downarrow(0) + b_{a,\uparrow}^\dagger P_\uparrow(0) P_\downarrow(1) 
  \nonumber \\
  &+ b_{a,\downarrow} P_\uparrow(1) P_\downarrow(1) + b_{a,\uparrow} P_\uparrow(1) P_\downarrow(0)
  \label{sigma_expansion}
  \\
  &= (b_{a,\downarrow}^\dagger + b_{a,\uparrow}) + (b_{a,\uparrow}^\dagger - b_{a,\uparrow}) n_{a,\downarrow} + (b_{a,\downarrow}-b_{a,\downarrow}^\dagger) n_{a,\uparrow}~.
  \nonumber
\end{align}

As described in Appendix~\ref{Inversion_Appendix}, Eqs.~\eqref{tau_expansion} and \eqref{sigma_expansion} can be inverted to yield
\begin{align}
  b_{a,\uparrow} &= \left[\sigma_a\frac{1-\tau_a^2}{4}+H.c.\right] +i\left[\sigma_a\frac{\tau_a^\dagger-\tau_a}{4}+H.c.\right]
  \label{b_up}
  \\
  b_{a,\downarrow} &= \left[\frac{1-\tau_a^2}{4}\sigma_a+H.c.\right] + i\left[\frac{\tau_a^\dagger-\tau_a}{4}\sigma_a+H.c.\right].
  \label{b_down}
\end{align}
We can now define spinful fermions
\begin{align}
  f_{a,\uparrow} &= e^{-i \frac{\pi}{4}} S_a b_{a,\uparrow}
  \label{f_up}
  \\
  f_{a,\downarrow} &= e^{-i \frac{\pi}{4}} S_a e^{i \pi n_{a,\uparrow}} b_{a,\downarrow}.
  \label{f_down}
\end{align}
The $e^{-i\frac{\pi}{4}}$ phases are introduced for later convenience, the factor $e^{i \pi n_{a,\uparrow}}$ in Eq.~\eqref{f_down} enforces anticommutation of spin-up and spin-down fermions on the same site \footnote{More generally, we could have inserted factors $e^{i \theta n_{a,\uparrow}}$ in Eq.~\eqref{f_up} and $e^{i (\theta+\pi)n_{a,\downarrow}}$ in Eq.~\eqref{f_down} to maintain on-site anticommutation.  The choice $\theta = 0$ that we adopted is particularly convenient for symmetries.}, and 
\begin{equation}
  S_a = e^{i \pi \sum_{b<a}(n_{b,\uparrow} + n_{b,\downarrow})} = \prod_{b<a}\tau_b^2 = \mu^2_{a-\frac{1}{2}}
  \label{string}
\end{equation}
is a Jordan-Wigner string that ensures off-site anticommutation.  Note the `doubled' string relative to the $\alpha_a$ operators, consistent with our heuristic picture above.  

Appendix \ref{symmetries} derives the action of $\mathbb{Z}_4, \mathcal{T}$, and $\mathcal{C}$ on the fermions; see Table~\ref{symmetry_table} for a summary.  With our conventions all three symmetries act nontrivially, in the sense that the fermions acquire a phase factor dependent on the occupation of the opposite spin species.  Combinations of these symmetries nevertheless correspond to familiar operations.  First, the generator $Q$ of $\mathbb{Z}_4$ symmetry squares to
\begin{equation}
  Q^2 = \prod_a \tau_a^2 = e^{i \pi\sum_a( n_{a,\uparrow} + n_{a,\downarrow})} = {\text{fermion parity}}.
  \label{parity}
\end{equation}
Thus $\mathbb{Z}_4^2$ sends $f_{a,\alpha} \rightarrow -f_{a,\alpha}$ and represents global fermion parity conservation---which can be broken neither explicitly nor spontaneously in a system of physical fermions.  By contrast, $\mathbb{Z}_4$ itself \emph{can} be readily broken (explicitly or spontaneously) provided $\mathbb{Z}_4^2$ remains intact.  Table~\ref{Z4_breaking_table} summarizes the varying robustness of $\mathbb{Z}_4$ symmetry in the clock, parafermionic, and fermionic representations. 

\begin{table}
\begin{center}
 \setlength\extrarowheight{2pt}
 \begin{tabular}{c | c | c | c} 
  & clock & parafermion & spinful fermion 
  \\ 
 \hline
 $\mathbb{Z}_4$ breakable & yes & no & yes \\ 
 explicitly? & & (locality) & 
 \\
 \hline
 $\mathbb{Z}_4$ breakable & yes & yes & yes \\ 
 spontaneously? & & ($\langle \alpha^2_a \rangle \neq 0$) & 
 \\
 \hline
 $\mathbb{Z}_4^2$ breakable & yes & no & no \\ 
 explicitly? & & (locality) & (locality)
 \\
 \hline
 $\mathbb{Z}_4^2$ breakable & yes & no & no \\ 
 spontaneously? & & (statistics) & (statistics)
 \\
 \hline
\end{tabular}
\end{center}
\caption{Comparison of $\mathbb{Z}_4$-symmetry robustness in various representations.  For the case of spinful fermions, the locality and statistics conditions listed in the right column reduce to the familiar statement that fermion-parity conservation can be broken neither spontaneously nor explicitly.}
\label{Z4_breaking_table}
\end{table}

Second, $\mathcal{T}_{\rm elec} \equiv \mathbb{Z}_4 \mathcal{T}$ acts according to
\begin{equation}
  \mathcal{T}_{\rm elec} f_{a,\alpha} \mathcal{T}^{-1}_{\rm elec} = i\sigma^y_{\alpha \beta} f_{a,\beta};
  \label{Telec}
\end{equation}
here and below $\sigma^{x,y,z}$ denote the usual Pauli matrices \footnote{We inserted the factors $e^{-i\frac{\pi}{4}}$ in Eqs.~\eqref{f_up} and \eqref{f_down} simply to recover the familiar form of electronic time-reversal in Eq.~\eqref{Telec}; without these factors the $i$ on the right side would be absent.}.  One can recognize $\mathcal{T}_{\rm elec}$ as electronic time-reversal symmetry that satisfies $\mathcal{T}_{\rm elec}^2 = -1$ when acting on single-particle states.  Third, $U_{\rm spin} \equiv \mathbb{Z}_4 \mathcal{C}$ corresponds to a $\pi$ spin rotation, i.e.,
\begin{equation}
  U_{\rm spin} f_{a,\alpha} U_{\rm spin}^\dagger = \sigma^y_{\alpha \beta} f_{a,\beta}.
  \label{Uspin}
\end{equation}

The set $\mathcal{T}_{\rm elec}, U_{\rm spin}$, and $\mathbb{Z}_4$ provides a convenient basis of symmetries in the fermionic representation.  While $\mathbb{Z}_4$ generally acts nontrivially on the fermions, a simplification is possible in the low-density limit where $\langle n_{a,\alpha} \rangle \ll 1$.  Here one can approximate $\mathbb{Z}_4$ by dropping the density-dependent phases acquired by the fermions.  The resulting operation, which we label $\overline{\mathbb{Z}}_4$, yields a simpler transformation
\begin{equation}
  \overline Q f_{a,\alpha} \overline Q^\dagger =  i \sigma^z_{\alpha\beta} f_{a,\beta}~, {\text{(low-density approx.~of $\mathbb{Z}_4$)}}
\end{equation} 
that represents $\pi$ spin rotation about a different axis.  Symmetry transformations under $\mathcal{T}_{\rm elec}, U_{\rm spin}$, and $\overline{\mathbb{Z}}_4$ appear in Table~\ref{symmetry_table2}.  

\begin{table}
\begin{center}
 \setlength\extrarowheight{2pt}
 \begin{tabular}{|c | c | c | c|} 
 \hline
\rule{0pt}{2.5ex}     & $\mathcal{T}_{\rm elec} = \mathbb{Z}_4 \mathcal{T}$ & $U_{\rm spin} = \mathbb{Z}_4 \mathcal{C}$ & $\overline{\mathbb{Z}}_4$ \\ [0.05ex] 
 \hline\hline
 $f\rightarrow$ & $i \sigma^y f$ & $\sigma^y f$ & $i\sigma^z f$
 \\
 \hline
\end{tabular}
\end{center}
\caption{Action of composite symmetries $\mathcal{T}_{\rm elec}$ and $U_{\rm spin}$ along with $\overline{\mathbb{Z}}_4$ on spinful fermions.  Remarkably, $\mathcal{T}_{\rm elec}$ implements electronic time-reversal symmetry with $\mathcal{T}_{\rm elec}^2 = -1$ while $U_{\rm spin}$ implements a $\pi$ spin rotation.  In the last column $\overline{\mathbb{Z}}_4$ is an approximation of the exact $\mathbb{Z}_4$ symmetry (see Table~\ref{symmetry_table}) valid in the low-fermion-density limit; this operation implements a $\pi$ spin rotation about a different axis.}
\label{symmetry_table2}
\end{table}

\subsection{Dual fermions}
\label{dualfermions}

One can of course straightforwardly generalize Eqs.~\eqref{b_up} through \eqref{string} to instead fermionize the \emph{dual} representation of the clock model.  To this end we first define dual hard-core bosons 
\begin{align}
  \tilde b_{\tilde a,\uparrow} &= \left[\mu_{\tilde a}\frac{1-\nu_{\tilde a}^2}{4}+H.c.\right] +i\left[\mu_{\tilde a}\frac{\nu_{\tilde a}^\dagger-\nu_{\tilde a}}{4}+H.c.\right]
  \label{tilde_b_up}
  \\
  \tilde b_{\tilde a,\downarrow} &= \left[\frac{1-\nu_{\tilde a}^2}{4}\mu_{\tilde a}+H.c.\right] +i\left[\frac{\nu_{\tilde a}^\dagger-\nu_{\tilde a}}{4}\mu_{\tilde a}+H.c.\right],
  \label{tilde_b_down}
\end{align}
where $\tilde a = a+\frac{1}{2}$ labels dual-lattice sites.  Dual fermions are then given by
\begin{align}
  \tilde f_{\tilde a,\uparrow} &= e^{-i \frac{\pi}{4}} \tilde S_{\tilde a} \tilde b_{\tilde a,\uparrow}~,
  \label{tilde_f_up}
  \\
  \tilde f_{\tilde a,\downarrow} &= e^{-i \frac{\pi}{4}} \tilde S_{\tilde a} e^{i \pi \tilde n_{\tilde a,\uparrow}} \tilde b_{\tilde a,\downarrow}~,
  \label{tilde_f_down}
\end{align}
with 
\begin{equation}
  \tilde S_{\tilde a} = e^{i \pi \sum_{\tilde b<\tilde a}(\tilde n_{\tilde b,\uparrow} + \tilde n_{\tilde b,\downarrow})} = \prod_{\tilde b<\tilde a}\nu_{\tilde b}^2 = \sigma^2_{a}\sigma^2_{-\infty}~.
  \label{tilde_string}
\end{equation}
Clock-model duality [Eq.~\eqref{duality}] non-locally transforms our original spinful fermions $f_{a,\alpha}$ into these dual fermions $\tilde f_{\tilde a,\alpha}$. The situation should be contrasted to the para\-fermion representation, where duality merely implements a spatial translation.  It is also worth contrasting to the Majorana-fermion representation of the Ising model, where Ising duality similarly corresponds to a spatial translation of the Majorana operators (as opposed to non-locally mapping to a new set of fermions).

The clock-operator fermionization described so far allows one to directly express lattice $\mathbb{Z}_4$ parafermions as non-local combinations of either fermions or dual fermions.  Interestingly, it is also possible to express parafermions in terms of a \emph{local} product of fermions and dual fermions---reflecting the roughly self-dual nature of the parafermion operators alluded to earlier.  The latter form resembles the factorization identified in Ref.~\onlinecite{Meidan} of $\mathbb{Z}_4$ parafermions into two sets of fermions that exhibit nontrivial commutation relations with one another.  We relegate explicit expressions linking parafermions and fermions to Appendix~\ref{explicit_map} (see also Sec.~\ref{long_wavelength_limit}).  

\subsection{Spin-1/2 representation and alternative fermionization schemes}
\label{sec.altferm}

There are numerous alternative mappings that relate $\mathbb{Z}_4$ clock operators to spin-1/2 or fermionic degrees of freedom. Among these, different choices may be convenient for revealing particular properties. This section briefly outlines an approach that yields the same spinful fermion operators as Sec.~\ref{fermions}, but through a very different route. Appendices \ref{app.spin} and \ref{app.altferm} present additional details about this mapping and several other schemes, including that of Refs.~\onlinecite{Kohmoto,Kohmoto2}. 

We begin by expressing the clock operators $\sigma_a,\tau_a$ in terms of spin-1/2 degrees of freedom via \cite{Kohmoto,Kohmoto2}
\begin{align}
&\sigma_a = \frac{1+i}{2}\left(s^z_{a+\frac{1}{4}}+ i  s^z_{a - \frac{1}{4}}\right)~,\\
&\tau_a=\frac{1}{2} \left(s^x_{a+\frac{1}{4}} + s^x_{a-\frac{1}{4}}\right)+\frac{1}{2}\left(s^x_{a+\frac{1}{4}} - s^x_{a-\frac{1}{4}}\right) s^z_{a+\frac{1}{4}} s^z_{a-\frac{1}{4}}~,
\end{align}
where $s^{x,y,z}$ denote Pauli matrices that reside at sites $a \pm \frac{1}{4}$.  
Next, we perform the familiar Ising-model duality mapping that trades in these variables for dual spins $t^{x,y,z}$ living on integer as well as half-integer sites,
\begin{align}&t^x_a =s^z_{a-\frac{1}{4}}s^z_{a+\frac{1}{4}}~, \qquad
&t^z_a =\prod_{a'<a}s^x_{a'}~.\label{Ising.duality}
\end{align} 
`Exchange' and `transverse-field' clock-model couplings take on a particular simple form in this language: 
 \begin{align}
\begin{split}
 -J(\sigma_a^\dagger \sigma_{a+1} + H.c.) = -J\left(t^x_{a}t^x_{a+\frac{1}{2}}+t^x_{a+\frac{1}{2}}t^x_{a+1}\right)~,\\
  -f(\tau_a + \tau_a^\dagger)    = -f\left(t^z_{a-\frac{1}{2}}t^z_{a}+t^z_{a}t^z_{a+\frac{1}{2}}\right) \label{Clocktospin1}~,
\end{split}
\end{align}
and in particular precisely coincide with couplings in the 1D XY model.  
(References~\onlinecite{Kohmoto,Kohmoto2} used a somewhat different mapping to a spin-$1/2$ model as discussed in Appendix~\ref{app.altferm}.)
Since clock-model duality interchanges the $J$ and $f$ terms, Eqs.~\eqref{Clocktospin1} naively suggest that such a duality transformation is implemented as a global $\pi/2$ rotation of $t$ spins around the $y$ axis. We caution, however, that this interpretation only holds for specific Hamiltonians and is not dictated by conditions of symmetry and locality; see Appendix~\ref{app.spin}.

Let us now employ a Jordan-Wigner transformation to define complex fermions
\begin{align}
&\mathtt{c}_{a} = \frac{1}{2} (t^y_a - i t^z_a)\prod_{a'<a} t^x_{a'}~\label{eq.JW1}
\end{align}
and then introduce spinful fermions $\mathtt{d}_{a,\alpha}$ via a Bogoliubov transformation: 
\begin{align}
 \mathtt{d}_{a,\alpha}=&\frac{i}{\sqrt{8}}
\left(-\mathtt{c}_{a}-\mathtt{c}^\dagger_{a}-\mathtt{c}_{a+\frac{1}{2}}+\mathtt{c}^\dagger_{a+\frac{1}{2}}\right)\nonumber\\
&+\frac{ \alpha}{\sqrt{8}}
\left(-\mathtt{c}_{a-\frac{1}{2}}-\mathtt{c}^\dagger_{a-\frac{1}{2}}-\mathtt{c}_{a}+\mathtt{c}^\dagger_{a}\right)~.
\label{JWferm1}
\end{align}
On the right side, $\alpha = +1$ for spin up and $-1$ for spin down.  
Somewhat lengthy but straightforward algebra sketched in Appendix~\ref{app.altferm} reveals that a local canonical transformation,
\begin{align}
\mathtt{f}_{a,\alpha} &= e^{-i \frac{\pi}{4}(1+\alpha)}\exp\left(-i \frac{\pi}{2} \mathtt{d}_{a,-\alpha}^\dagger \mathtt{d}_{a,-\alpha}\right)\mathtt{d}_{a,\alpha}~,\label{JWferm2}
\end{align}
yields operators that are identical to $f_{a,\alpha}$ up to a boundary term that squares to unity.

An alternative set of fermions can be formed by defining $\tilde{\mathtt{c}}_a = U \mathtt{c}_a U^\dagger$, where $U$ implements a global $\pi/2$ spin rotation around $t^y$.  Note that $\mathtt{c}_a$ and $\tilde{\mathtt{c}}_a$ are nonlocally related---the Jordan-Wigner string consists solely of $t^x$ operators in the former but $t^z$ operators in the latter.  Since $U$ is precisely the spin rotation that swaps the two lines of Eq.~\eqref{Clocktospin1}, it is natural to expect that $\tilde{\mathtt{c}}_a$ fermions closely relate to the dual fermions $\tilde f_{a,\alpha}$ of Sec.~\ref{dualfermions}.  Let $\tilde{\mathtt{d}}_{a,\alpha}$ and $\tilde{\mathtt{f}}_{a,\alpha}$ denote spinful fermions defined analogously to Eqs.~\eqref{JWferm1} and \eqref{JWferm2}.  On the level of single-fermion operators, $\tilde f_{a,\alpha}$ and $\tilde{\mathtt{f}}_{a,\alpha}$ are related nonlocally.  Nevertheless, Hamiltonians for which clock-model duality corresponds to a spin rotation take on an identical form when expressed in terms of either set of operators, though this relation breaks down for more generic models.

\section{Mappings Between Phases}
\label{Phases}

\subsection{Hamiltonians}

The remainder of this paper primarily explores \emph{translationally invariant} fermionic phases and their clock/parafermion counterparts.  
All of the phases that we will discuss can be accessed microscopically from limits of (or in some cases weak perturbations to) the Hamiltonian
\begin{align}
H &= -J \sum_{a = 1}^{N-1} (\sigma_a^\dagger \sigma_{a+1} +\sigma_{a+1}^\dagger \sigma_a - \lambda \sigma_a^2\sigma_{a+1}^2) 
\nonumber \\
  &- f \sum_{a = 1}^N (\tau_a + \tau_a^\dagger - \lambda \tau_a^2)
  \label{AshkinTeller}
\end{align}
for an $N$-site clock chain.  Equation~\eqref{AshkinTeller} corresponds to the well-studied Ashkin-Teller model \cite{ATmodel}, which exhibits a variety of ordered and disordered gapped phases, novel critical points, and extended critical phases (see, e.g., Refs.~\onlinecite{Kohmoto,Alcaraz,Yang,Kohmoto2}).  Throughout we assume non-negative $J,f$ couplings and take open boundary conditions to highlight nontrivial edge physics that arises in certain regimes.  Since duality interchanges the $J$ and $f$ terms, the Hamiltonian is self-dual at $J = f$ for any $\lambda$. 

In terms of parafermions, the model becomes
\begin{align}
  H &= -J \sum_{a = 1}^{N-1} [(e^{i\frac{\pi}{4}} \alpha_{2a}^\dagger \alpha_{2a+1} + H.c.) + \lambda \alpha_{2a}^2 \alpha_{2a+1}^2] 
  \nonumber \\ 
  &-f \sum_{a = 1}^N [(e^{i\frac{\pi}{4}} \alpha_{2a-1}^\dagger \alpha_{2a} + H.c.) + \lambda \alpha_{2a-1}^2 \alpha_{2a}^2]~.
  \label{H_parafermion}
\end{align}
The first and second lines favor competing dimerization patterns for the parafermion operators.  

For spinful fermions it is useful to partition the Hamiltonian as $H = H_0 + H_\lambda$, where $H_\lambda$ contains the terms proportional to $\lambda$ in the Ashkin-Teller model.  Implicitly summing repeated spin indices and neglecting unimportant overall constants, $H_0$ can be expressed as
\begin{align}
H_0 =& -J \sum_{a = 1}^{N-1}\left( \hat t^{\alpha,\beta}_{a} f_{a,\alpha}^\dagger f_{a+1,\beta}+i \hat \Delta^{\alpha,\beta}_{a}f_{a,\alpha}^\dagger f_{a+1,\beta}^\dagger+H.c.\right)\nonumber \\
 &+ 2f \sum_{a = 1}^N f^\dagger_{a,\alpha}f_{a,\alpha}~.
 \label{H0}
\end{align}
The $f$ coupling simply yields a chemical potential for the fermions.  In the $J$ term, $\hat t^{\alpha,\alpha'}_{a}$ and $\hat \Delta^{\alpha,\alpha'}_{a}$ encode spin- and density-dependent hoppings and triplet pairings, respectively.  We explicitly have
\begin{align}
\begin{split}
  \hat t^{\alpha,\alpha}_{a} &= 1-n_{a,-\alpha} - n_{a+1,-\alpha}~,  \\
  \hat t^{\alpha,-\alpha}_{a} &= \alpha[2n_{a,-\alpha} n_{a+1,\alpha}-n_{a,-\alpha} -n_{a+1,\alpha}]~,\\
  \hat \Delta^{\alpha,\alpha}_{a} &= \alpha [n_{a,-\alpha} - n_{a+1,-\alpha}]~,  \\ 
  \hat \Delta^{\alpha,-\alpha}_{a} &= [n_{a,-\alpha} + n_{a+1,\alpha}-2n_{a,-\alpha} n_{a+1,\alpha} -1]~. 
  \end{split} \label{couplings} 
\end{align}
The $\lambda$ terms yield nontrivial four-fermion interactions:
\begin{align}
H_\lambda &= \lambda J \sum_{a = 1}^{N-1} (if_{a,\uparrow}^\dagger + f_{a,\uparrow})(f_{a,\downarrow}^\dagger + i f_{a,\downarrow})
\nonumber \\
&\qquad\qquad\times (if_{a+1,\uparrow}^\dagger + f_{a+1,\uparrow})(f_{a+1,\downarrow}^\dagger + i f_{a+1,\downarrow}) \nonumber \\
 &+ \lambda f \sum_{a = 1}^N (2n_{a,\uparrow} - 1)(2n_{a,\downarrow} - 1)~.
 \label{Hlambda}
\end{align}

\subsection{View from the long-wavelength limit}
\label{long_wavelength_limit}

It will prove exceedingly useful to obtain a bosonized description of $H$ that filters out all but the long-wavelength modes needed to describe the phases of interest.  To this end we focus on the spinful-fermion representation and assume the low-density limit $n_{a,\alpha} \approx 0$ where $\mathbb{Z}_4$ symmetry is well-approximated by $\overline{\mathbb{Z}}_4$.  Consider first the $\lambda = 0$ limit.  Upon retaining only the density-independent pieces from Eqs.~\eqref{couplings}, $H_0$ reduces to a free-fermion Hamiltonian
\begin{align}
  \overline{H}_0 &= -J \sum_{a = 1}^{N-1}\left(f_{a,\alpha}^\dagger f_{a+1,\alpha}-i f_{a,\alpha}^\dagger \sigma^x_{\alpha\beta}f_{a+1,\beta}^\dagger+H.c.\right)
  \nonumber \\
  &+  2f \sum_{a = 1}^N f^\dagger_{a,\alpha}f_{a,\alpha}.
  \label{barH0}
\end{align}
When $f = J$ the spectrum becomes gapless at zero momentum; low-energy excitations are captured by one right- and one left-moving fermion mode, $\psi_{R/L}$.  

A bosonized description of this critical point arises from the identification
\begin{align}
 \begin{split}
 i(f_\uparrow - f_\downarrow^\dagger) &\sim \psi_{R} \sim e^{i (\phi + \theta)}
 ~,\\
  f_\uparrow + f_\downarrow^\dagger &\sim \psi_{L} \sim e^{i (\phi - \theta)}~,
   \end{split}
  \label{bosonization}
\end{align}
where $\phi,\theta$ are continuum fields satisfying
\begin{equation}
  [\phi(x),\theta(x')] = i \pi \Theta(x'-x)~.
  \label{phi_theta_commutator}
\end{equation}
(Our bosonization recipe closely follows that employed by Ref.~\onlinecite{FisherBosonization}.)
For later use we note that $\partial_x\theta/\pi$ yields the spin density since 
\begin{equation}
  f_{\uparrow}^\dagger f_{\uparrow} - f_{\downarrow}^\dagger f_{\downarrow} \sim \psi_R^\dagger \psi_R + \psi_L^\dagger \psi_L \sim \partial_x\theta/\pi~,
  \label{spindensity}
\end{equation}
while
\begin{align}
 &e^{i\pi \sum_{a,\alpha} f_{a,\alpha}^\dagger f_{a,\alpha} }=e^{i\pi \sum_{a} [ f_{\uparrow}^\dagger f_{\uparrow} - f_{\downarrow}^\dagger f_{\downarrow}]} = e^{i\int_x \partial_x\theta}
  \label{ParityDef}  
\end{align}
correspondingly specifies the total fermion parity in a region of the chain.  

\begin{table}
\begin{center}
 \setlength\extrarowheight{2pt}
 \begin{tabular}{|c | c | c | c| c|} 
 \hline
  & $\mathbb{Z}_4$ & $\mathcal{C}$ & $\mathcal{T}$ \\ [0.05ex] 
 \hline\hline
 $\phi \rightarrow$ & $\phi + \pi/2$ & $-\phi$ & $\phi$ \\
 \hline
 $\theta \rightarrow$ & $\theta$ & $-\theta$ & $-\theta$ \\
 \hline 
\end{tabular}
\end{center}
\caption{Symmetry properties of bosonized fields used to construct long-wavelength expansions of clock operators, parafermions, and fermions. }
\label{symmetry_table_bosonized}
\end{table}

Table~\ref{symmetry_table_bosonized} catalogues symmetry properties of the bosonized fields inferred from Eq.~\eqref{bosonization}.  [Technically, Eq.~\eqref{bosonization} yield the action of $\overline{\mathbb{Z}}_4$ instead of $\mathbb{Z}_4$, though as we will see below this distinction is immaterial in the long-wavelength limit.  We caution, however, that Eq.~\eqref{bosonization} can be used to relate microscopic fermion operators to continuum fields only in the low-density limit; outside of this regime one must exploit symmetry to find the bosonized form of a given lattice operator.]  With these symmetries in hand we can deduce the low-energy expansion for operators in various other representations.  Order and disorder operators correspond to
\begin{equation}
  \sigma_a \sim e^{i \phi}~,\qquad\mu_{a+\frac{1}{2}} = \prod_{b<a+\frac{1}{2}} \tau_b \sim e^{-i \theta/2}~.
  \label{sigma_mu_expansion}
\end{equation}
Note that the right-hand sides not only yield consistent symmetry properties, but are also faithful to the clock-operator commutation relations.  Similarly expanding our two parafermion representations---which again arise from attaching either a string of $\tau$ or $\tau^\dagger$ to $\sigma$---gives
\begin{equation}
  \alpha_a \sim e^{i(\phi - \theta/2)}~,\qquad \alpha'_a \sim e^{i(\phi + \theta/2)}~.
  \label{alpha_expansion}
\end{equation}
As a useful sanity check, doubling the string yields precisely the continuum limit of spinful fermions derived in Eqs.~\eqref{bosonization}; cf.~the lattice picture provided in Sec.~\ref{fermions}.  

From a dual perspective, one essentially views $\mu$ as the elementary spin operator and $\sigma$ as the string.  The dual analogue of Eq.~\eqref{sigma_mu_expansion} is then
\begin{equation}
  \mu_{a+\frac{1}{2}} \sim e^{i \tilde \phi}~,\qquad\sigma_a \sigma_{-\infty}^\dagger= \prod_{b<a} \nu_{b+\frac{1}{2}} \sim e^{-i \tilde\theta/2}
\end{equation}
with $[\tilde \phi(x),\tilde \theta(x')] = i \pi \Theta(x'-x)$ as in Eq.~\eqref{phi_theta_commutator}.  Clearly the original continuum $\phi,\theta$ fields and their duals are related by 
\begin{equation}
  \tilde \phi(x) = -\theta(x)/2~,\qquad\tilde \theta(x) = -2[\phi(x)-\phi(-\infty)]~.
  \label{dual_relation}
\end{equation}
Attaching a string of $\nu$ or $\nu^\dagger$ to $\mu$ yields essentially the same long-wavelength limit of parafermion operators as before.  Doubling this string, however, generates the continuum limit of our dual fermions: 
\begin{equation}
 \tilde \psi_R \sim e^{i (\tilde \phi + \tilde \theta)}~,\qquad \tilde \psi_L \sim e^{i (\tilde \phi - \tilde \theta)}~.
\end{equation}
In Sec.~\ref{fermions} we noted that parafermions can be expressed as local combinations of fermions and dual fermions on the lattice.  This relation becomes particularly simple in the long-wavelength limit.  Using Eq.~\eqref{dual_relation} one immediately obtains
\begin{equation}
  \alpha_a \sim \psi_R^\dagger \tilde \psi_L^\dagger~,\qquad\alpha_a' \sim \psi_L^\dagger \tilde \psi_R~,
  \label{fermion_dualfermion}
\end{equation}
very similar to Ref.~\onlinecite{Meidan}.

Returning to the critical Hamiltonian, the bosonized form of Eq.~\eqref{barH0} reads $\overline{\mathcal{H}}_0 = \int_x\frac{v_0}{2\pi}[(\partial_x \phi)^2 + (\partial_x\theta)^2]$ with $v_0 \propto J$.  
Turning on $\lambda \neq 0$ and resurrecting interaction terms from $H_0$ that were neglected in Eq.~\eqref{barH0} generically modifies the low-energy Hamiltonian to 
\begin{align}
  \mathcal{H} = \int_x \bigg{\{}&\frac{v}{2\pi}[g(\partial_x \phi)^2 +g^{-1} (\partial_x\theta)^2] 
  \nonumber \\
  &- \kappa_1 \cos(4\phi)- \kappa_2 \cos(2\theta)\bigg{\}}~.\label{eqn.abel.boson}
\end{align}
Here $v$ is a renormalized velocity; $g$ is the Luttinger parameter characterizing the interaction strength ($g = 1$ corresponds to free fermions, while $g <1$ and $g>1$ respectively indicate repulsive and attractive interactions); and the $\kappa_{1,2}$ terms are the leading harmonics consistent with symmetries and locality.  Effective Hamiltonians of this form have been studied in related contexts in Refs.~\onlinecite{ZhangKane,Orth,ScriptaPoorMansParafermions,PoorMansParafermions}.  We can appeal to self-duality of the microscopic Hamiltonian at $J = f$ to further constrain $\mathcal{H}$.  In particular, here the continuum Hamiltonian must take the same form in terms of either $\phi,\theta$ or their duals $\tilde \phi,\tilde \theta$.  Using Eq.~\eqref{dual_relation} we thus obtain $\kappa_1 = \kappa_2$ and $g = 2$. The latter constraint guarantees that the two cosines---which swap under duality---are both marginal at the self-dual critical point.  Upon rescaling $\phi \rightarrow \phi/\sqrt{2}$ and $\theta \rightarrow \sqrt{2} \theta $, $\mathcal{H}$ maps onto one of the manifestly self-dual theories analyzed in Ref.~\onlinecite{LECHEMINANT}. There, non-Abelian bosonization techniques showed that the self-dual model exhibits a `hidden' continuous $U(1)$ symmetry.  

Breaking self-duality spoils these relations and can drive the system into various possible gapped phases that we explore next, both from a continuum and microscopic viewpoint.  The phases that arise depend sensitively on the signs of $\kappa_{1}$ and $\kappa_2$.  In the $\lambda = 0$ limit we must have $\kappa_1,\kappa_2>0$ so that the familiar ferromagnetic and paramagnetic phases of the clock model are `nearby' (see below).  We will show, however, that turning on $\lambda$ provides access to phases driven by $\kappa_1,\kappa_2<0$ as well.

\begin{figure*}
\includegraphics[width=2\columnwidth]{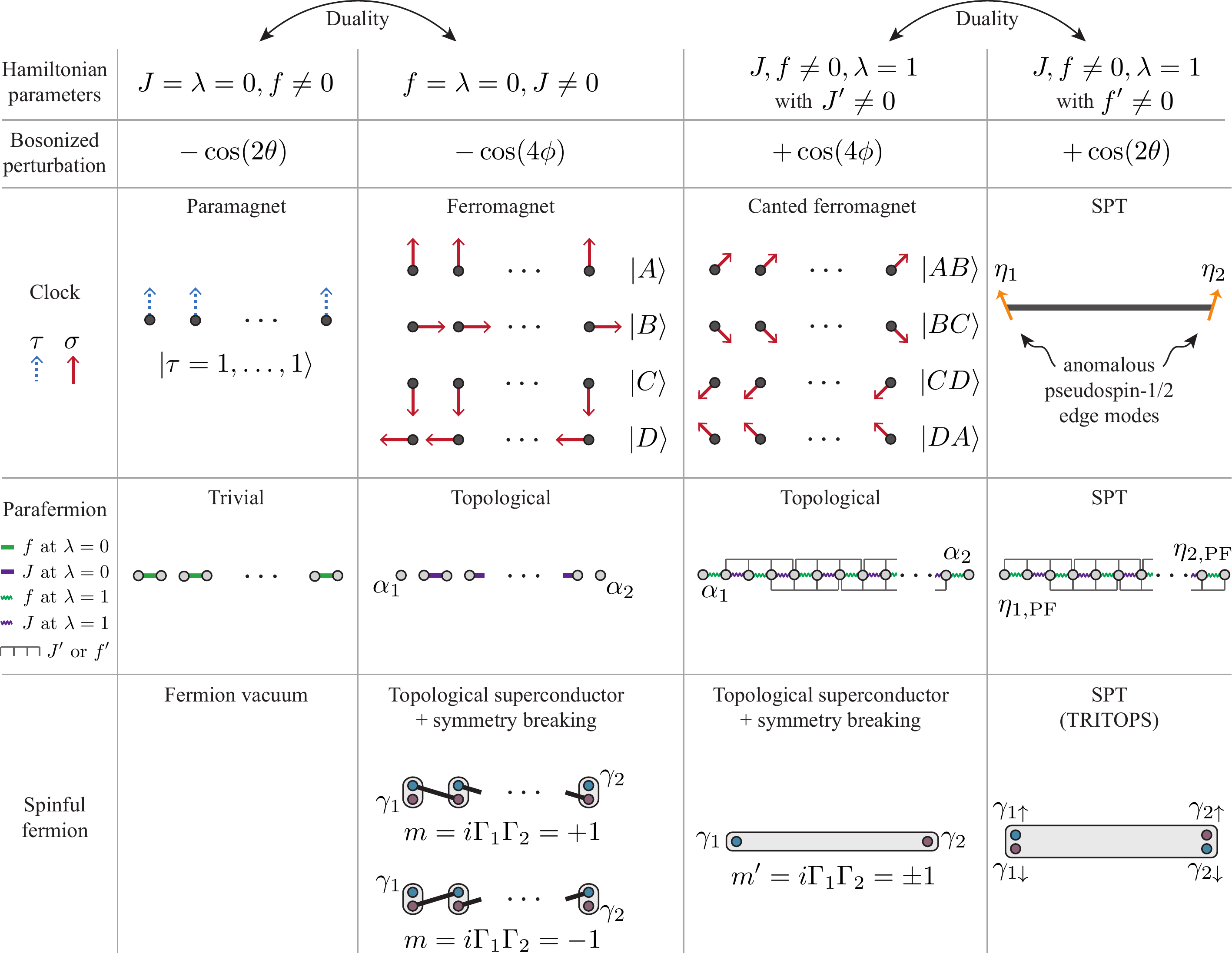}
\caption{Correspondence between gapped phases in the clock, $\mathbb{Z}_4$ parafermion, and spinful fermion representations.  The first and second rows respectively indicate the microscopic Hamiltonian parameters and associated bosonized perturbations that generate the phases summarized in each column.   Phases in the first and second columns are dual to one another, as are the phases in the third and fourth columns.  }
\label{Phases_fig}
\end{figure*}

\subsection{Phases driven by $\kappa_2>0$}  
\label{trivial_phase}

With relevant $\kappa_2>0$ the $\cos(2\theta)$ term pins $\theta$ to $0$ modulo $\pi$.  In terms of clock spins, the disorder operator then condenses ($\langle \mu\rangle \neq 0$), yielding a trivial paramagnet.  
Microscopically, the paramagnetic state arises most simply from the Ashkin-Teller model at $J = \lambda = 0$, where the unique ground state is $|\tau = 1,\ldots,1\rangle$.  One sees from Eq.~\eqref{H_parafermion} that the corresponding parafermion system dimerizes in a trivial manner that gaps out the entire chain, including the ends.  Finally, according to Eq.~\eqref{H0} spinful fermions realize the vacuum with no fermions present.  The first column of Fig.~\ref{Phases_fig} summarizes the properties of this regime in all three representations.  

\subsection{Phases driven by $\kappa_1>0$}  
\label{FM_phase}

When $\kappa_1$ is relevant and positive, the $\cos(4\phi)$ term pins $\phi$ to 0 modulo $\pi/2$.  Implications of the pinning depend strongly on which degrees of freedom are regarded as physical. According to Eq.~\eqref{sigma_mu_expansion}, a system of clock spins spontaneously breaks $\mathbb{Z}_4$ symmetry and realizes a four-fold-degenerate ferromagnetic state characterized by the local order parameter $\langle \sigma \rangle = \pm 1$ or $\pm i$.  Such ferromagnetic order can be accessed straightforwardly from the $f = \lambda = 0$ limit of the Ashkin-Teller model, which admits broken-symmetry ground states 
\begin{align}
  &|A\rangle = |\sigma = 1,\ldots,1\rangle~,\quad&&|B\rangle = |\sigma = i,\ldots,i \rangle~,
  \nonumber \\
  &|C\rangle = |\sigma = -1,\ldots,-1\rangle~,\quad&&|D\rangle = |\sigma = -i,\ldots,-i \rangle~ .
  \label{FMstates}
\end{align}

A parafermion chain, by contrast, realizes the topological phase introduced by Fendley \cite{Fendley:2012}.  From Eq.~\eqref{H_parafermion} and Fig.~\ref{Phases_fig} one sees that at $f = \lambda = 0$ the parafermions dimerize in a pattern that gaps the interior but leaves behind an `unpaired' zero-energy mode at each edge.  These parafermion zero modes encode a four-fold degeneracy that can not be lifted by any perturbation that is local from the parafermion viewpoint.  Physical ground states in this representation correspond to $\mathbb{Z}_4$-preserving Schr\"odinger-cat superpositions of clock states defined in Eq.~\eqref{FMstates}.

\begin{figure}[h!]
\includegraphics[width=\columnwidth]{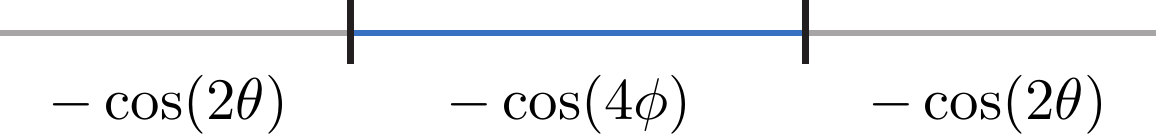}
\caption{Domain configuration used to extract zero-mode operators from the bosonized theory. }
\label{DomainWall_fig}
\end{figure}

Although the parafermion zero-mode operators are easily identified from the microscopic Hamiltonian, it is instructive to recover their form also from the low-energy bosonized point of view.  Figure~\ref{DomainWall_fig} sketches a domain configuration in which trivial phases gapped by $-\cos(2\theta)$ (recall Sec.~\ref{trivial_phase}) flank a central region gapped by $-\cos(4\phi)$.  For compactness we choose a gauge where $\theta$ pins to $0$ in the left domain, but parametrize $\phi = \pi \hat a/2$ in the central domain and $\theta = \pi \hat b$ in the right domain.  Here $\hat a, \hat b$ are integer-valued operators that obey the commutator $[\hat a,\hat b] = 2i/\pi$ inherited from Eq.~\eqref{phi_theta_commutator}.  Using Eq.~\eqref{alpha_expansion}, parafermion operators acting at the left and right domain walls respectively project to
\begin{equation}
  \alpha_1 = e^{i \frac{\pi}{2} \hat a}~,\qquad\alpha_2 = e^{i \frac{\pi}{2}(\hat a - \hat b)}~, 
  \label{alphas_bosonized}
\end{equation}
which are the continuum counterpart of the lattice parafermion zero modes.

A system of spinful fermions splits the difference between the clock and parafermion realizations: half of the degeneracy has a topological origin, while the other half is encoded in the local order parameter 
\begin{equation}
  m \equiv \langle i \psi_R \psi_L + H.c. \rangle  \sim \langle \cos(2\phi)\rangle = \pm 1~,
  \label{mdef}
\end{equation}
signaling spontaneous breaking of electronic time-reversal $\mathcal{T}_{\rm elec}, U_{\rm spin}$, and $\mathbb{Z}_4$.  
Similar phases have been captured previously in both 1D systems \cite{Stoudenmire,Pedder,PoorMansParafermions}---most notably Fe chains proximitized by a Pb superconductor \cite{Choy,Yazdani,Nadj-Perge,Ruby,Pawlak,Feldman,Jeon}---and proximitized quantum-spin-Hall edges \cite{ZhangKane,Orth}.  
Even at $f = \lambda = 0$, the surviving pieces of the microscopic fermion Hamiltonian in Eq.~\eqref{H0} appear nontrivial due to the interactions implicit in the $J$ term.  (Recall the density dependence in $\hat t, \hat \Delta$.)  In terms of dual fermions $\tilde f_{a,\alpha}$, the $f = \lambda = 0$ model is of course quadratic. Changing from fermions to dual fermions, however, requires a non-local change of basis. Alternatively, one can tame these interactions with a judicious \textit{local} basis change,
\begin{align}
  f_{a,\uparrow} &= \frac{e^{-i\frac{\pi}{4}}}{2}(c_a + c_a^\dagger + d_a-d_a^\dagger)~,
  \label{f_up_transformation}
  \\
  f_{a,\downarrow} &= \frac{e^{-i\frac{\pi}{4}}}{2}(d_a + d_a^\dagger + c_a-c_a^\dagger)~,
  \label{f_down_transformation}
\end{align}
where $c_a, d_a$ are canonical fermions with symmetry properties given in Table~\ref{symmetry_table3}.  In this basis the Hamiltonian becomes
\begin{align}
  H_{f = \lambda = 0} &= -J \sum_{a = 1}^{N-1}(m_a c_a^\dagger + c_a)(c_{a+1} - m_{a+1} c_{a+1}^\dagger) + H.c. \nonumber \\ 
  \label{Hf}
\end{align}  
with
\begin{equation}
  m_a = e^{i \pi d_a^\dagger d_a} = -f_a^\dagger \sigma^x f_a + (i f_{a,\uparrow}^\dagger f_{a,\downarrow}^\dagger + H.c.)
  \label{m_lattice}
\end{equation}
operators that commute with the Hamiltonian for any $a$ [see Appendix~\ref{explicit_map} for an alternate derivation of Eq.~\eqref{Hf}].  Note that in clock language we have $m_a = \sigma_a^2$.  

By symmetry, $m_a$ is the lattice analogue of the continuum order parameter in Eq.~\eqref{mdef}.  We note that one cannot obtain this microscopic order parameter by using Eqs.~\eqref{bosonization} in conjunction with Eq.~\eqref{mdef} because the former relations holds only in the low-density limit, which is not relevant here; recall the discussion below Eq.~\eqref{ParityDef}. In terms of the original spinful fermions, $m_a$ receives contributions from the magnetization along $x$ and singlet pairing with an imaginary coefficient---both of which share common symmetry properties.  For simplicity we will refer to $m_a$ as just `magnetization' in what follows.  The energy is minimized by choosing either $m_a = +1$  or $-1$ uniformly across the entire chain.  Focusing on such uniform configurations and replacing $m_a\rightarrow m$, the Hamiltonian further simplifies to
\begin{align}
  H_{f = \lambda = 0} \rightarrow -2J \sum_{a = 1}^{N-1}(m c_a^\dagger + c_a)(c_{a+1} - m c_{a+1}^\dagger)~.
  \label{Hf_limit}
\end{align}  

\begin{table}
\begin{center}
 \setlength\extrarowheight{2pt}
 \begin{tabular}{|c | c | c | c|} 
 \hline
\rule{0pt}{2.5ex}     & $\mathcal{T}_{\rm elec} = \mathbb{Z}_4 \mathcal{T}$ & $U_{\rm spin} = \mathbb{Z}_4 \mathcal{C}$ & $\mathbb{Z}_4$ \\ [0.05ex] 
 \hline\hline
 $c\rightarrow$ & $i c^\dagger$ & $i c^\dagger$ & $i e^{i \pi d^\dagger d} c^\dagger$
 \\
 \hline
 $d\rightarrow$ & $-i d^\dagger$ & $-i d^\dagger$ & $-i e^{i \pi c^\dagger c} d^\dagger$
 \\
 \hline
 $ \gamma_1 \rightarrow$ & $m \gamma_1$ & $\gamma_1$ & $m \gamma_1$
 \\
 \hline
 $ \gamma_2 \rightarrow$ & $m \gamma_2$ & $-\gamma_2$ & $-m \gamma_2$
 \\
 \hline
 $ \Gamma_1 \rightarrow$ & $p \Gamma_1$ & $\Gamma_1$ & $p \Gamma_1$
 \\
 \hline
 $ \Gamma_2 \rightarrow$ & $p \Gamma_2$ & $-\Gamma_2$ & $-p \Gamma_2$
 \\
  \hline
\end{tabular}
\end{center}
\caption{Symmetry properties for the microscopic fermions $c_a, d_a$ defined through the basis change in Eqs.~\eqref{f_up_transformation} and \eqref{f_down_transformation}.  The middle two lines summarize the transformations for the symmetry-enriched Majorana zero mode operators [Eq.~\eqref{gamma1} and \eqref{gamma2}] that arise in the fermionic representation of the Ashkin-Teller model at $f = \lambda = 0$.  The quantity $m = i \Gamma_1\Gamma_2 = \pm 1$, which is odd under all three symmetries in the table, is the order parameter whose condensation catalyzes the topological phase.  Finally, the last two lines list the transformations for $\Gamma_{1,2}$.  The factor $p = i \gamma_1 \gamma_2$ is required to preserve anticommutation between $\Gamma_j$ and $\gamma_j$.  }
\label{symmetry_table3}
\end{table}

Equation~\eqref{Hf_limit} can be recognized as the trivially solvable limit of the Kitaev chain in the topological phase \cite{Kitaev:2001}, but with one crucial distinction:  In our case the model arose from \emph{spontaneous} breaking of symmetries, most notably electronic time reversal.  Consequently, the phase of matter realized here is distinct from that of the Kitaev chain.  (See, e.g., Ref.~\onlinecite{ChenSPT2} for a general discussion of the classification of short-range entangled phases with spontaneous symmetry breaking.) The Hamiltonian supports `symmetry-enriched edge Majorana zero modes' described by 
\begin{align}
  \gamma_1 &= e^{i \frac{\pi}{4}(m + 1)}c_1^\dagger + e^{-i \frac{\pi}{4}(m + 1)}c_1~,
  \label{gamma1}
  \\
  \gamma_2 &= e^{i \frac{\pi}{4}(m - 1)}c_N^\dagger + e^{-i \frac{\pi}{4}(m - 1)}c_N~,
  \label{gamma2}
\end{align}  
whose form depends on the magnetization order parameter.  These zero modes satisfy the usual Majorana algebra $\gamma_i^2 = 1, \gamma_i = \gamma_i^\dagger$, and $\{\gamma_1,\gamma_2\} = 0$, but transform nontrivially under electronic time-reversal symmetry,
\begin{equation} 
  \mathcal{T}_{\rm elec}: \gamma_j \rightarrow m \gamma_j,
  \label{gamma_transformation}
\end{equation}
reflecting the intertwined symmetry-breaking order and topological physics.  One can not sweep away the $m$ in Eq.~\eqref{gamma_transformation} by any redefinition of the Majorana operators that preserves their algebra.  More physically, since each edge hosts only one Majorana mode, the $m$ factor is required by the fact that $\mathcal{T}_{\rm elec}^2$ must send $\gamma_j \rightarrow - \gamma_j$.  In Sec.~\ref{Discussion} we will argue on general grounds that proximitized Fe chains provide a concrete physical realization of our modified Kitaev-chain Hamiltonian.
  
Projecting the total-fermion-parity operator [Eq.~\eqref{parity}] into the ground-state manifold yields
\begin{align}
 P_{\rm tot} \equiv e^{i \pi\sum_a(n_{a,\uparrow} + n_{a,\downarrow})} \rightarrow m p~,
 \label{parity_projection}
\end{align}  
where we defined
\begin{equation}
  p = i \gamma_1\gamma_2~.
  \label{pdef}
\end{equation}
Equation~\eqref{parity_projection} further illustrates the intertwinement of symmetry and topology: Flipping $m$ while leaving $p$ constant changes the total parity.  
This type of magnetization reversal is thus naturally implemented by fermionic operators, which one can efficiently obtain by decomposing 
\begin{equation}
  m = i\Gamma_1\Gamma_2.
  \label{mdecomposition}
\end{equation}
Here $\Gamma_{1,2}$ are Majorana operators that we take to additionally obey $\{\Gamma_i,\gamma_j\} = 0$; they simultaneously flip the magnetization \emph{and} total parity as desired.  Together, $\gamma_j$ and $\Gamma_{j}$ form a complete set of low-energy operators describing this fermionic phase (see Table~\ref{symmetry_table3} for their symmetry properties).  We emphasize that $\Gamma_{1,2}$, in contrast to $\gamma_{1,2}$, are generally \emph{not} local operators since they change the magnetization across the entire system.  Locality therefore dictates that $\Gamma_j$ can only appear in the Hamiltonian when the system becomes sufficiently small that the magnetization becomes a fluctuating quantum degree of freedom.  We will encounter such `small' systems in Sec.~\ref{ExpImplications}.

It is worth noting that while the factor of $m$ in Eq.~\eqref{gamma_transformation} is unavoidable, the form of the parity operator above depends on our specific definition of $\gamma_{1,2}$.  One could instead define $\gamma_1' = \gamma_1$ and $\gamma_2' = m \gamma_2$, yielding a more standard expression $P_{\rm tot} = i \gamma_1'\gamma_2'$.  Magnetization flips would then more naturally be implemented by bosonic operators.  This alternate convention is, however, less convenient for understanding hybridization of symmetry-enriched Majorana modes that will be discussed later.  

Interestingly, one can reassemble the four Majorana operators characterizing the low-energy subspace into a single pair of $\mathbb{Z}_4$ parafermion zero modes:
\begin{align}
  \alpha_1 &= -e^{i \frac{\pi}{4}(m - 1)}\gamma_1~,
  \label{alpha1}
  \\
  \alpha_2 &= -e^{-i \frac{\pi}{4}[p(m+1)+1]}\Gamma_2~.
  \label{alpha2}
\end{align}
These expressions arise upon translating the microscopic zero-mode operators from the parafermion representation into fermionic language and projecting into the low-energy subspace. Such a reorganization is always possible for \emph{any} quartet of Majorana operators.  Some caution is thus warranted when invoking a parafermion interpretation of the physics, particularly when the operators are non-local (as is the case for $\alpha_2$ above when the fermion system is `large').  Section~\ref{ExpImplications} elaborates on the issue.  

Here too we can recover the zero-mode structure from the low-energy bosonized theory.  Consider again the setup from Fig.~\ref{DomainWall_fig}, and respectively write $\theta = 0$, $\phi = \pi \hat a/2$, and $\theta = \pi \hat b$ in the left, central, and right domains.  In the present context $\hat a,\hat b$ determine  the central domain's magnetization and total fermion parity according to
\begin{equation}
  m = e^{i \pi \hat a},~~~ P_{\rm tot} = e^{i \pi \hat b}~,
  \label{mPtot}
\end{equation}
where we used Eq.~\eqref{ParityDef} for the parity operator.  
The bosonized analogue of Eqs.~\eqref{gamma1} and \eqref{gamma2} are 
\begin{align}
  \gamma_1 &= \sqrt{2} \cos\left[{\frac{\pi}{2}\left(\hat a - \frac{1}{2}\right)}\right]~, 
  \label{gamma1bosonized}
  \\
  \gamma_2 &= -i\sqrt{2} \cos\left[{\frac{\pi}{2}\left(\hat a + \frac{1}{2}\right)}\right] e^{i \pi \hat b}~.
\end{align}
Both operators are local in the sense that $\gamma_1$ involves only projections of physical fermions $\psi_{R/L} \sim e^{i(\phi \pm \theta)}$ evaluated at the left domain wall, while $\gamma_2$ similarly involves fermions evaluated at the right domain wall.  Moreover, using Eq.~\eqref{mPtot} we have $p = i \gamma_1 \gamma_2 = m P_{\rm tot}$, in harmony with Eqs.~\eqref{parity_projection} and \eqref{pdef}.  The remaining pair of Majorana operators can be written
\begin{align}
  \Gamma_1 &= \cos\left[ \frac{\pi}{2} \left( \hat a - \hat b + \frac{1}{2} \right) \right] - \cos\left[ \frac{\pi}{2} \left( \hat a + \hat b + \frac{1}{2} \right) \right]
  \label{Gamma1bosonized}
  \\
  \Gamma_2 &= \cos\left[ \frac{\pi}{2} \left( \hat a - \hat b - \frac{1}{2} \right) \right] + \cos\left[ \frac{\pi}{2} \left( \hat a + \hat b - \frac{1}{2} \right) \right],
  \label{Gamma2bosonized}
\end{align}
which involve not only domain-wall fermions, but also the operator $e^{i \int_{x \in {\rm{central~domain}}} \partial_x \theta/2} \sim e^{i \frac{\pi}{2} \hat b}$ that flips the central domain's magnetization.  This definition of $\Gamma_{1,2}$ reflects a gauge choice and is certainly not unique: Any rotation among $\Gamma_1$ and $\Gamma_2$ that preserves the magnetization constitutes an equally valid set of operators. 
Equations~\eqref{Gamma1bosonized} and \eqref{Gamma2bosonized} yield $i\Gamma_1\Gamma_2 = m$, consistent with the decomposition in Eq.~\eqref{mdecomposition}.  Using Eqs.~\eqref{alpha1} and \eqref{alpha2} to repackage the bosonized form of the Majorana operators into $\mathbb{Z}_4$ parafermion zero modes precisely reproduces the parafermion operators from Eq.~\eqref{alphas_bosonized}. 

The Majorana representation of the zero modes is far less compact compared to the parafermion representation; cf. Eqs.~\eqref{alphas_bosonized} and \eqref{gamma1bosonized} through \eqref{Gamma2bosonized}.  Nevertheless, the former provides a much more natural description for an electronic system as it clearly partitions the topological and non-topological parts of the degeneracy.  A similar viewpoint was very recently stressed by Mazza et al.~\cite{PoorMansParafermions}.  We also note while some references (e.g., the review in Ref.~\onlinecite{AliceaFendleyReview}) discussed domain walls in quantum-spin-Hall edges with spontaneously broken time-reversal in terms of $\mathbb{Z}_4$ parafermions, it is now clear that the physics is more accurately described in terms of symmetry-enriched Majorana modes.  

The form of the Hamiltonians in Eqs.~\eqref{Hf} and \eqref{Hf_limit} implies that the ground states, and in fact all energy eigenstates, have a free-fermion character  despite the obviously interacting nature of the original fermionic Hamiltonian in Eq.~\eqref{barH0}.  (More precisely, for any fixed configuration of $m_a$'s the Hamiltonian is quadratic.)  This observation connects with the recent work of Meichanetzidis et al.~\cite{Pachos} that inferred free-fermion eigenstates from an analytic solution of the $f=0$ fixed point combined with an interesting numerical diagnostic for the general case \cite{Turner}.  In terms of the clock-model states in Eq.~\eqref{FMstates}, the total-even-parity fermionic ground states correspond to $|A\rangle + |C\rangle$, $|B\rangle + |D\rangle$ while the odd-parity states are $|A\rangle - |C\rangle$, $|B\rangle - |D\rangle$ (to see this, recall that $P_{\rm tot} = Q^2 = \prod_{a} \tau_a^2$).  

Figure~\ref{Phases_fig}, second column, summarizes the results from this subsection.  

\subsection{Phases driven by $\kappa_1<0$}
\label{FM_phase2}

When $\kappa_1$ is relevant and negative, $\phi$ locks to $\pi/4$ modulo $\pi/2$, leading to physics similar to what we encountered in Sec.~\ref{FM_phase} for positive $\kappa_1$.  Clock spins once again realize a broken-symmetry phase with four degenerate ground states, parafermions form a topological phase where the degeneracy is fully protected, and fermions enter a topological state hosting a partially protected degeneracy encoded through symmetry-enriched Majorana zero modes.  These states are distinct, however, from those of Sec.~\ref{FM_phase}, at least in the presence of $\mathcal{C}$ symmetry.  The bosonized theory encodes this distinction as follows.  To smoothly interpolate between phases driven by $\kappa_1>0$ and $\kappa_1<0$, one could in principle replace $-\kappa_2\cos(4\phi) \rightarrow -\kappa_2\cos(4\phi-\phi_0)$ and then continuously sweep $\phi_0$ between 0 and $\pi$.  However, $\mathcal{C}$ symmetry permits \emph{only} $\phi_0 = 0$ or $\pi$, thereby obstructing the interpolation; similar arguments appear in Ref.~\onlinecite{MontorsiHaldane} in the context of symmetry-protected topological phases.

\begin{figure}
\includegraphics[width=\columnwidth]{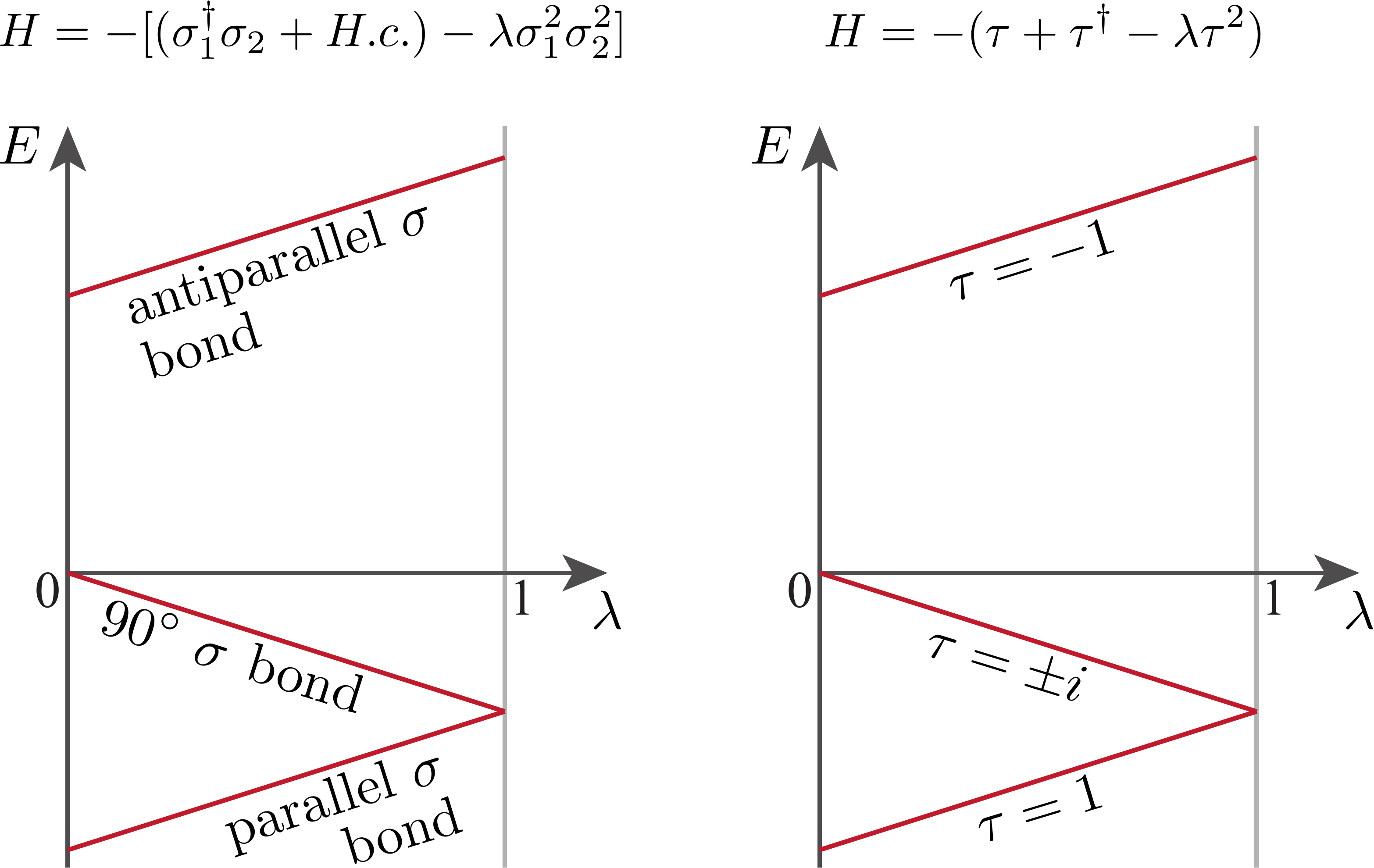}
\caption{Energies versus $\lambda$ obtained from the Hamiltonians shown at the top of the figure.  The left plot represents the energy for a single $J$ bond in the Ashkin-Teller model, Eq.~\eqref{AshkinTeller}.  As $\lambda$ increases from zero, the energy difference between parallel $\sigma$ bonds (i.e., $\sigma_1^\dagger \sigma_2 = 1$) and $90^\circ$ $\sigma$ bonds ($\sigma_1^\dagger \sigma_2 = \pm i$) decreases.  At $\lambda = 1$ these states become degenerate; the Hamiltonian then penalizes antiparallel $\sigma$ bonds ($\sigma_1^\dagger \sigma_2 = -1$) but does not distinguish other configurations.  
The right plot similarly represents the energy for a single $f$ term in the Ashkin-Teller model.  Here the energy difference between $\tau = 1$ and $\tau = \pm i$ states diminishes with $\lambda$ until they become degenerate at $\lambda = 1$; the Hamiltonian then penalizes $\tau = -1$ states but does not differentiate other configurations.     
As discussed in Secs.~\ref{FM_phase2} and \ref{SPT}, the $\lambda =1$ limit is useful for accessing canted-ferromagnet and symmetry-protected topological phases for clock spins, and by extension the analogous phases for parafermions and spinful fermions.}
\label{AT_fig}
\end{figure}

Pinning of $\phi$ to $\pi/4$ modulo $\pi/2$ implies that clock spins spontaneously break $\mathbb{Z}_4$ symmetry by developing a canted ferromagnetic polarization $\langle \sigma \rangle = (1 \pm i)/2$ or $(-1 \pm i)/2$.  By modifying the `root states' $|A,B,C,D\rangle$ defined in Eq.~\eqref{FMstates}, we can construct trial wavefunctions 
\begin{align} 
\begin{split}
  |AB\rangle &= \prod_a\frac{1+\tau_a}{\sqrt{2}}|A\rangle~,\qquad |BC\rangle = \prod_a\frac{1+\tau_a}{\sqrt{2}}|B\rangle
   \\
  |CD\rangle &= \prod_a\frac{1+\tau_a}{\sqrt{2}}|C\rangle~,\qquad |DA\rangle = \prod_a\frac{1+\tau_a}{\sqrt{2}}|D\rangle
  \label{CantedStates}
  \end{split}
\end{align}
with precisely these expectation values \footnote{These trial states do not form an orthogonal set on a finite chain, though any nontrivial overlaps vanish as $1/2^N$.}.  For example, in $|AB\rangle$ any site is equally likely to be found with $\sigma = 1$ or $i$ (and similarly for $|BC\rangle$, etc.).  Two closely related properties are worth noting: $(i)$ these trial states involve no antiparallel $\sigma$ bonds at any distance and $(ii)$ the $(1+\tau_a)$ factors ensure that the wavefunctions contain no $\tau = -1$ components.  States with these characteristics are exact ground states of the Ashkin-Teller model [Eq.~\eqref{AshkinTeller}] at $\lambda = 1$, independent of $f/J$.  At $\lambda = 1$ the $J$ term penalizes antiparallel nearest-neighbor $\sigma$ bonds but does not distinguish parallel and $90^{\circ}$ bonds, while the $f$ term penalizes $\tau = -1$ but does not differentiate other $\tau$ states.  See Fig.~\ref{AT_fig} for an illustration.  Trial states in Eq.~\eqref{CantedStates} incur no such penalties, and are thus indeed ground states.

\begin{figure}
\includegraphics[width=0.49\columnwidth]{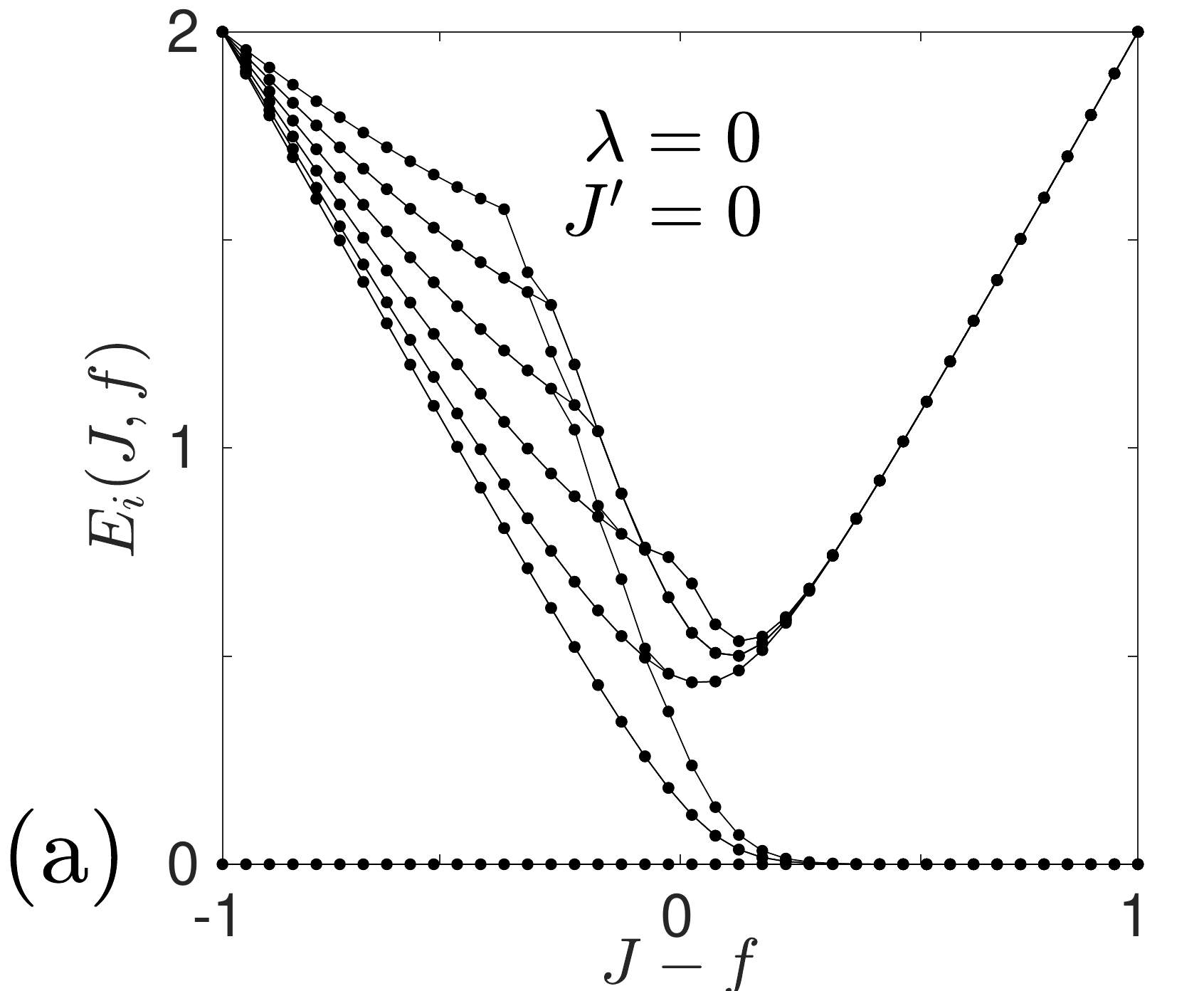}
\includegraphics[width=0.49\columnwidth]{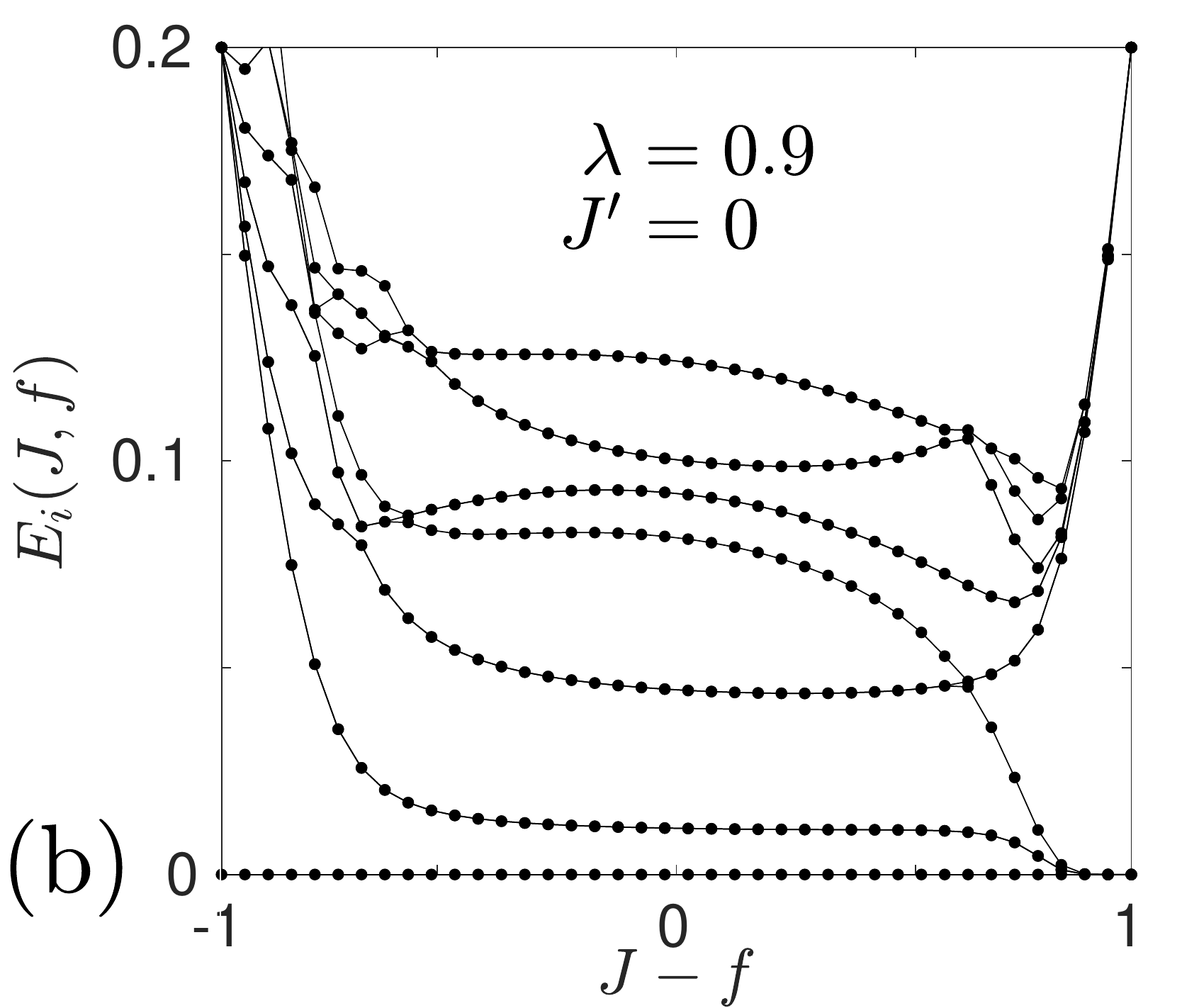}
\includegraphics[width=0.49\columnwidth]{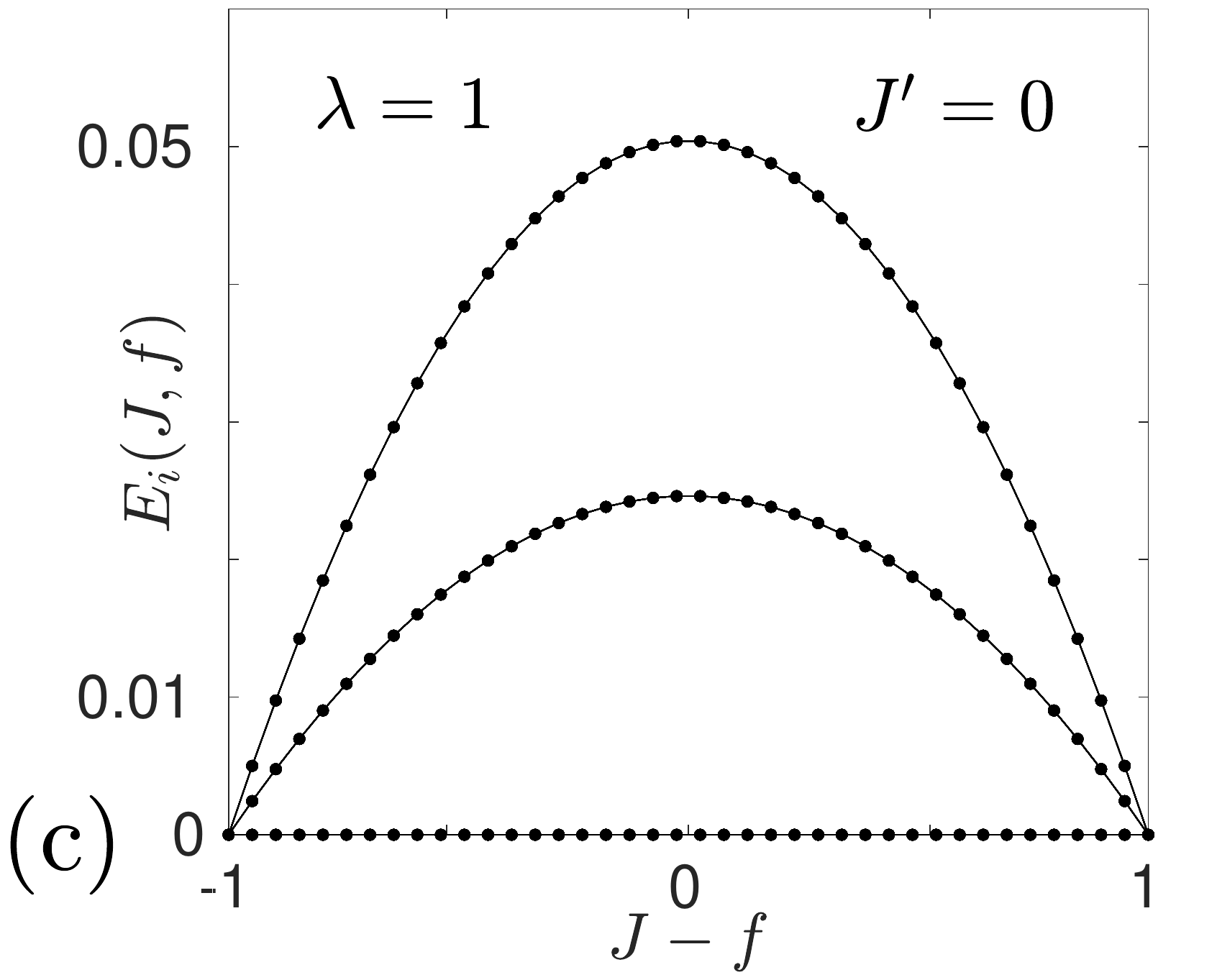}
\includegraphics[width=0.49\columnwidth]{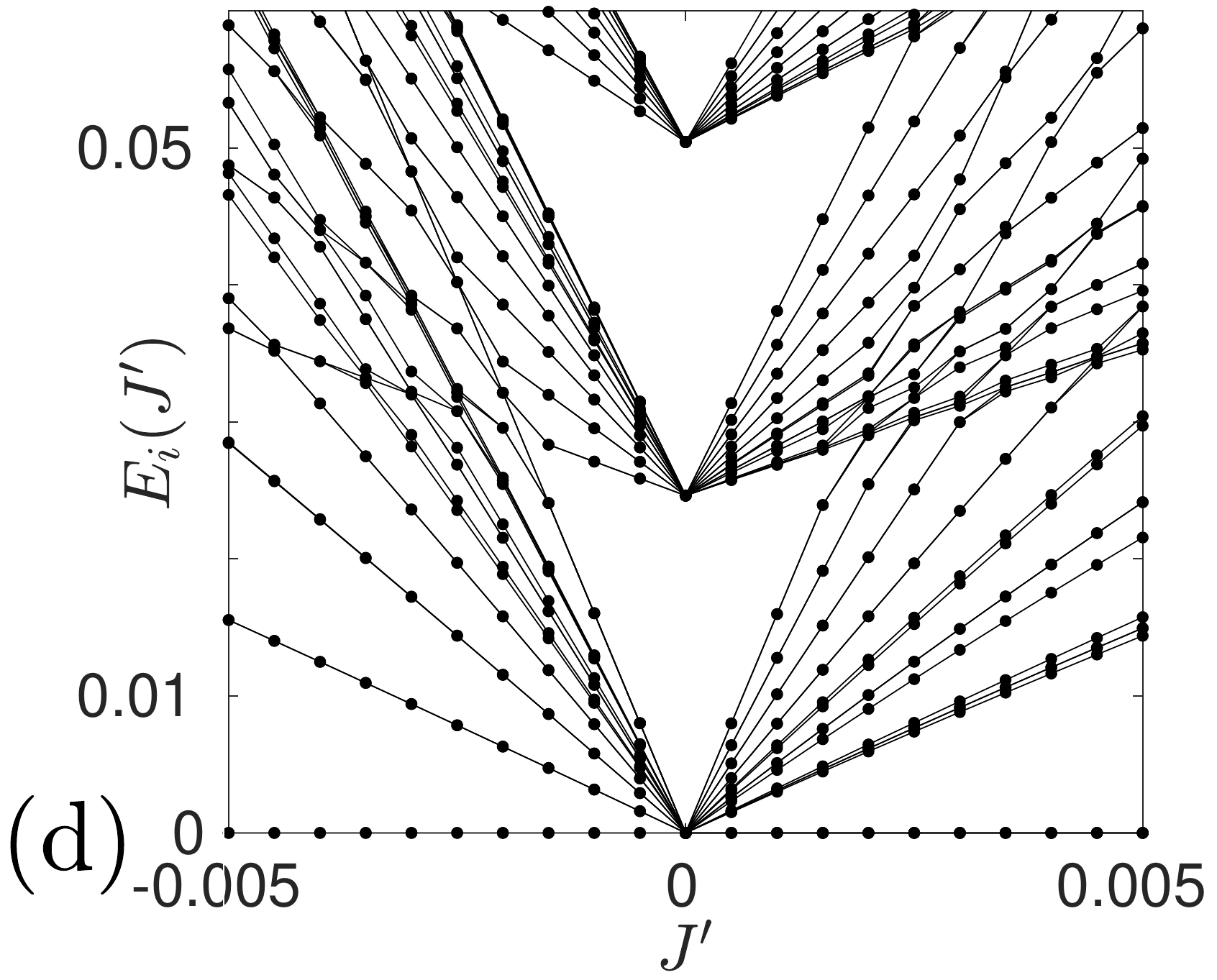}
\caption{Low-energy spectra of the perturbed Ashkin-Teller model $H + \delta H$ given in Eqs.~\eqref{AshkinTeller} and \eqref{deltaH} for a chain of $N=10$ sites with open boundary conditions. All spectra are shifted such that the ground states sit at zero energy, independent of parameters.  (a) The `vanilla' clock model corresponding to $\lambda=J'=0$ undergoes a phase transition at $J=f$ separating the paramagnetic ($f >J$) from the ordered ($J>f$) phase. In a finite system, we find a unique ground state in the former and an (approximately) four-fold-degenerate ground state in the latter. (b) For non-zero $\lambda$, there is a finite region around $J=f$ where the spectrum remains relatively flat, and which we interpret as a finite-size avatar of the critical fan \cite{Kohmoto}. (c) At $\lambda=1$ the spectrum is highly degenerate for arbitrary $J$ and $f$. For $N \in [2,10]$ the ground-state degeneracy grows as $2N+1$. (d) Turning on non-zero $J'$ immediately lifts this degeneracy; for $J'>0$ only a four-fold-degenerate ground state remains as expected for the canted-ferromagnet phase.  
 }
\label{fig.ed}
\end{figure}

Other ground states exist as well---a consequence of an `accidental' U(1) symmetry supported by the Ashkin-Teller model in this limit \cite{Kohmoto}.  In fact at $\lambda = 1$ the Ashkin-Teller model is known to reside at the edge of an extended `critical fan' in the phase diagram \cite{Kohmoto}.  To move away from criticality we therefore additionally incorporate a second-neighbor interaction
\begin{equation}
  \delta H = -J' \sum_{a = 1}^{N-2} (\sigma_a^\dagger \sigma_{a+2} + \sigma_{a+2}^\dagger \sigma_a - \sigma_a^2 \sigma_{a+2}^2)
  \label{deltaH}
\end{equation}
with $J'>0$.  The above perturbation spoils the accidental U(1) by penalizing second-neighbor antiparallel $\sigma$ bonds (similar to the $J$ term), leaving our trial canted ferromagnet states as unique ground states.  Exact diagonalization numerics summarized in Fig.~\ref{fig.ed} support this scenario; see caption for details. 
As a further check, DMRG calculations were performed on a 400-site system using ITensor \footnote{Calculations performed using the ITensor C++ library, http://itensor.org/}.  With $J' = 0$, DMRG exhibited characteristics of a gapless system, predicting a gap several orders of magnitude below the $J,f$ couplings.  When a small $J'$ perturbation was added, DMRG instead converged to the expected canted ground states \footnote{More precisely, with $\mathbb{Z}_4$ symmetry enforced, DMRG returns Schrodinger-cat superpositions of the states in Eq.~\eqref{CantedStates}.  Adding a small $\mathbb{Z}_4$-breaking perturbation of the form $e^{i\frac{\pi}{4}} \sigma_j + H.c.$ to a single site $j$, however, yields one of the physical canted product states.} while predicting a gap of order $J'$.  These results strongly suggest that the Hamiltonian is indeed gapped so long as $J' > 0$.  

Translating into parafermion language, $\delta H$ becomes
\begin{align}
  \delta H = -J' \sum_{a = 1}^{N-2}\big[&(i\alpha_{2a}^\dagger\alpha_{2a+1}\alpha_{2a+2}^\dagger\alpha_{2a+3} + H.c.)
  \nonumber \\
  &-\alpha_{2a}^2 \alpha_{2a+1}^2 \alpha_{2a+2}^2 \alpha_{2a+3}^2\big].
  \label{deltaHparafermion}
\end{align}
See Fig.~\ref{Phases_fig} for an illustration of the full set of couplings for the parafermion chain arising from both $\delta H$ and the Ashkin-Teller model at $\lambda = 1$.  Our prior analysis allows us to deduce some general features of the parafermion phase realized here:  First, ground states necessarily correspond to $\mathbb{Z}_4$-preserving superpositions of clock states in Eq.~\eqref{CantedStates}, and second, the chain must host edge $\mathbb{Z}_4$ parafermion zero modes.  (Upon breaking $\mathcal{C}$ this phase smoothly connects to the topological phase discussed in Sec.~\ref{FM_phase}; since parafermion zero modes obviously exist in the latter case, they must also survive in the former by continuity.  Restoring $\mathcal{C}$ can not change this conclusion.)  Explicitly constructing lattice zero-mode operators is nevertheless nontrivial given that the Hamiltonian no longer consists of a sum of commuting terms \footnote{We expect that localized `strong zero mode' operators that commute with the full microscopic Hamiltonian---and thus guarantee at least four-fold degeneracy of \emph{all} eigenstates---do not actually exist, similar to the situations encountered in Refs.~\onlinecite{Fendley:2012,Jermyn,Alexandradinata,MazzaWeakZeroModes,Moran}.  `Weak zero modes', which arise from projections of local operators and ensure degeneracy only among ground states, certainly exist and are captured by the bosonization description that follows.}.  

We will content ourselves with capturing the zero modes within a bosonized framework.  Let us take a domain configuration akin to Fig.~\ref{DomainWall_fig}, with outer regions again gapped by $-\cos(2\theta)$ but with the central region gapped by $+\cos(4\phi)$ instead of $-\cos(4\phi)$.  We parametrize the low-energy sector with integer-valued operators $\hat a, \hat b$ by writing $\theta = 0$, $\phi = \pi/4 + \pi \hat a/2$, and $\theta = \pi \hat b$ in the left, middle, and right regions.  Note in particular the $\pi/4$ shift in $\phi$ compared to the parametrization adopted in Sec.~\ref{FM_phase}.  The zero modes we seek follow from projecting parafermions evaluated at domain walls, and then introducing phase factors to ensure that the resulting low-energy operators fourth to unity; this procedure yields parafermion zero modes $\alpha_{1,2}$ given precisely by Eq.~\eqref{alphas_bosonized}.  What, then, is the distinction between the parafermion analogue of the conventional ferromagnetic and canted ferromagnetic phases?  The answer lies in the symmetry properties of the zero modes.  In particular, under $\mathcal{C}$ the zero modes obtained in Sec.~\ref{FM_phase} transform as $\alpha_j \rightarrow \alpha_j^\dagger$, while in the present case they transform as $\alpha_j \rightarrow -i \alpha_j^\dagger$---a consequence of the $\pi/4$ shift mentioned above.  Without $\mathcal{C}$ symmetry this distinction vanishes, consistent with our earlier arguments.    

For spinful fermions, two ground states arise from Majorana zero modes while the other two reflect spontaneous symmetry breaking.  A more obvious distinction from Sec.~\ref{FM_phase} emerges here: The local order parameter 
\begin{equation}
  \bar m \equiv \langle \psi_R \psi_L + H.c. \rangle  \sim \langle \sin(2\phi)\rangle = \pm 1
  \label{barmdef}
\end{equation}
again breaks $\mathcal{T}_{\rm elec}$ and $\mathbb{Z}_4$ but, contrary to Eq.~\eqref{mdef}, preserves $U_{\rm spin}$.  We can readily obtain the zero-mode structure from the continuum bosonized theory, following exactly the same procedure as for parafermions above.  Within this framework our four Majorana zero mode operators once again take the form in Eqs.~\eqref{gamma1bosonized} through \eqref{Gamma2bosonized} and similarly satisfy $p = i\gamma_1\gamma_2 = \bar m P_{\rm tot}$ and $\bar m = i \Gamma_1\Gamma_2$.  Moreover, the Majorana operators transform under $\mathcal{T}_{\rm elec}$ and $\mathbb{Z}_4$ precisely as in Table~\ref{symmetry_table3} (with $m \rightarrow \bar m$); they are invariant under $U_{\rm spin}$, however, because the ground states now preserve that symmetry.

We can again interpret the physics in terms of a Kitaev-chain-like model arising from spontaneous symmetry breaking.  The microscopic order parameter can be written as
\begin{equation}
  \bar m_a = i(c_a d_a - d_a^\dagger c_a^\dagger) = -f_a^\dagger \sigma^y f_a,
  \label{barm_lattice}
\end{equation}
corresponding to a magnetization along $y$.  
The above expression arises from fermionizing $\sigma^2(\tau-\tau^\dagger)/2$, which has the same symmetry properties as Eq.~\eqref{barmdef}.  Because $\bar m_a$ no longer commutes with the lattice Hamiltonian, an exact microscopic analysis is nonetheless more nontrivial than in Sec.~\ref{FM_phase} and will not be pursued here.  

The canted phase and its parafermionic and fermionic counterparts are summarized in the third column of Fig.~\ref{Phases_fig}; note the close relation to the phases from the second column.

\subsection{Phases driven by $\kappa_2<0$}  
\label{SPT}

With relevant $\kappa_2<0$ the $\cos(2\theta)$ term pins $\theta$ to $\pi/2$ modulo $\pi$.  It is tempting to conclude that clock spins then form a trivial, symmetric gapped phase as found in Sec.~\ref{trivial_phase} for $\kappa_2>0$, since the pinning once again condenses the disorder operator $\mu$.  However, one can not smoothly interpolate between phases driven by $\kappa_2>0$ and $\kappa_2<0$ without violating symmetries.  Let us first apply the same logic as in the previous subsection: A term of the form $-\kappa_2 \cos(2\theta-\theta_0)$ can only have $\theta_0 = 0$ or $\pi$ unless both $\mathcal{C}$ and $\mathcal{T}$ are explicitly broken, which precludes symmetrically bridging the two phases via continuous evolution of $\theta_0$ \cite{MontorsiHaldane}.  We could alternatively connect the phases by $(i)$ starting from the trivial regime gapped by $\kappa_2>0$, $(ii)$ ramping up a `large' $\cos(\phi-\phi_0)$ perturbation for some constant $\phi_0$, $(iii)$ sweeping $\kappa_2$ from positive to negative, and $(iv)$ turning off the $\cos(\phi-\phi_0)$ term.  The system follows a unique 
ground state throughout this path, yet along the way maximally breaks $\mathbb{Z}_4$ and possibly other symmetries depending on $\phi_0$.  By `maximally', we mean that $\mathbb{Z}_4$ and $\mathbb{Z}_4^2$ are both violated.  To better understand this second scenario, suppose that we replace $\cos(\phi-\phi_0)$ with $\cos(2\phi)$---which also breaks $\mathbb{Z}_4$ but preserves $\mathbb{Z}_4^2$.  Here, passing from $(i)$ to $(ii)$ incurs an Ising-type phase transition at which the order parameter $e^{i \phi}$ condenses into one of two spontaneously chosen values.  The $\cos(\phi-\phi_0)$ term, by contrast, circumvents criticality by favoring a unique state.  An identical distinction arises between the $\beta^2 = 2\pi$ and $4\pi$ theories discussed in Ref.~\onlinecite{LECHEMINANT}; in our conventions, the self-dual Sine-Gordon models described there model the deformation from the $\cos(2\theta)$-dominated phase to the $\cos(q\phi)$-dominated phase, where $q$ is an integer.   

The observations above suggest that $\kappa_2<0$ germinates a symmetry-protected topological phase (SPT).  We will show that this is indeed the case not only for clock spins, but also for parafermions and fermions.   

Recall that phases driven by $\kappa_2<0$ and $\kappa_1<0$ are dual to one another, and that the $\kappa_1<0$ state arises microscopically from the Ashkin-Teller Hamiltonian at $\lambda = 1$ supplemented by $\delta H$ in Eq.~\eqref{deltaH}.  Dualizing the perturbed Ashkin-Teller model thus immediately yields a parent Hamiltonian for the phases of interest here.  In the Ashkin-Teller parts, dualizing merely swaps $J\leftrightarrow f$.  At $\lambda = 1$ the swap is inconsequential insofar as ground states are concerned, since these pieces merely penalize $\tau = -1$ configurations and antiparallel nearest-neighbor $\sigma$ bonds for any $f/J$ (see again Fig.~\ref{AT_fig}).  The dual of $\delta H$ takes the form
\begin{equation}
  \widetilde{\delta H} = - f' \sum_{a = 1}^{N-1} (\tau_a\tau_{a+1} + \tau_a^\dagger\tau_{a+1}^\dagger - \tau_a^2\tau_{a+1}^2).
\end{equation}
For $f'>0$, which we assume throughout, $\widetilde{\delta H}$ additionally penalizes nearest-neighbor configurations with $(\tau_a,\tau_{a+1}) = (1,-1), (-1,1), (i, i)$, or $(-i,-i)$.  

We can modify the `root state' $|\tau = 1,\ldots, 1\rangle$ to construct an exact ground state of our new perturbed Ashkin-Teller model.  For reasons that will become clear shortly, we label the wavefunction
\begin{equation}
  |\downarrow\uparrow\rangle = \prod_{a = 1}^{N-1} \frac{1+ \sigma_a^\dagger \sigma_{a+1}}{\sqrt{2}}|\tau = 1,\ldots,1\rangle~;
  \label{psiSPT}
\end{equation}
note the dual relation to the canted-ferromagnet states defined in Eq.~\eqref{CantedStates}.  The $(1+\sigma_a^\dagger \sigma_{a+1})$ product generates an entangled state that, by construction, projects away all antiparallel $\sigma$ bonds.  Nontrivial elements in the product take the form $\sigma_{a_1}^\dagger \sigma_{a_1+1} \sigma_{a_2}^\dagger \sigma_{a_2+1} \cdots \sigma_{a_m}^\dagger \sigma_{a_m+1}$ where all $a_i$'s are distinct.  Crucially, such terms produce neither $\tau = -1$ configurations nor $(\tau_a,\tau_{a+1}) = (i,i)$ or $(-i,-i)$ pairs.  (Obtaining $\tau = -1$ contributions would require $\sigma^2_a$ factors, while the latter pairs would require $\sigma_a^\dagger \sigma_{a+1}^\dagger$ or $\sigma_a \sigma_{a+1}$; none of these appear.)  So $|\downarrow\uparrow\rangle$ maximally satisfies both the $\lambda = 1$ Ashkin-Teller model and $\widetilde {\delta H}$, and hence is a ground state as claimed.  

For any site away from the edges, configurations with $\tau = 1, i$, and $-i$ all occur in $|\downarrow\uparrow\rangle$.  Acting with $\sigma$ or $\sigma^\dagger$ in the bulk thus necessarily takes the system out of the ground state, e.g., by mixing in $\tau = -1$ components penalized by the Ashkin-Teller terms.  Boundaries behave differently.  
The leftmost two sites involve only $(\tau_1,\tau_2) = (1,1), (1,i), (i,1)$, and $(i,-i)$ pairs, and the rightmost two sites involve only $(\tau_{N-1},\tau_N) = (1,1), (1,-i), (-i,1)$, and $(i,-i)$ pairs.  We can therefore twist the edge spins without energy cost, yielding three additional ground states
\begin{equation}
  |\uparrow\uparrow\rangle = \sigma_1 |\downarrow\uparrow\rangle~,\quad|\downarrow\downarrow\rangle = \sigma_N^\dagger |\downarrow\uparrow\rangle~,\quad|\uparrow\downarrow\rangle = \sigma_1\sigma_N^\dagger |\downarrow\uparrow\rangle~.
  \label{other_states}
\end{equation}
For later use, observe that the generator $Q$ of $\mathbb{Z}_4$ symmetry acts in the ground-state subspace as follows:
\begin{equation}
\begin{aligned}
  Q|\uparrow\uparrow\rangle &= i|\uparrow\uparrow\rangle~,\quad &&Q|\downarrow\downarrow\rangle = -i|\downarrow\downarrow\rangle~,
  \\
  Q|\downarrow\uparrow\rangle &= |\downarrow\uparrow\rangle~,\quad &&Q|\uparrow\downarrow\rangle = |\uparrow\downarrow\rangle~.
  \label{Qaction}
\end{aligned}
\end{equation}

Our construction shows that each boundary of the clock chain hosts a degenerate pseudospin-1/2 degree of freedom, which we describe with Pauli matrices $\eta^\mu_1$ and $\eta^{\mu}_2$.  (Arrows in the kets above designate $\eta_{1,2}^z$ eigenvalues.)  The pseudospins are locally distinguishable by Hermitian operators $i(\tau-\tau^\dagger)$ since $\langle\eta^z_1\eta^z_2|i(\tau_1-\tau_1^\dagger)|\eta^z_1\eta^z_2\rangle  = \eta^z_1$ and $\langle\eta^z_1\eta^z_2|i(\tau_N-\tau_N^\dagger)|\eta^z_1\eta^z_2\rangle = \eta^z_2$.  These expectation values, together with Eqs.~\eqref{other_states} and \eqref{Qaction}, enable us to relate pseudospins and microscopic operators projected into the ground-state subspace with a projector $\mathcal{P}$:
\begin{align}
  &\mathcal{P}i(\tau_1-\tau_1^\dagger)\mathcal{P} = \eta^z_1,\quad&&\mathcal{P}i(\tau_N-\tau_N^\dagger)\mathcal{P} = \eta^z_2~,
  \label{etaz} \\
  &\mathcal{P}\sigma_1\mathcal{P} = (\eta^x_1 + i \eta^y_1)/2,\quad &&\mathcal{P}\sigma_N\mathcal{P} = (\eta^x_2 + i \eta^y_2)/2~,
  \label{etaxy} \\
  & \mathcal{P}Q\mathcal{P} = e^{i\frac{\pi}{4}(\eta^z_1 + \eta^z_2)}. 
  \label{Q_proj}  
\end{align}
Table~\ref{SPT_symmetry_table} summarizes the pseudospin symmetry properties that follow from these relations.  

Abandoning $\mathcal{C}$ and $\mathcal{T}$ allows the boundary degeneracy to be lifted through local edge perturbations of the form $h_z(\eta^z_1 + \eta^z_2)$, while discarding $\mathbb{Z}_4$ permits a perturbation  $h_x(\eta^x_1 + \eta^x_2)$ that likewise spoils the degeneracy.  The symmetry-protection of the edge degeneracy seen here fully corroborates the analysis of the bulk given in the beginning of this subsection.  In Appendix~\ref{ProjectiveRepsAppendix} we further show that the edge modes are anomalous (in all representations) in the presence of either $\mathbb{Z}_4\mathcal{T}$, or $\mathbb{Z}_4$ and $\mathcal{C}$, thus proving that the system forms an SPT.  

\begin{table}
\begin{center}
 \setlength\extrarowheight{2pt}
 \begin{tabular}{|c | c | c | c| c|} 
 \hline
  & $\mathbb{Z}_4$ & $\mathcal{C}$ & $\mathcal{T}$ \\ [0.05ex] 
 \hline\hline
 $\eta_j^x \rightarrow$ & $-\eta_j^y$ & $\eta_j^x$ & $\eta_j^x$ \\
 \hline
  $\eta_j^y \rightarrow$ & $\eta_j^x$ & $-\eta_j^y$ & $\eta_j^y$ \\
 \hline
  $\eta_j^z \rightarrow$ & $\eta_j^z$ & $-\eta_j^z$ & $-\eta_j^z$ \\
 \hline
  \hline
  & $\mathcal{T}_{\rm elec} = \mathbb{Z}_4\mathcal{T}$ & $U_{\rm spin} = \mathbb{Z}_4\mathcal{C}$ & $\mathbb{Z}_4$ \\ [0.05ex] 
 \hline 
 \hline
 $\gamma_{j\uparrow} \rightarrow$ & $\gamma_{j\downarrow}$ & $\gamma_{j\downarrow}$ & $\gamma_{j\downarrow}$ \\
 \hline
  $\gamma_{j\downarrow} \rightarrow$ & $-\gamma_{j\uparrow}$ & $\gamma_{j\uparrow}$ & $-\gamma_{j\uparrow}$ \\
 \hline
\end{tabular}
\end{center}
\caption{Symmetry transformations for the SPT edge degrees of freedom in the clock realization (top) and spinful-fermion realization (bottom).  Here $j = 1$ and 2 respectively correspond to the left and right boundaries. }
\label{SPT_symmetry_table}
\end{table}

Suppose next that parafermions form the physical degrees of freedom.  Figure~\ref{Phases_fig} sketches the parafermion-chain couplings for this case [including $\widetilde {\delta H}$, which takes the same form as Eq.~\eqref{deltaHparafermion} but translated by one site].  The ground states in Eq.~\eqref{Qaction} are already eigenstates of the $\mathbb{Z}_4$ generator $Q$, and so form a physical basis also in this realization.  Physical low-energy operators should, however, now derive from projections of parafermionic rather than clock degrees of freedom.  Specifically, the microscopic operators to be projected become 
\begin{align}
  & -e^{-i\frac{\pi}{4}}\alpha_1^\dagger \alpha_2 + H.c. = i(\tau_1-\tau_1^\dagger), 
  \nonumber \\
  & -e^{-i\frac{\pi}{4}}\alpha_{2N-1}^\dagger \alpha_{2N} + H.c. = i(\tau_N-\tau_N^\dagger)
  \nonumber \\
  & \alpha_1 = \sigma_1,~~~~\alpha_{2N} = e^{i \frac{\pi}{4}} Q^\dagger \sigma_N,
  \label{alphas_to_project}
\end{align}
which give rise to edge operators that we label $\eta^\mu_{j,\rm PF}$.
At the left boundary the projection is unmodified compared to the clock case; hence $\eta^{\mu}_{1,\rm PF} = \eta^{\mu}_1$.  The factor of $Q^\dagger$ appearing in $\alpha_{2n}$ does modify the structure of the edge mode at the right boundary, yielding
\begin{equation}
  \eta^z_{2,\rm PF} = \eta^z_2~,\qquad\eta^{x,y}_{2,\rm PF} = e^{-i \frac{\pi}{4}\eta^z_1} \eta^{x,y}_2~.
  \label{PFetas}
\end{equation}
Notice that $\eta^{x,y}_{1,\rm PF}$ and $\eta^{x,y}_{2,\rm PF}$ do not commute---a remnant of the nonlocal parafermionic commutation relations.  We stress that locality prevents these operators from appearing in the Hamiltonian by themselves.  The \emph{only} local operators that can remove the edge degeneracy in the parafermion SPT realization take the form $h_{z,1}\eta^z_{1, \rm PF}$ and $h_{z,2}\eta^z_{2, \rm PF}$, which require breaking $\mathcal{C}$ and $\mathcal{T}$.  In other words, the $\mathbb{Z}_4$-breaking route to connecting the trivial and SPT phases discussed earlier for clock spins is inaccessible because $\mathbb{Z}_4$ can never be broken explicitly in a parafermion system.  

We treat the spinful-fermion realization analogously.  Since fermions arise from attaching a `doubled' string to clock operators (Fig.~\ref{ops_fig}), the edge modes take the same form as for the parafermion chain but with $e^{-i\frac{\pi}{4}\eta^z_1} \rightarrow e^{-i\frac{\pi}{2}\eta^z_1} = -i \eta^z_1$ in Eq.~\eqref{PFetas}.  [That is, the fermionic counterpart of Eq.~\eqref{alphas_to_project} involves $Q^2$ instead of $Q^\dagger$.]  One can conveniently parametrize the resulting edge modes as follows,\begin{align}
  &\eta^z_1 = i\gamma_{1\downarrow}\gamma_{1\uparrow},~~~\eta^z_2 = i\gamma_{2\downarrow}\gamma_{2\uparrow}
  \\
  &\eta^x_1 + i\eta^y_1 = \gamma_{1\uparrow} - i \gamma_{1\downarrow}
  \\
  &-i\eta^z_1(\eta^x_2 + i \eta^y_2) = \gamma_{2\downarrow} - i \gamma_{2\uparrow},
\end{align}
where $\gamma_{j\alpha}$ are Majorana-fermion operators.  Table~\ref{SPT_symmetry_table} lists their transformation properties under the symmetry generators $\mathcal{T}_{\rm elec}, U_{\rm spin}$, and $\mathbb{Z}_4$ that are natural for the fermionic representation.  Most importantly, we see that \emph{the pair of Majorana modes at each end form a Kramers doublet under electronic time reversal}---which immediately implies that the SPT in this representation corresponds to a time-reversal-invariant topological superconductor (TRITOPS) \cite{QiTRITOPS,ChungTRITOPS,ZhangTRITOPS,KeselmanTRITOPS,Haim1TRITOPS,Haim2TRITOPS,CamjayiTRITOPS}.  

Two additional observations further illuminate the edge physics.  First, our fermionization algorithm yields the relation 
\begin{equation}
  S^z_a \equiv \frac{\hbar}{2} (f^\dagger_{a,\uparrow} f_{a,\uparrow}-f^\dagger_{a,\downarrow} f_{a,\downarrow}) = -i\frac{\hbar}{4}(\tau_a-\tau_a^\dagger)~,
\end{equation}
where $S^z_a$ denotes the $z$-component of the electronic spin at site $a$.  Upon combining with Eq.~\eqref{etaz} we obtain
\begin{align}
  \mathcal{P}S^z_1\mathcal{P} = -\frac{\hbar}{4}(i\gamma_{1\downarrow}\gamma_{1\uparrow}),\quad\mathcal{P}S^z_N\mathcal{P} = -\frac{\hbar}{4}(i\gamma_{2\downarrow}\gamma_{2\uparrow})~.
\end{align}
Thus each edge hosts a fractional spin $\pm \hbar/4$, which is another known signature of a TRITOPS phase \cite{KeselmanTRITOPS,CamjayiTRITOPS}.  It is illuminating to view this result also in bosonization.  In our bosonized theory the edge can be modeled by taking a TRITOPS phase gapped by $+\cos(2\theta)$ bordered by trivial phases gapped by $-\cos(2\theta)$.  Using Eq.~\eqref{spindensity}, we see that the resulting $\pi/2$ kinks in $\theta$  at the domain walls bind fractional spin in agreement with our lattice calculation.  Interestingly, an identical domain structure arises in the bosonized description of a quantum-spin-Hall edge gapped by regions with opposite magnetization. In that context the domain walls bind $e/2$ fractional charge \cite{FractionalChargeQSH}, which we now see is a precise analogue of fractional spin at a TRITOPS edge.  Yet another instance in which fractional spin binds to the edge of a 1D model occurs in the Haldane phase \cite{HaldaneChain, Affleck1988, denNijsHaldane}, which was analyzed using similar bosonization methods in Ref.~\onlinecite{MontorsiHaldane}.  Note that the status of the Haldane phase as an SPT is subtle when viewed as arising from electrons; see Refs.~\onlinecite{AnfusoHaldane, MoudgalyaHaldane}. By contrast, time-reversal-symmetry alone protects TRITOPS as a nontrivial SPT.

Second, the total fermion parity operator obeys 
\begin{equation}
  P_{\rm tot} = Q^2\rightarrow \gamma_{1\downarrow}\gamma_{1\uparrow}\gamma_{2\downarrow}\gamma_{2\uparrow}.
\end{equation} 
Equation~\eqref{Qaction} then implies that $|\downarrow\uparrow\rangle$ and $|\uparrow\downarrow\rangle$ have even parity while $|\uparrow\uparrow\rangle$ and $|\downarrow\downarrow\rangle$ have odd parity.  It is now clear that the edge Majorana modes cycle through the ground states by simultaneously flipping the total fermion parity and fractional edge spins.  Electronic time-reversal by itself suffices to preserve the boundary degeneracy and SPT order; in principle $U_{\rm spin}$ can also protect the topological phase but is a less natural symmetry to impose on an electronic system.  Finally, as in the parafermion realization breaking $\mathbb{Z}_4$ does not destroy the SPT, in this case because $\mathbb{Z}_4^2$ can never be broken explicitly.  

The final column of Fig.~\ref{Phases_fig} summarizes the SPT's in each representation.  

\subsection{Hybrid order}  
\label{HybridOrder}

It is also possible to stabilize phases with both $\langle e^{2i \phi} \rangle\neq 0$ \emph{and} $\langle e^{i \theta} \rangle \neq 0$.  In clock language such `hybrid order' translates into the square of order and disorder operators condensing simultaneously---i.e., $\langle \sigma^2 \rangle \neq 0$ and $\langle \mu^2 \rangle \neq 0$---while $\sigma$ itself fluctuates wildly.  Clock spins thus spontaneously break $\mathbb{Z}_4$ but preserve $\mathbb{Z}_4^2$, yielding only two degenerate ground states.  For simplicity, we will concentrate on hybrid orders that preserve $\mathcal{C}$ and $\mathcal{T}$ symmetries, which admit a particularly simple microscopic parent Hamiltonian given by  
\begin{equation}
  H_{\rm hybrid~order} = -J_2 \sum_{a = 1}^{N-1} \sigma_a^2 \sigma_{a+1}^2 - f_2 \sum_{a = 1}^N \tau^2_a.
  \label{Hhybrid}
\end{equation}
We assume $J_2,f_2>0$ throughout this subsection.  Equation~\eqref{Hhybrid} corresponds to the Ashkin-Teller model with only the $\lambda$ terms retained, and is trivially solvable since $\sigma^2$ and $\tau^2$ commute.  

For any $f_2/J_2$, ground states have $\tau_a^2 = 1$ for all $a$ and Ising-like ferromagnetic order with either $\sigma_a^2 = +1$ or $-1$ uniformly across the chain.  The ground-state wavefunctions can be written in a similar form as the canted-ferromagnet states in Eq.~\eqref{CantedStates}:
\begin{align}
\ket{+} &= \prod_a \dfrac{1+\tau_a^2}{\sqrt{2}} \ket{\sigma = 1,\ldots,1} \\
\ket{-} &= \prod_a \dfrac{1+\tau_a^2}{\sqrt{2}} \ket{\sigma = i,\ldots,i}.
\end{align}
The $(1+\tau_a^2)$ factors simultaneously disorder $\sigma$ and project out $\tau_a^2 = -1$ configurations.  
As desired, both states are $\mathcal{C},\mathcal{T}$-symmetric and yield $\langle \sigma \rangle = 0$, while $\tau^2|\pm\rangle = |\pm\rangle$ and $\sigma^2|\pm\rangle = \pm |\pm\rangle$.  When $J_2 = f_2$ the Hamiltonian is self-dual; one can also view the phase itself as self-dual for general $f_2/J_2$, in the sense that swapping $f_2 \leftrightarrow J_2$ yields exactly the same order.  We show in Appendix~\ref{SelfDualityHybridOrder} that duality indeed leaves the above states invariant, modulo a trivial basis transformation.  

\begin{figure}
\includegraphics[width=0.6\columnwidth]{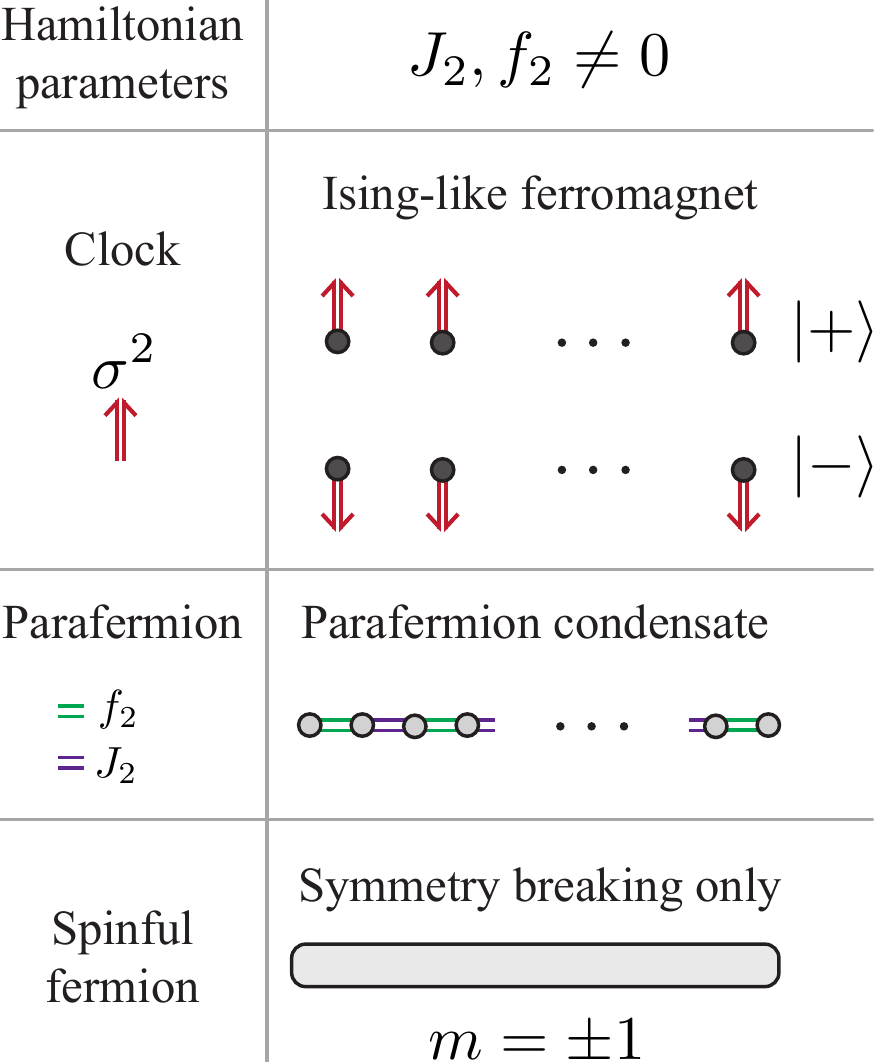}
\caption{Summary of phases stabilized by the equivalent Hamiltonians of Eqs.~\eqref{Hhybrid}, \eqref{Hhybrid2}, and \eqref{Hhybrid3}.  In the spinful-fermion realization, the system forms a topologically trivial strong-pairing superconductor with spontaneous symmetry breaking.}
\label{Hybrid_Phases_fig}
\end{figure}

Equation~\eqref{alpha_expansion} implies that the associated parafermion system realizes a `parafermion condensate' phase with $\langle \alpha^2\rangle \neq 0$ \cite{Motruk:2013,Bondesan:2013,Alexandradinata}.  The parent Hamiltonian in this representation becomes
\begin{equation}
  H_{\rm hybrid~order} = J_2 \sum_{a = 1}^{N-1}\alpha_{2a}^2\alpha_{2a+1}^2 + f_2 \sum_{a = 1}^N\alpha_{2a-1}^2\alpha_{2a}^2~,
  \label{Hhybrid2}
\end{equation}
which is an example of the commuting-projector models from Ref.~\onlinecite{Bondesan:2013}, and can also be viewed as a simpler variant of the parafermion-condensate model introduced by Motruk et al.~\cite{Motruk:2013}.  The two ground states correspond to $\mathbb{Z}_4$-preserving superpositions $|\tilde +\rangle = \left(|+\rangle + |-\rangle\right )/\sqrt{2}$ and $|\tilde -\rangle = \left(|+\rangle - |-\rangle\right)/\sqrt{2}$ that are locally indistinguishable and satisfy $Q|\tilde \pm \rangle = \pm |\tilde \pm \rangle$.  Operators $\alpha^2 \propto \sigma^2 \mu^2$ acting \emph{anywhere} in the chain toggle between the ground states.  As emphasized in Refs.~\onlinecite{Motruk:2013,Bondesan:2013,Alexandradinata}, the system exhibits a protected degeneracy yet lacks edge zero modes.

In the spinful-fermion realization, the two-fold degeneracy arises entirely from spontaneous symmetry breaking.  The order parameter $m$ from the bosonized theory in fact takes the same form given in Eq.~\eqref{mdef}.  
This symmetry breaking emerges transparently from the fermionic representation of Eq.~\eqref{Hhybrid}, which can be conveniently expressed as
\begin{align}
  H_{\rm hybrid~order} =& - J_2 \sum_{a = 1}^{N-1} m_a m_{a+1} 
  \nonumber \\
  &- f_2\sum_{a = 1}^N (2n_{a,\uparrow} - 1)(2n_{a,\downarrow} - 1).
  \label{Hhybrid3}
\end{align}
Here $m_a = \sigma_a^2 = -f_a^\dagger \sigma^x f_a + (i f_{a,\uparrow}^\dagger f_{a,\downarrow}^\dagger + H.c.)$ is the microscopic clock order parameter re-expressed in terms of fermions [cf.~Eq.~\eqref{m_lattice}].  To maximally satisfy the $f_2$ term we project into the sector where both spin species on a given site are either occupied or unoccupied.  In effect, the projection strongly pairs the fermions into bosons that can be conveniently described with spin-singlet Cooper-pair operators $b_a = f_{a,\uparrow} f_{a,\downarrow}$.  Within this low-energy subspace, the order parameter projects to $m_a \rightarrow i(b_a^\dagger -  b_a)$.  Clearly the system can now also maximally satisfy the $J_2$ term by condensing $\langle i(b_a^\dagger - b_a)\rangle = \pm 1$.  We thereby obtain a strong-pairing superconductor in which the fermions spontaneously develop an $s$-wave pairing potential with imaginary coefficient, thus breaking electronic time-reversal as well as $U_{\rm spin}$ and $\mathbb{Z}_4$.

Figure~\ref{Hybrid_Phases_fig} summarizes the phases highlighted in this subsection.

\section{Experimental Implications}
\label{ExpImplications}

\subsection{How much non-Abelian-anyon physics survives in 1D electronic systems?}
\label{nonAbelian1D}

At this point we have studied in detail the exact mapping between parafermions and spinful fermions, relating symmetries and various phases of matter in these representations.  In Secs.~\ref{FM_phase} and \ref{FM_phase2} we found that a parafermion chain with unpaired $\mathbb{Z}_4$ parafermion zero modes translates into an electronic topological superconductor that hosts symmetry-enriched Majorana zero modes and spontaneously breaks time-reversal symmetry.  A natural question arises in light of this connection: To what extent does the non-Abelian-anyon physics encoded through parafermion zero modes survive in the latter strictly 1D fermionic setting?
We will specifically address the survival of the three signature properties of non-Abelian anyons highlighted in the introduction: $(i)$ the existence of \emph{locally indistinguishable} ground states produced by the anyons, $(ii)$ non-Abelian braiding that `rigidly' rotates the system within the ground-state manifold, and $(iii)$ nontrivial fusion rules that specify the different types of quasiparticles that the anyons can form when they coalesce.  To bolster connection to experiment, in our treatment of the electronic setting below we will at most enforce $\mathcal{T}_{\rm elec} = \mathbb{Z}_4 \mathcal{T}$ symmetry and \emph{not} separately enforce $\mathbb{Z}_4$ (which is unnatural in that realization).

Concerning property $(i)$, a pair of $\mathbb{Z}_4$ parafermion zero modes yields four locally indistinguishable ground states.  The corresponding electron system certainly does not preserve this characteristic; Majorana modes generate two locally indistinguishable ground states, but the other two ground states reflect order-parameter configurations that local measurements readily distinguish.  We note that this point is well-appreciated by previous works on related electronic systems; see, e.g., Refs.~\onlinecite{ZhangKane,ScriptaPoorMansParafermions,Pedder,PoorMansParafermions}.

\begin{figure*}
\includegraphics[width=1.8\columnwidth]{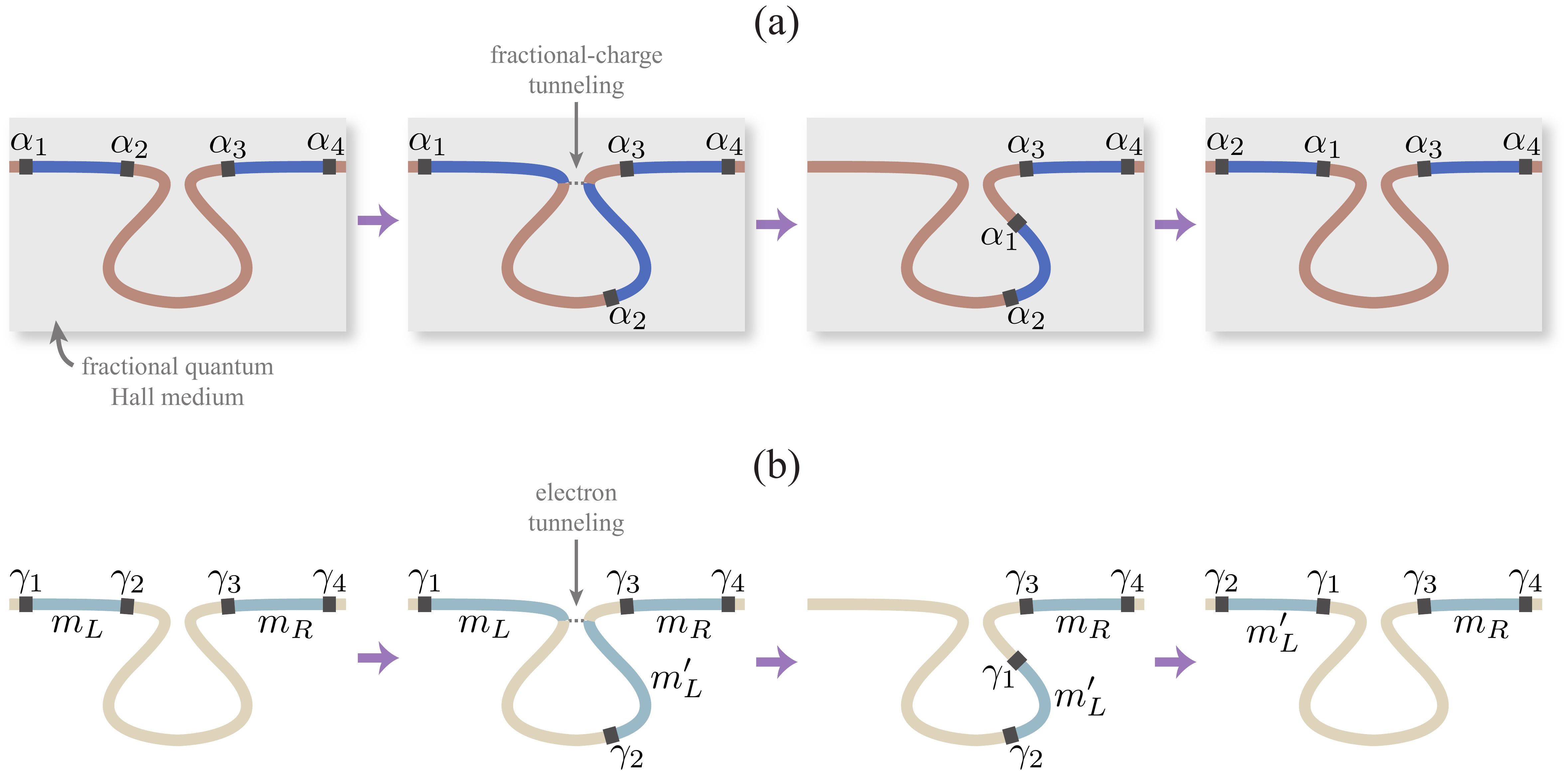}
\caption{Sample braiding protocol in (a) a $\mathbb{Z}_4$ parafermion platform and (b) its electronic counterpart.  In (a) $\mathbb{Z}_4$ parafermion zero modes $\alpha_{1,\ldots,4}$ arise at line defects in a parent fractional-quantum-Hall medium.  The sequence shown braids $\alpha_{1,2}$ (other braids proceed similarly).  The electron equivalent in (b) hosts two strictly 1D topological superconductors with spontaneously chosen magnetizations $m_{L/R}$ and symmetry-enriched Majorana zero modes $\gamma_{1,\ldots,4}$.  Here the panels sketch a braid of $\gamma_{1,2}$---which is \emph{not} described by parafermionic braid matrices.  Differences in braiding properties can be traced to the second panels above: in (a) the dashed line represents a parafermion coupling that is non-local when mapped to fermions.  Thus the Hamiltonian implementing parafermionic braid transformations is unphysical in the electronic realization.  Braiding $\gamma_{1,2}$ does nevertheless allow for additional freedom compared to conventional Majorana platforms, since the initial and final magnetizations, $m_L$ and $m_L'$, need not coincide.  }
\label{Braiding_fig}
\end{figure*}

To address property $(ii)$ we will first summarize non-Abelian braiding in the parafermion realization, which is known to be richer than in conventional Majorana systems [\onlinecite{Lindner:2012,  Clarke:2013a, ChengBraiding, Vaezi:2013, Barkeshli:2013a, Hastings:2013, LossBraiding}].  Imagine four $\mathbb{Z}_4$ parafermion zero modes $\alpha_{1,\ldots,4}$ realized at defects in a parent fractional-quantum-Hall fluid; see Fig.~\ref{Braiding_fig}(a).  For a given fixed overall $\mathbb{Z}_4$ charge, the system admits four degenerate ground states, and arbitrary superpositions of these states are physically permissible.  Braiding, as implemented, e.g., in Fig.~\ref{Braiding_fig}(a), rotates the system within this manifold.  One specifically finds that swapping $\alpha_j$ and $\alpha_{j+1}$ sends $\alpha_j \rightarrow \alpha_{j+1}$ and $\alpha_{j+1} \rightarrow i\alpha_j^\dagger \alpha_{j+1}^2$, \footnote{We focused on one particular chirality for the braid here.  Moreover, in the more general case the operators could transform as $\alpha_j \rightarrow e^{-i \frac{\pi}{2} k} \alpha_{j+1}$, $\alpha_{j+1} \rightarrow e^{i\frac{\pi}{2} (1-k)} \alpha_j^\dagger \alpha_{j+1}^2$ for integer $k$ \cite{Lindner:2012,Clarke:2013a}.  We have taken $k = 0$ for simplicity.} which is implemented by the unitary braid operator
\begin{equation}
  U_{j,j+1} = \exp\left\{\frac{i\pi}{8}[2(e^{i \frac{\pi}{4}}\alpha_j^\dagger \alpha_{j+1} + H.c.) - i (\alpha_j^\dagger \alpha_{j+1})^2]\right\}.
  \label{PFbraid}
\end{equation}

The equivalent 1D electronic setup, sketched in Fig.~\ref{Braiding_fig}(b), features a pair of topological superconductors each with spontaneous time-reversal symmetry breaking.  The left superconductor hosts symmetry-enriched Majorana zero modes $\gamma_{1,2}$, magnetization $m_L = i \Gamma_1\Gamma_2$, and fermion parity $P_{{\rm tot},L} = m_L(i\gamma_1\gamma_2)$; the right superconductor similarly hosts Majorana modes $\gamma_{3,4}$, magnetization $m_R = i \Gamma_3 \Gamma_4$, and parity $P_{{\rm tot},R} = m_R(i\gamma_3\gamma_4)$.  In this realization, physical wavefunctions---i.e., non-Schr\"odinger-cat states with fixed global fermion parity---take the form
\begin{align}
  \ket{\psi} &= a \ket{m_L,P_{{\rm tot},L};m_R,P_{{\rm tot},R}} 
  \nonumber \\
  &+ b \ket{m_L,-P_{{\rm tot},L};m_R,-P_{{\rm tot},R}}
\end{align}
for some complex $a,b$.  Compared to the parafermion case, we now have eight states instead of four, since only global $\mathbb{Z}_4^2$ charge needs to be fixed, though the allowed superpositions are strongly restricted by the need to avoid cat states.  

Braiding symmetry-enriched Majorana zero modes can induce rotations that are forbidden in conventional Majorana platforms yet still differ fundamentally from those in the parafermion realization.  Consider adiabatically swapping $\gamma_{i,j}$ such that the instantaneous Hamiltonian $H(t)$ does not explicitly break time-reversal symmetry at any point during the exchange.  The time-evolution operator implementing the braid is $U_{i,j}^{\rm elec}(t_i,t_f) = T e^{i \int_{t_i}^{t_f} dt H(t)}$.  Here $T$ denotes time ordering, and we take $t_i = -\infty$ and $t_f = +\infty$ as appropriate for an adiabatic process.  Applying time reversal yields $\mathcal{T}_{\rm elec} U_{i,j}^{\rm elec}(t_i,t_f) \mathcal{T}_{\rm elec}^{-1} = [U_{i,j}^{\rm elec}(t_f,t_i)]^\dagger$.  On the right side, Hermitian conjugation reverses the braid chirality but so does swapping $t_i \leftrightarrow t_f$.  These factors thus `cancel', so that the braid operator satisfies
\begin{equation}
\mathcal{T}_{\rm elec} U_{i,j}^{\rm elec} \mathcal{T}_{\rm elec}^{-1} = U_{i,j}^{\rm elec}~.
\label{TimeReversalBraidingConstraint}
\end{equation}
For a similar analysis see Ref.~\onlinecite{LiuTRITOPSBraiding}.  Equation~\eqref{TimeReversalBraidingConstraint} together with parity conservation allow us to infer the braiding properties of symmetry-enriched Majorana modes.  All results below have been verified by explicit calculations similar to those in Ref.~\onlinecite{Clarke:2013a}.  

Figure~\ref{Braiding_fig}(b) sketches an exchange of $\gamma_1$ and $\gamma_2$.  The first step of the braid extends the left magnetized region into the lower loop.  Crucially, the magnetization $m_L'$ in the loop segment can either align or anti-align with the original magnetization $m_L$ depending on details of the junction Hamiltonian.  If $m_L' = m_L$ then the braid preserves the magnetization, and we obtain a standard Majorana exchange that acts as 
\begin{equation}
  U_{1,2}^{\rm elec}\gamma_1 (U_{1,2}^{\rm elec})^\dagger = -s\gamma_2,~~U_{1,2}^{\rm elec}\gamma_2(U_{1,2}^{\rm elec})^\dagger = s\gamma_1
  \label{U12action}
\end{equation}
for some sign $s$ \cite{Ivanov, AliceaBraiding, ClarkeBraiding}.  As usual, the extra minus sign acquired by one of the Majorana operators is necessary to ensure conservation of parity $P_{{\rm tot},L}$ for the left topological region.  By applying time reversal to Eq.~\eqref{U12action} using Table~\ref{symmetry_table3} and Eq.~\eqref{TimeReversalBraidingConstraint}, one finds that the left and right sides are consistent only if $s$ does not depend on magnetization.  
Taking $s = +1$ for concreteness, the associated braid matrix then reads
\begin{equation}
  U_{1,2}^{\rm elec} = e^{\frac{\pi}{4}\gamma_1\gamma_2}~~({\text{mag.-preserving~braid}})~.
  \label{U12a}
\end{equation} 

If $m_L' = -m_L$ then the braid flips the magnetization.  In this case conservation of $P_{{\rm tot},L}$ dictates that the Majorana operators (written in our conventions) transform slightly differently from above: 
\begin{equation}
  U_{1,2}^{\rm elec}\gamma_1 (U_{1,2}^{\rm elec})^\dagger = s' \gamma_2,~~U_{1,2}^{\rm elec}\gamma_2(U_{1,2}^{\rm elec})^\dagger = s' \gamma_1
\end{equation}
with some sign $s'$.  Consistency with time reversal now requires $s' = m_L$ (or $s' = -m_L$, but we focus on the former for simplicity).  This transformation is implemented by
\begin{equation}
  U_{1,2}^{\rm elec} = \frac{1}{\sqrt{2}}(e^{-i \frac{\pi}{4}} \Gamma_1\gamma_1 + e^{i \frac{\pi}{4}}\Gamma_2\gamma_2)~~({\text{mag.-flipping~braid}})~.
  \label{U12b}
\end{equation}
Note that in addition to transforming $\gamma_{1,2}$, $U_{1,2}^{\rm elec}$ also sends $\Gamma_1 \rightarrow p\Gamma_2$ and $\Gamma_2 \rightarrow p\Gamma_1$, yielding the required magnetization flip $m_L \rightarrow -m_L$.  
We stress that Eq.~\eqref{U12b} can not describe an adiabatic closed cycle in a Majorana system with explicit time-reversal symmetry breaking, for which the initial and final magnetizations would necessarily coincide.  

Other braids can be analyzed similarly.  The braid matrix $U_{3,4}^{\rm elec}$ governing the exchange of $\gamma_3$ and $\gamma_4$ clearly conforms to a straightforward generalization of Eqs.~\eqref{U12a} and \eqref{U12b}.  Swapping zero modes $\gamma_{2,3}$ that reside on different topological segments, however, naturally preserves both magnetizations.  We find that consistency with time reversal yields
\begin{equation}
U_{2,3}^{\rm elec} = \exp\left[\frac{\pi}{8}(1+m_L-m_R+m_Lm_R)\gamma_2\gamma_3\right]~.
\label{U23}
\end{equation}
One can readily verify using Table~\ref{symmetry_table3} that Eqs.~\eqref{U12a}, \eqref{U12b}, and \eqref{U23} all satisfy Eq.~\eqref{TimeReversalBraidingConstraint}.

To directly compare the parafermion and electronic braid matrices, we will now recast Eq.~\eqref{PFbraid} in terms of Majorana operators $\gamma_j$ and $\Gamma_j$ using exact mappings that generalize Eqs.~\eqref{alpha1} and \eqref{alpha2} to the case with four parafermion zero modes.  Appendix~\ref{BraidingAppendix} sketches this exercise.  For $U_{1,2}$ we obtain
\begin{equation}
  U_{1,2} = \exp\left\{ \frac{i \pi}{4}\left[i(\gamma_2\Gamma_1 + \gamma_1\Gamma_2)-\frac{1}{2}P_{{\rm tot},L}\right]\right\}~,
  \label{U12fermionic}
\end{equation} 
which is clearly very different from $U_{1,2}^{\rm elec}$.  The first two pieces in the exponent swap local Majorana operators $\gamma_{1,2}$ with the non-local operators $\Gamma_{1,2}$; consequently, when acting on generic physical fermion wavefunctions $\ket{\psi}$, $U_{1,2}$ generates cat states that superpose $m_L = \pm 1$ configurations (see Appendix~\ref{BraidingAppendix}).  A similar conclusion holds for $U_{3,4}$.  
For $U_{2,3}$ we find
\begin{equation}
  U_{2,3} = \exp\left\{\frac{i\pi}{4}\left[(m_L + m_R)i\gamma_2 \gamma_3 - \frac{1}{2}m_L m_R\right]\right\}~,
  \label{U23fermionic}
\end{equation}
which preserves the magnetizations and thus does not generate cat states.  Nevertheless $U_{2,3}$ and $U_{2,3}^{\rm elec}$ still differ qualitatively, and in fact the latter generates a finer protected rotation of the $\gamma_{2,3}$ zero-mode operators compared to the former.  

The stark contrast between parafermion and electronic braid matrices seen here may appear surprising given that exact mappings bridge the two representations.  This difference originates from the fact that the physical Hamiltonian governing the exchange in the parafermion realization becomes non-local when translated into fermion language.  Specifically, the dashed line from Fig.~\ref{Braiding_fig}(a), second panel, represents a coupling between parafermions at opposite edges of the loop, which microscopically arises from tunneling of fractional charge through the intervening quantum-Hall fluid.  Mapping this term to spinful fermions generates an `uncanceled' string across the entire loop below---yielding an unphysical process in this representation.  Instead the analogous physical coupling in the electronic realization arises from ordinary electron tunneling (along with coupling between the magnetizations) across the upper part of the loop; see Fig.~\ref{Braiding_fig}(b) \footnote{Our discussion here applies equally well to the braiding scheme proposed in Ref.~\onlinecite{Orth} in the quantum-spin-Hall setting.}.

\begin{figure}
\includegraphics[width=\columnwidth]{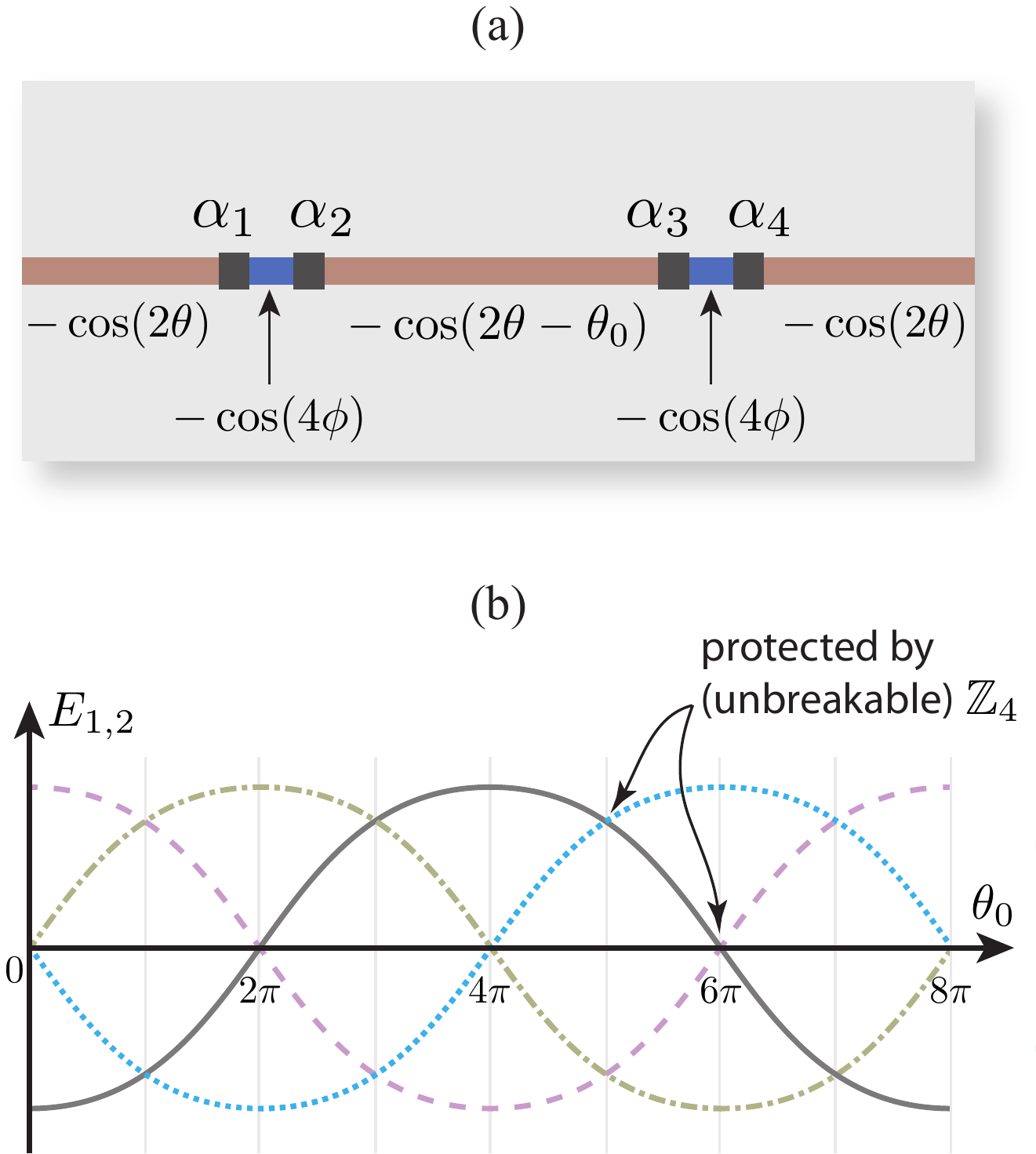}
\caption{(a) Setup used for fusion in a $\mathbb{Z}_4$ parafermion platform.  Parafermions $\alpha_1$ and $\alpha_2$ hybridize on the left, and similarly for $\alpha_3$ and $\alpha_4$ on the right.  We label the bosonized perturbations gapping each region; note in particular the shift $\theta_0$ in the central region, which modulates the parafermion couplings.  (b) Energies $E_{1,2}$ for the hybridized parafermions $\alpha_{1,2}$ versus $\theta_0$.  All level crossings are protected by the parafermion platform's unbreakable $\mathbb{Z}_4$ symmetry.  For a given $\theta_0$ the different energy levels correspond to the four possible fusion channels for the non-Abelian defects binding the parafermions.  Adiabatically winding $\theta_0$ cycles the system among these four fusion channels, leading to an anomalous $8\pi$-periodic response even though the underlying Hamiltonian is $2\pi$ periodic. }
\label{Fusion_fig_PF}
\end{figure}

The situation for fusion, property $(iii)$, is different.  Fusion brings two zero modes together, thereby intentionally removing any topologically protected degeneracies that arise when the zero modes are far apart.  In the context of fusion properties, the distinction between the parafermion and electronic realizations is thus naturally blurred.  Consider a parafermion platform and let $X$ denote a domain-wall defect that binds a $\mathbb{Z}_4$ parafermion zero mode.  These non-Abelian defects obey the fusion rule
\begin{equation}
  X\times X \sim I + q_1 + q_2 + q_3~,
  \label{FusionRule}
\end{equation}
indicating that two defects can annihilate, corresponding to the identity fusion channel $I$, or form three different nontrivial quasiparticle types $q_{1,2,3}$.  We will explore this fusion rule further by examining the setup from Fig.~\ref{Fusion_fig_PF}(a) that hybridizes the pair $\alpha_{1,2}$ as well as the pair $\alpha_{3,4}$.  The figure indicates the bosonized perturbation gapping out each region; most importantly, the central domain is gapped by $-\cos(2\theta-\theta_0)$, where $\theta_0$ represents a `knob' that we will use to probe the parafermionic fusion characteristics.  

In Appendix~\ref{FusionHamiltonian} we show that hybridization between $\alpha_{1,2}$ can be described by the Hamiltonian
\begin{equation}
  H_{1,2} = -t \left[e^{i\frac{\pi-\theta_0}{4}}\alpha_1^\dagger \alpha_2 + H.c.\right]
  \label{FusionH}
\end{equation}
for some real coupling $t$ that we take to be postive (hybridization between $\alpha_{3,4}$ can be treated similarly).  When $\theta_0 = 0$ $H_{1,2}$ admits a unique ground state with $e^{i \frac{\pi}{4}} \alpha_1^\dagger \alpha_2 = 1$, corresponding to the identity fusion channel in Eq.~\eqref{FusionRule}; excited states with $e^{i \frac{\pi}{4}} \alpha_1^\dagger \alpha_2 = \pm i,-1$ correspond to the three possible nontrivial quasiparticles $q_{1,2,3}$.  Figure~\ref{Fusion_fig_PF}(b) plots the energy spectrum for $H_{1,2}$ as a function of $\theta_0$.  Crucially, all level crossings are protected by the (unbreakable) $\mathbb{Z}_4$ symmetry exhibited by the parafermion platform, thus  strongly constraining the system's response to $\theta_0$ sweeps.  As an example, imagine starting from the ground state with $\theta_0 = 0$ and then adiabatically winding $\theta_0$ by $2\pi$.  This cycle returns the Hamiltonian to its original form---which is clear from Fig.~\ref{Fusion_fig_PF}(a) \footnote{Shifting $\theta_0$ by $2\pi$ also returns the hybridization Hamiltonian $H_{1,2}$ to its original form, when followed by a gauge transformation $\alpha_2 \rightarrow i \alpha_2$.  The key point is that $e^{i \frac{\pi}{4}} \alpha_1^\dagger \alpha_2$ is a conserved quantity; once fixed, the eigenvalue thus can not readjust to accommodate shifts in $\theta_0$.}---but maps the ground state into an excited state.  The system re-enters its ground state only after sweeping $\theta_0$ by a total of $8\pi$.  This anomalous periodicity reflects the fact that \emph{winding $\theta_0$ cycles the system among the four possible fusion channels in Eq.~\eqref{FusionRule}.}  The pumping cycle reviewed here is a cousin of the generalized fractional Josephson effect discussed for parafermions in fractional-quantum-Hall systems in Refs.~\onlinecite{Lindner:2012, Clarke:2013a, ChengBraiding, ChengLutchyn}.

Below we will turn to the equivalent electronic setup and identify an analogous $8\pi$-periodic pumping cycle that, remarkably, represents a purely 1D manifestation of nontrivial parafermionic fusion rules.  We will also draw connections with closely related work in Refs.~\onlinecite{ZhangKane,Orth} in the context of interacting quantum-spin-Hall edges, viewed from a new perspective in light of our mappings.  

\subsection{Imprint of parafermionic fusion rules in a 1D electron system}

\begin{figure*}
\includegraphics[width=2\columnwidth]{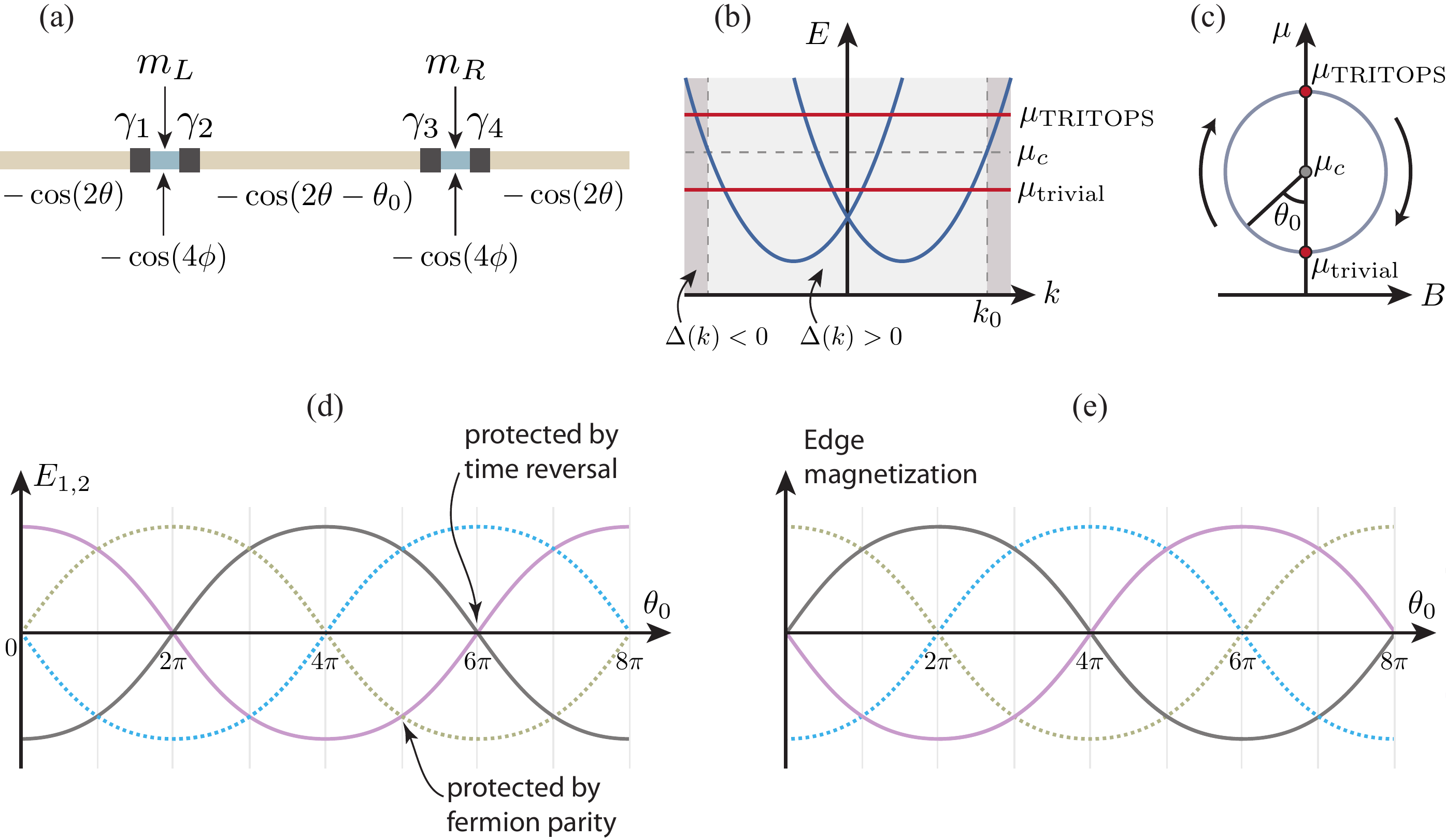}
\caption{(a) Electronic counterpart of the fusion setup from Fig.~\ref{Fusion_fig_PF}(a).  Outer regions form a trivial phase smoothly connected to the electron vacuum. The central region interpolates between a trivial phase at $\theta_0 = 0$ and TRITOPS phase at $\theta_0 = \pi$, and can be realized experimentally by a spin-orbit-coupled wire with an $s$-wave pair potential $\Delta(k)$ that changes sign at some momentum $k_0$.  (b) Band structure for such a wire along with chemical potentials corresponding to trivial and TRITOPS phases.  In this realization one can wind $\theta_0$ by $2\pi$ by varying the chemical potential $\mu$ and an applied magnetic field $B$ along the cycle shown in (c).  Hybridization of the symmetry-enriched Majorana operators $\gamma_{1,2}$ and fluctuating quantum magnetization degree of freedom $m_L$ yields the energy spectrum versus $\theta_0$ sketched in (d).  The levels are similar to those in the parafermion platform [Fig.~\ref{Fusion_fig_PF}(b)] except that crossings at $\theta_0 = \pi \mod 2\pi$ are protected by fermion parity whereas those at $\theta_0 = 0 \mod 2\pi$ are protected by electronic time-reversal symmetry.  Provided these crossings are maintained, the system inherits the parafermion platform's $8\pi$-periodic pumping cycle---an imprint of nontrivial parafermionic fusion rules in our strictly 1D electron setting.  The pumping cycle can be detected experimentally by measuring the magnetization at the edge, which as (e) illustrates is also $8\pi$ periodic.  Magnetization for a given curve in (d) is shown with the same line type in (e).}
\label{Fusion_fig_elec}
\end{figure*}

Figure~\ref{Fusion_fig_elec}(a) shows the strictly 1D electronic counterpart of the parafermion platform from Fig.~\ref{Fusion_fig_PF}(a).  Recall that the outer segments gapped by $-\cos(2\theta)$ form trivial, $\mathcal{T}_{\rm elec}$-invariant gapped phases that smoothly connect to the fermion vacuum.  The central region gapped by $-\cos(2\theta-\theta_0)$ interpolates between a trivial phase (at $\theta_0 = 0$) and a $\mathcal{T}_{\rm elec}$-invariant TRITOPS phase (at $\theta_0 = \pi$) via a path that explicitly breaks $\mathcal{T}_{\rm elec}$.  For a practical implementation of this region, we envision a spin-orbit-coupled wire with a momentum-dependent $s$-wave pairing potential that changes sign at some momentum $k_0$.  As Fig.~\ref{Fusion_fig_elec}(b) illustrates, trivial and TRITOPS phases arise depending on 
whether the outer Fermi momentum is smaller or larger than $k_0$ \cite{FermiSurfaceInvariant,WongTRITOPS,NakosaiTRITOPS,ZhangTRITOPS, HaimTRITOPS}.  One can, moreover, smoothly tune between these phases by varying the chemical potential $\mu$ and a $\mathcal{T}_{\rm elec}$-breaking magnetic field $B$ along the path shown in Fig.~\ref{Fusion_fig_elec}(c)---which in bosonized language winds $\theta_0$ by $2\pi$.  Note that the $B$ field generically induces both a Zeeman term \emph{and} an imaginary component to the $s$-wave pair potential, thus preempting a phase transition.  

The `small' adjacent $-\cos(4\phi)$ regions in Fig.~\ref{Fusion_fig_elec}(a) exhibit magnetizations that now form fluctuating quantum degrees of freedom. 
Consequently, the Majorana operators $\Gamma_j$ that we used to decompose the magnetizations become physical operators that can appear in the Hamiltonian, in addition to the symmetry-enriched Majorana operators $\gamma_j$.  Focusing on the left region, we describe hybridization of these operators by an effective Hamiltonian
\begin{equation}
  H_{1,2}^{\rm elec} = H_t + H_{\mathbb{Z}_4{\rm {-breaking}}}~.
\end{equation}
The first term,
\begin{align}
  H_t = -t\bigg{[}&\cos\left(\frac{\theta_0}{4}\right)i (\gamma_1\Gamma_2 + \gamma_2\Gamma_1) 
  \nonumber \\
  - &\sin\left(\frac{\theta_0}{4}\right)i (\gamma_1\Gamma_1 + \gamma_2\Gamma_2)\bigg{]}~,
  \label{Ht}
\end{align}
represents Eq.~\eqref{FusionH} rewritten in terms of fermions using Eqs.~\eqref{alpha1} and \eqref{alpha2}.  At both $\theta_0 = 0$ and $\theta_0 = \pi$, $H_t$ preserves $\mathcal{T}_{\rm elec}$ symmetry \footnote{At $\theta_0 = 0$ the Majorana operators transform under $\mathcal{T}_{\rm elec}$ precisely as in Table~\ref{symmetry_table3} from Sec.~\ref{FM_phase}.  At $\theta_0 = \pi$, however, the domain configuration differs from that analyzed in Sec.~\ref{FM_phase}, so here one obtains the modified transformations $\gamma_1 \rightarrow m \gamma_1$, $\gamma_2 \rightarrow - m \gamma_2$, $\Gamma_1 \rightarrow - p \Gamma_2$, and $\Gamma_2 \rightarrow p \Gamma_1$ under $\mathcal{T}_{\rm elec}$.}.  The second term, $H_{\mathbb{Z}_4{\rm {-breaking}}}$, encodes additional allowed couplings that violate $\mathbb{Z}_4$ symmetry and hence are unphysical in the parafermion context; we assume that this piece also preserves $\mathcal{T}_{\rm elec}$ at $\theta_0 = 0$.  For any $\theta_0$ the Hamiltonian commutes with $P_{{\rm tot},L} = (i \gamma_1\gamma_2)(i\Gamma_1\Gamma_2)$.  

Suppose for now that $H_{\mathbb{Z}_4{\rm {-breaking}}} = 0$.  Figure~\ref{Fusion_fig_elec}(d) sketches the resulting energies $E_{1,2}$ versus $\theta_0$; solid and dashed curves respectively correspond to states with $P_{{\rm tot},L} = +1$ and $-1$.  By construction the energies are identical to those in Fig.~\ref{Fusion_fig_PF}(b), though the nature of the eigenstates changes.  At $\theta_0 = 0$, $H_t$ energy eigenstates have $i \gamma_1 \Gamma_2 = \pm 1$ and $i \gamma_2 \Gamma_1 = \pm 1$.  The many-body spectrum correspondingly features non-degenerate states with energies $\pm 2t$ along with a degenerate Kramers doublet of states at zero energy.  Increasing $\theta_0$ breaks $\mathcal{T}_{\rm elec}$ and eliminates the degeneracy until time-reversal symmetry is revived at $\theta_0 = \pi$.  To understand the $\theta_0 = \pi$ spectrum it is convenient to employ a rotated basis $\gamma_\pm = (\gamma_1 \pm \gamma_2)/\sqrt{2}$ and $\Gamma_\pm = (\Gamma_1 \pm \Gamma_2)/\sqrt{2}$.  The $t$ term then becomes 
\begin{equation}
  H_t(\theta_0 = \pi) = \sqrt{2}t i \gamma_-\Gamma_-~.
\end{equation}
Notice that $i\gamma_+ \Gamma_+$---which is odd under $\mathcal{T}_{\rm elec}$---does not appear in the Hamiltonian, i.e., the system supports a fermionic zero mode corresponding to the hallmark Majorana Kramers pair for a TRITOPS phase \footnote{While it is illuminating to describe the Majorana Kramers pair using our effective Hamiltonian that couples $\gamma_i$ and $\Gamma_i$, its existence more fundamentally arises from the TRITOPS state. That is, the `small' $-\cos(4\phi)$ region functions as a quantum dot that houses the Majorana Kramers pair that is guaranteed to exist due to the adjacent TRITOPS region.}. The many-body spectrum thus contains levels at $\pm \sqrt{2} t$, each with two degenerate states carrying opposite fermion parity.  Further increasing $\theta_0$ to $2\pi$ yields a spectrum identical to that at $\theta_0 = 0$, except with the $P_{{\rm tot},L}$ eigenvalues reversed.  

As a technical aside, the opposite $P_{{\rm tot},L}$ eigenvalues at $\theta_0 = 0$ and $2\pi$ may seem surprising.  Clearly the bosonized Hamiltonian is identical at $\theta_0 = 0$ and $2\pi$, so the energies and eigenstates must also be identical at these points.  The resolution is that in our conventions the bosonized fermion-parity operator $e^{i \int_x \partial_x\theta}$ across the left $-\cos(4\phi)$ region projects to $P_{{\rm tot},L}$ at $\theta_0 = 0$ but $-P_{{\rm tot},L}$ at $\theta_0 = 2\pi$ \footnote{For an explicit example, at either $\theta_0 = 0$ or $2\pi$, the ground state is unique and must have $\theta$ pinned to the same value on both sides of the `small' $-\cos(4\phi)$ region (twists in $\theta$ cost energy in such geometries). Thus ground-state projection yields $e^{i \int_x \partial_x\theta}\rightarrow 1$.  From Eq.~\eqref{Ht}, however, one can readily see that $P_{{\rm tot},L} = (i\gamma_1\gamma_2)(i\Gamma_1\Gamma_2)$ projects to $+1$ at $\theta_0 = 0$ but $-1$ at $\theta_0 = 2\pi$.}; thus opposite $P_{{\rm tot},L}$ eigenvalues are actually required.  The virtue of this convention is that tracking the evolution of states in response to $\theta_0$ sweeps becomes particularly transparent. 

Turning on $H_{\mathbb{Z}_4{\rm {-breaking}}} \neq 0$ of course non-universally modifies the energies in Fig.~\ref{Fusion_fig_elec}(d).  Nevertheless, 
the level crossings at $\theta_0 = 0 \mod 2\pi$ remain protected by $\mathcal{T}_{\rm elec}$, whereas the crossings at $\theta_0 = \pi \mod 2\pi$ are unbreakable due to fermion-parity protection.  (Breaking $\mathcal{T}_{\rm elec}$ can only shift the the latter degeneracy points to different $\theta_0$ values but can not turn them into avoided crossings.)  Consequently, despite the obliteration of $\mathbb{Z}_4$ symmetry, our 1D electronic system inherits the parafermion platform's anomalous $8\pi$-periodic pumping cycle, so long as $\mathcal{T}_{\rm elec}$ is preserved at $\theta_0 = 0 \mod 2\pi$.  

We can understand the pumping process physically as follows.  Suppose the system starts in its unique ground state at $\theta_0 = 0$.  Due to conservation of $P_{{\rm tot},L}$, adiabatically winding $\theta_0$ to $2\pi$ necessarily evolves the system to an excited state in which the fermion parity in the left $-\cos(4\phi)$ region has flipped (recall the relation between fermion parity and $P_{{\rm tot},L}$ noted above).  That is, the $0\rightarrow 2\pi$ sweep pumps a fermion between the left and right $-\cos(4\phi)$ regions, producing a state that generically exhibits a non-zero magnetization even though the ending Hamiltonian preserves $\mathcal{T}_{\rm elec}$.  Subsequently sweeping $\theta_0$ from $2\pi$ to $4\pi$ restores the original fermion parities.  Time-reversal symmetry, however, now prevents the system from returning to the ground state.  Restoring the ground state requires winding $\theta_0$ by a total of $8\pi$.  One can experimentally probe this anomalous pumping cycle by measuring the magnetization at the edge of the wire, which exhibits the same $8\pi$ periodicity.  Figure~\ref{Fusion_fig_elec}(e) sketches possible magnetization curves colored according to the corresponding branch in Fig.~\ref{Fusion_fig_elec}(d).  Note that dispensing with $\mathcal{T}_{\rm elec}$ still yields a nontrivial $4\pi$-periodic cycle; in this case the pumping process becomes very similar to that introduced in Ref.~\onlinecite{KeselmanTRITOPS} (see also Ref.~\onlinecite{BergPumping}).  

The specific electronic setup examined so far makes the connection to parafermions explicit and also allows one to capture the $8\pi$-periodic pumping cycle within a very simple effective Hamiltonian.  However, the requirements for implementing the cycle in practice can be distilled into a few basic ingredients shared by a much broader class of superconducting systems: 
\begin{itemize}
\item A generic family of electron Hamiltonians $H(\theta_0)$, where $\theta_0$ is an adiabatic parameter such that $H(\theta_0 + 2\pi) = H(\theta_0)$.  By `generic' we mean that $H(\theta_0)$ should contain no accidental degeneracies.  
\item $H(\theta_0)$ describes a phase that preserves electronic time-reversal symmetry if and only if $\theta_0 = 0 \mod \pi$.  At these $\theta_0$ points time reversal guarantees Kramers degeneracy for states with odd electron number.  
\item A single fermion zero mode---or equivalently, a pair of Majorana zero modes---at each end of the system when $\theta_0 = \pi \mod 2\pi$.  Due to time-reversal invariance at this point, the zero mode must be anomalous.  
\item A set of four many-body sub-gap states whose evolution is constrained by the first three items above.  These sub-gap states must be separated from the continuum for any value of $\theta_0$ so that an adiabatic pumping cycle is well-defined.
\end{itemize}
(Once these items are satisfied, one can actually break time-reversal symmetry at $\theta_0 = \pi \mod 2\pi$ without spoiling the $8\pi$ periodicity, consistent with the preceding discussion.)  Perhaps most importantly, the `small' $-\cos(4\phi)$ regions [Fig.~\ref{Fusion_fig_elec}(a)] bordering our spin-orbit-coupled wire are inessential.   Any source of sub-gap states---e.g., band bending at the edges---can satisfy the last item in this list.  In this modified picture, the symmetry-enriched Majorana modes and fluctuating magnetization degree of freedom are simply adiabatically deformed into a pair of fermions encoding those sub-gap states.  Further intuition can be gained by comparing with the more familiar $4\pi$-periodic fractional Josephson effect \cite{Kitaev:2001} arising in junctions formed by a pair of topological superconductors with explicitly broken time-reversal symmetry.  There, the nontrivial $4\pi$-periodic cycle is conveniently understood as arising from Majorana modes that hybridize across a finite-width barrier in the junction; the effect survives equally well, however, if the barrier width shrinks to zero---so long as a sub-gap localized state persists.  The latter sub-gap state is continuously connected to the hybridized Majorana modes in the finite-barrier situation, just as the sub-gap states in our problem are connected to the symmetry-enriched Majorana modes and magnetization degree of freedom.

\begin{figure}
\includegraphics[width=\columnwidth]{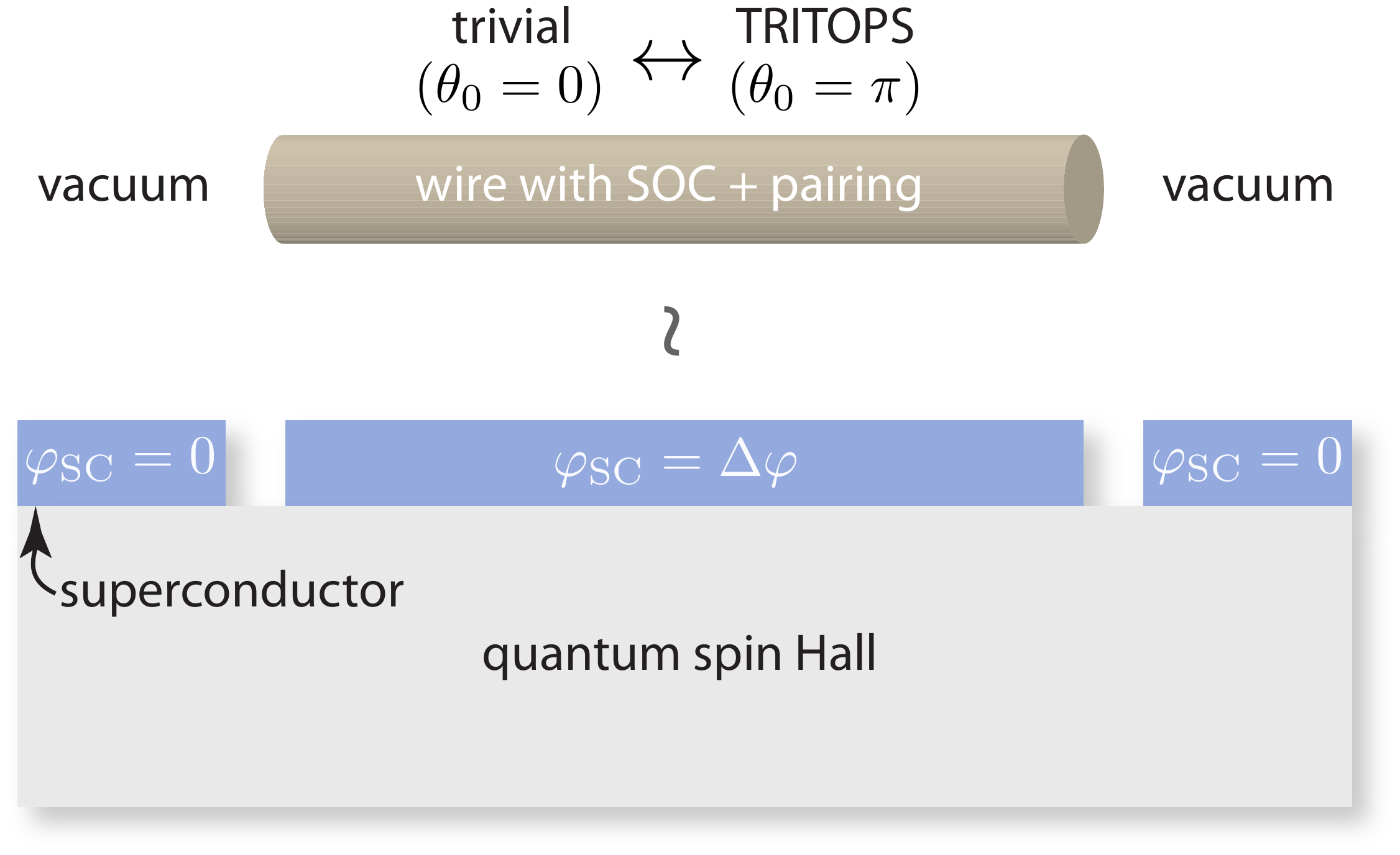}
\caption{Connection between our strictly 1D electronic system (top) and a quantum-spin-Hall Josephson junction (bottom). The outer vacuum regions in the 1D setting correspond to segments of the Josephson junction with superconducting phase $\varphi_{\rm SC} = 0$.  The wire with spin-orbit-coupling (SOC) and momentum-dependent $s$-wave pairing corresponds to the central part of the junction with phase $\varphi_{\rm SC} = \Delta \varphi$.  Varying the adiabatic parameter $\theta_0$ in the 1D system yields an $8\pi$-periodic edge magnetization, while varying $\Delta \varphi$ yields an $8\pi$-periodic Josephson current.}
\label{Comparison_fig}
\end{figure}

References \onlinecite{ZhangKane,Orth} introduced a quite different platform satisfying the above properties, namely a Josephson junction realized at a quantum-spin-Hall edge.  The quantum-spin-Hall setup is described by the same bosonized perturbations from Fig.~\ref{Fusion_fig_elec}(a), but with $\theta \leftrightarrow \phi$ (in the notation of Ref.~\onlinecite{ZhangKane}) and the adiabatic parameter $\theta_0$ replaced by the superconducting-phase difference $\Delta\varphi$ across the junction.  The $\cos(4\theta)$ terms in the barrier regions of the Josephson junction reflect two-particle backscattering; when relevant, these perturbations catalyze spontaneous time-reversal symmetry breaking with a magnetization order parameter $\cos(2\theta)$---very similar to the order parameter in our problem.  Electronic time-reversal symmetry is present at $\Delta \varphi = 0$ and $\pi$, and at the latter value the barrier binds a single fermionic zero mode.  Moreover, the necessary sub-gap levels can arise from Andreev bound states in a `wide' junction.  These properties, in conjunction with arbitrarily weak interactions, conspire to yield an $8\pi$-periodic Josephson current.  

Figure~\ref{Comparison_fig} summarizes the relation between our strictly 1D realization and the analogous quantum-spin-Hall setup.  In the latter setting, the anomalous Josephson effect can also be naturally viewed as arising from hybridization of symmetry-enriched Majorana modes with a quantum magnetization degree of freedom---similar to Refs.~\onlinecite{Peng,Hui,Vinkler} which analyzed the junction coupled to an impurity spin.  Our exact mappings clarify the precise connection between these electronic setups and a system hosting bona fide $\mathbb{Z}_4$ parafermions: the hybridized sub-gap states mediating the anomalous pumping cycles are in one-to-one correspondence with fusion channels of non-Abelian defects binding $\mathbb{Z}_4$ parafermion zero modes.  Given our general discussion in Sec.~\ref{nonAbelian1D}, which applies equally well to the strict 1D and quantum-spin-Hall platforms, we expect that this is the maximal extent to which non-fractionalized electron systems inherit non-Abelian $\mathbb{Z}_4$-parafermion physics.  

\begin{figure}
\includegraphics[width=1\columnwidth]{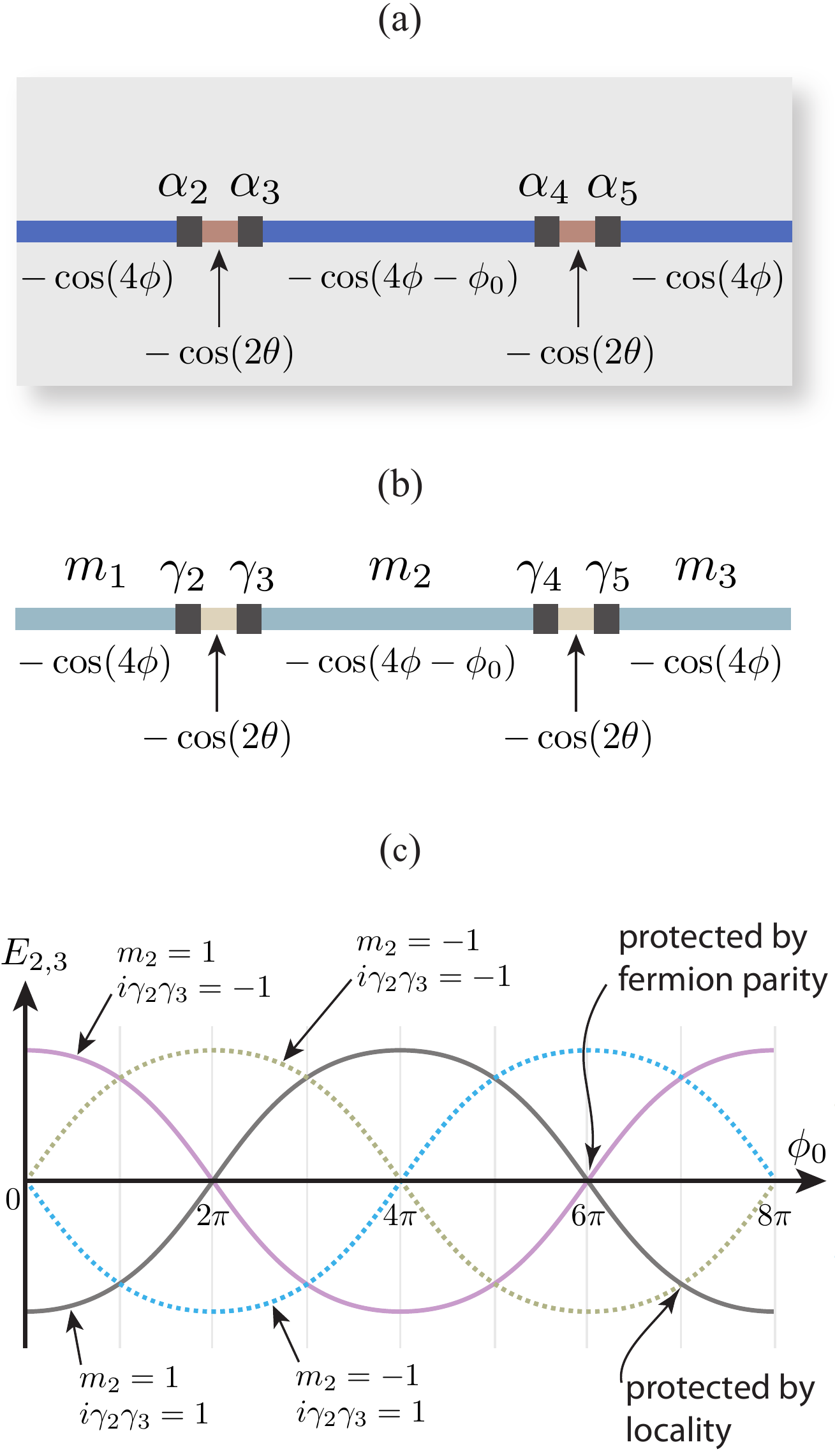}
\caption{Variation of Figs.~\ref{Fusion_fig_PF}(a) and \ref{Fusion_fig_elec}(a) for (a) a parafermion platform and (b) the corresponding electron system.  Here a pumping process is carried out by varying the parameter $\phi_0$ in the interactions governing the central region.  In (b), $m_{1,2,3}$ denote spontaneously chosen magnetizations for the adjacent domains.  (c) Energy spectrum describing hybridization of symmetry-enriched Majorana modes $\gamma_{2,3}$ at the left junction in (b), assuming fixed $m_1 = +1$.  All level crossings are protected by either locality or fermion-parity considerations.  The electronic system therefore exhibits an anomalous $8\pi$-periodic response to $\phi_0$ even when all symmetries are abandoned.   }
\label{Fusion_fig_elec_2}
\end{figure}

We conclude this section with a discussion of the alternative fusion setup shown in Fig.~\ref{Fusion_fig_elec_2}.  Compared to our previous setups, the $\cos(2\theta)$ and $\cos(4\phi)$ regions have essentially swapped roles.  Note especially that the central domain in the figure is gapped by $-\cos(4\phi-\phi_0)$, where $\phi_0$ is the control parameter that we wish to vary.  For the parafermion realization in Fig.~\ref{Fusion_fig_elec_2}(a), Appendix \ref{FusionHamiltonian} shows that parafermions $\alpha_{2,3}$ at the left junction hybridize according to
\begin{equation}
  H_{2,3} = -t \left[ e^{i \frac{\pi + \phi_0}{4}} \alpha_2^\dagger \alpha_3 + H.c.\right],
\end{equation} 
which takes a nearly identical form to Eq.~\eqref{FusionH}.  Modulating $\phi_0$ thus also cycles the system among the four possible fusion channels in Eq.~\eqref{FusionRule}, in turn generating a robust $8\pi$-periodic response even though the microscopic Hamiltonian is $2\pi$ periodic.  Fusing parafermions across regions gapped by $\cos(4\phi)$ versus $\cos(2\theta)$ evidently makes little difference.  

The electronic realization in Fig.~\ref{Fusion_fig_elec_2}(b) nevertheless differs starkly from Fig.~\ref{Fusion_fig_elec}(a) because pairs of symmetry-enriched Majorana modes now hybridize across \emph{trivial} domains.  The outer regions gapped by $-\cos(4\phi)$ exhibit spontaneously chosen magnetizations $m_1$ and $m_3$ determined by $\langle \cos(2\phi)\rangle = \pm 1$, while the central region gapped by $-\cos(4\phi-\phi_0)$ exhibits a magnetization
\begin{equation}
  m_2 = \langle \cos(2\phi-\phi_0/2)\rangle = \pm 1
  \label{m2}
\end{equation}
whose microscopic meaning evolves with $\phi_0$. For example, $m_2$ corresponds to a magnetization along $x$ at $\phi_0 = 0$ but along $y$ at $\phi_0 = \pi$; see Eqs.~\eqref{m_lattice} and \eqref{barm_lattice}.  

Converting $H_{2,3}$ into fermionic language using the dictionary from Appendix~\ref{BraidingAppendix} yields \footnote{Despite appearances, the hybridization Hamiltonian $H_{2,3}$ is also $2\pi$ periodic in $\phi_0$, both in the parafermionic and fermionic representations.  In the fermionic case, the periodicity reflects the fact that sending $\phi_0 \rightarrow \phi_0 + 2\pi$ shifts $m_2 \rightarrow -m_2$ and $\gamma_3 \rightarrow m_2 \gamma_3$.}, 
\begin{align}
  H_{2,3} = -t\bigg{[}&(m_1+m_2)\cos\left(\frac{\phi_0}{4}\right) 
  \nonumber \\
  &+ (1-m_1m_2)\sin\left(\frac{\phi_0}{4}\right)\bigg{]}i \gamma_2\gamma_3~.
  \label{H23m}
\end{align}
For simplicity let us fix the magnetization in the leftmost region to $m_1 = +1$.  
Figure~\ref{Fusion_fig_elec_2}(c) sketches the energy levels $E_{2,3}$ versus $\phi_0$ for the four remaining sectors labeled by $m_2 = \pm 1$ and $i \gamma_2 \gamma_3 = \pm1$.  The level crossings at $\phi_0 = 0\mod 2\pi$ arise from states with opposite fermion parity and are therefore protected. Furthermore, the crossings at $\phi_0 = \pi \mod 2\pi$ arise from macroscopically distinct states carrying opposite $m_2$ magnetizations, and can not be lifted by virtue of locality.  Thus all level crossings are protected, implying that the electronic system \emph{automatically} inherits the parafermion platform's $8\pi$-periodic response without any symmetry enforcement required (as long as the microscopic Hamiltonian remains invariant under $\phi_0 \rightarrow \phi_0 + 2\pi$).  An alternative way of viewing the resilience of the $8\pi$ periodicity is to observe that $m_1,m_2,$ and $i\gamma_2\gamma_3$ are conserved quantities in $H_{2,3}$, and must remain so even in the presence of arbitrary physical perturbations: The magnetization-flipping operators $\Gamma_j$ are non-local in the present setting, and there are no other sources of low-energy fermions that can flip $i\gamma_2\gamma_3$.  Spoiling the $8\pi$ periodicity requires shrinking the magnetized domains so that the order parameters become fluctuating quantum degrees of freedom and additional Majorana operators can provide a mechanism for fermion-parity switching.  

The $8\pi$-periodic cycle proceeds as follows: Start from the unique ground state at $\phi_0 = 0$.  Winding $\phi_0$ by $4\pi$ rotates the central domain's magnetization by a full $2\pi$ around the $z$ axis [recall Eq.~\eqref{m2} and the comments just below], but also pumps a fermion to the junction---yielding an excited state.  One must wind $\phi_0$ by $4\pi$ a second time to recover the original ground state.  The absolute robustness of this process is not without a price: implementing the cycle requires strong correlation together with interactions that can be tuned to twist $\phi_0$.  

Reference \onlinecite{Pedder} examined a somewhat similar setup consisting of a Josephson junction formed by topological superconductors with spontaneous time-reversal symmetry breaking.  These authors predicted an $8\pi$-periodic Josephson effect protected by time-reversal symmetry.  We would like to point out, however, that \emph{within a fixed order-parameter sector}, time reversal does not protect level crossings in the spectrum.  We expect that in such systems anomalous periodicity should either be protected by a symmetry that is present within a given order-parameter sector, or enjoy absolute protection due to locality constraints as found above.

\section{Extension to higher parafermions}
\label{Z2Msec}

Our results for the $\mathbb{Z}_4$ case can be efficiently extended to arbitrary $\mathbb{Z}_{2M}$ parafermions, where $M$ is any positive integer.  In this section we outline a general fermionization scheme, then posit models that capture analogues of the four types of phases summarized in Fig.~\ref{Phases_fig}, and finally develop anomalous pumping cycles that reveal nontrivial fusion properties for higher parafermions.  

\subsection{Fermionization procedure}

It is useful to introduce bosonic $\mathbb{Z}_{2M}$ clock variables $\sigma_a,\tau_a$ as an intermediary between parafermions and fermions.  These unitary operators are now taken to satisfy 
\begin{equation}
\sigma^{2M}_a = \tau^{2M}_a = 1,~~~~\sigma_a \tau_a = e^{i\frac{\pi}{M}} \tau_a \sigma_a~.
\label{clock2Mproperties}
\end{equation} 
We will consider a $\mathbb{Z}_{2M}$ symmetry that sends
\begin{equation}
  \sigma_a \rightarrow e^{i\frac{\pi}{M}} \sigma_a~,\qquad\tau_a \rightarrow \tau_a
\end{equation}
along with $\mathcal{T}$ and $\mathcal{C}$ symmetries that act exactly as in Eqs.~\eqref{T} and \eqref{C}.  
These clock variables can be nonlocally combined to form unitary $\mathbb{Z}_{2M}$ parafermion operators
\begin{equation}
  \alpha_{2a-1} = \sigma_a \mu_{a-\frac{1}{2}}~,\qquad \alpha_{2a} = e^{-i\frac{\pi}{2M}}\sigma_a \mu_{a+\frac{1}{2}}~,
\label{eqn.Z2Mparaferm}
\end{equation}
where $\mu_{a+\frac{1}{2}} = \prod_{b<a+\frac{1}{2}} \tau_b$ as before.  These para\-fermions obey
\begin{equation}
\alpha_{a}^{2M} = 1~, \qquad \alpha_a \alpha_{b>a} = e^{i\frac{\pi}{M}}\alpha_b \alpha_a~.
\end{equation}

Next, we would like to relate $\mathbb{Z}_{2M}$ clock variables to fermions.  In analogy with Sec.~\ref{fermions}, fermion anticommutation at long separation can be obtained by binding $\sigma$ to the $M^{\text{th}}$ power of the string $\mu$, but the local structure requires some additional work.  Each clock site now hosts a $2M$-dimensional Hilbert space.  For $\mathbb{Z}_4$, the dimension matches that for two species of fermions, facilitating a complete fermionization of the clock operators as carried out in Sec.~\ref{fermions}.  A similar matching occurs when $M = 2^{k-1}$ ($k$ is an integer), which in principle allows a complete fermionization into $k$ species of fermions.  At other $M$ values, however, this relation breaks down.  

To cover all $M$'s in one formalism, we will thus follow a variant of the route adopted for the $\mathbb{Z}_4$ case in Sec.~\ref{FM_phase}.  In particular, there we utilized an explicit separation into a fermionic sector (described by a single species $c_a$) coupled to a $\mathbb{Z}_2$ magnetization order parameter $m_a = e^{i \pi d_a^\dagger d_a} = \sigma_a^2$.  When generalizing to the $\mathbb{Z}_{2M}$ case, we will again employ a single fermion species $C_a$, but promote the magnetization $m_a$ to a unitary $\mathbb{Z}_{M}$ order parameter $\mathcal{O}_a = \sigma_a^2$ whose eigenvalues are cycled by a conjugate unitary operator ${\cal D}_a$, i.e.,
\begin{align}
  \mathcal{O}_a^M = \mathcal{D}_a^M = 1~,\qquad \mathcal{O}_a \mathcal{D}_a = e^{i\frac{2\pi}{M}} \mathcal{D}_a \mathcal{O}_a~.
  \label{ODproperties}
\end{align}
In this way the clock-spin Hilbert-space dimension is faithfully recovered for all $M$.
The explicit mapping to these variables follows from
\begin{align}
  \sigma_a &= B_a + \mathcal{O}_a B_a^\dagger ~,
  \label{sigma2M} \\
  \tau_a &= e^{i\frac{\pi}{M} B_a^\dagger B_a} \mathcal{D}_a~,
  \label{tau2M}
\end{align} 
where $B_a$ are hard-core bosons that commute with $\mathcal{O}_a,\mathcal{D}_a$.  One can readily verify that the decomposition above preserves the properties in Eq.~\eqref{clock2Mproperties}.  Finally, we introduce spinless fermions via
\begin{equation}
    C_a \equiv B_a e^{i\pi \sum_{b<a} B_{b}^\dagger B_{b}}= B_a \prod_{b<a} \tau_b^M~.
  \label{Cfermions}
\end{equation} 
The order parameter $\mathcal{O}_a$ can also be rewritten in terms of fermions, as in the case for $Z_4$ parafermions, though if $M$ is not a power of $2$ we will need to project out the excess states in the Hilbert space.

Appendix~\ref{HigherDictionaryAppendix} inverts Eqs.~\eqref{sigma2M} and \eqref{tau2M} and, in the special case of $\mathbb{Z}_4$, relates the operators above to the $c_a$ and $d_a$ fermions used in Sec.~\ref{FM_phase}; specifically, we show that
\begin{align}
  \mathcal{D}_a &= (d_a+d_a^\dagger)(c_a^\dagger - c_a)~,
  \\
  C_a &= \frac{1-m_a}{2} c_a + \frac{1+m_a}{2} c_a^\dagger~. 
\end{align}

Table~\ref{Z_2M_symmetry_table} enumerates the symmetry properties for the original clock variables along with operators defined in Eqs.~\eqref{sigma2M} through \eqref{Cfermions}.  From the table one sees that in the fermionic representation, $(\mathbb{Z}_{2M})^M$ is the $\mathbb{Z}_2$ symmetry associated with conservation of global fermion parity.  We also observe that the composite anti-unitary symmetry $\mathcal{T}' \equiv \mathbb{Z}_{2M}\mathcal{T}$ is a generalization of electronic time-reversal symmetry for which $(\mathcal{T'})^2$ has eigenvalues $e^{i\frac{2\pi}{M}l}$ ($l$ is an integer).     

With this general fermionization algorithm in hand, we will now explore the correspondence between various phases in the clock, parafermion, and fermion representations.  It is worth keeping in mind, however, that many different fermionization schemes are possible as alluded to above and will generally yield different fermionic phases from what we describe below; pursuing such alternative representations is certainly interesting but left to future work.

\begin{table}
\begin{center}
 \setlength\extrarowheight{2pt}
 \begin{tabular}{|c | c | c | c| c|} 
 \hline
  & $\mathbb{Z}_{2M}$ & $\mathcal{C}$ & $\mathcal{T}$ \\ [0.05ex] 
 \hline\hline
 $\sigma \rightarrow$ & $e^{i\frac{\pi}{M}}\sigma$ & $\sigma^\dagger$ & $\sigma^\dagger$ \\
 \hline
 $\tau \rightarrow$ & $\tau$ & $\tau^\dagger$ & $\tau$ \\
 \hline\hline
 $B \rightarrow$ & $e^{i\frac{\pi}{M}} B$ & $B \mathcal{O}^\dagger$ & $B \mathcal{O}^\dagger$ \\
 \hline
 $\mathcal{O} \rightarrow$ & $e^{i\frac{2\pi}{M}} \mathcal{O}$ & $\mathcal{O}^\dagger$ & $\mathcal{O}^\dagger$ \\
 \hline
 $\mathcal{D} \rightarrow$ & $\mathcal{D}$ & $\mathcal{D}^\dagger e^{-i\frac{2\pi}{M} C^\dagger C}$ & $\mathcal{D} e^{i\frac{2\pi}{M} C^\dagger C}$ \\
 \hline
 $C \rightarrow$ & $e^{i\frac{\pi}{M}} C$ & $C \mathcal{O}^\dagger$ & $C \mathcal{O}^\dagger$ \\
 \hline
 
\end{tabular}
\end{center}
\caption{Transformation properties for $\mathbb{Z}_{2M}$ clock variables and the operators used to decompose them through Eqs.~\eqref{sigma2M}, \eqref{tau2M}, and \eqref{Cfermions}. }
\label{Z_2M_symmetry_table}
\end{table}

\subsection{Paramagnetic and ferromagnetic phases}

It is simplest to first examine the $\mathbb{Z}_{2M}$ clock model 
\begin{equation}
H = -J \sum_{a = 1}^{N-1} (\sigma_a^\dagger \sigma_{a+1} + H.c.) - f \sum_{a = 1}^N (\tau_a + H.c.)~.
\end{equation}
The $J = 0$ and $f = 0$ limits provide trivially solvable realizations of the non-degenerate paramagnetic state and $2M$-fold degenerate ferromagnetic phase, respectively.  
In terms of $\mathbb{Z}_{2M}$ parafermions $H$ becomes \cite{Fendley:2012}
\begin{align}
H =& -J \sum_{a = 1}^{N-1} (e^{i\frac{\pi}{2M}}\alpha_{2a}^\dagger \alpha_{2a+1} + H.c.) \nonumber \\
&- f \sum_{a = 1}^N (e^{i\frac{\pi}{2M}}\alpha_{2a-1}^\dagger \alpha_{2a} + H.c.)~.
\end{align} 
At $J = 0$ all parafermions dimerize yielding a unique ground state; at $f = 0$ the parafermions form a topological phase with unpaired parafermion zero modes that encode a robust degeneracy consisting of $2M$ locally indistinguishable states.  

The fermionized Hamiltonian reads
\begin{align}
H =& -J \sum_{a = 1}^{N-1}[(C_a^\dagger - \mathcal{O}_a^\dagger C_a)(C_{a+1}+\mathcal{O}_{a+1}C_{a+1}^\dagger) + H.c.]
\nonumber \\
&- f \sum_{a = 1}^N (e^{i\frac{\pi}{M} C_a^\dagger C_a} \mathcal{D}_a + H.c.)~.
\end{align}
In the $J = 0$ limit, the ground state has $\mathcal{D}_a = 1$ and $C_a^\dagger C_a = 0$ for all sites.  Hence, the trivial parafermion phase corresponds to the fermionic vacuum with a vanishing order parameter $\langle \mathcal{O}_a \rangle = 0$.  The $f = 0$ Hamiltonian closely resembles Eq.~\eqref{Hf}, though recall that the $c_a$ and $C_a$ fermions do not coincide at $M = 2$.  
 Here the energy is minimized by uniformly condensing the order parameter, i.e., taking $\langle \mathcal{O}_a\rangle = e^{i \frac{2\pi}{M} n}$ for some arbitrary integer $n$.  The fermions then enter a topological phase with symmetry-enriched Majorana end states whose wavefunctions again depend on the precise order-parameter configuration.  Just like the $\mathbb{Z}_4$ case, topological degeneracy encoded by parafermion zero modes becomes a mixture of $2$-fold topological degeneracy and $M$-fold symmetry-breaking degeneracy.

We can also appeal to a long-wavelength approach to recover these phases, following a straightforward generalization of Sec.~\ref{long_wavelength_limit}.  Using bosonized variables $\phi,\theta$ that satisfy the commutator in Eq.~\eqref{phi_theta_commutator}, clock order and disorder operators can now be expanded as $\sigma \sim e^{i\phi}, \mu \sim e^{-i\theta/M}$.  Inserting these expansions into Eqs.~\eqref{eqn.Z2Mparaferm} and \eqref{Cfermions} yields $\alpha \sim e^{i(\phi - \theta/M)}$ for long-wavelength parafermions and $\psi_{R/L} \sim e^{i(\phi \pm \theta)}$ for long-wavelength fermions. The bosonized Hamiltonian takes the form
\begin{align}
  \mathcal{H} = \int_x \bigg{\{}&\frac{v}{2\pi}[g(\partial_x \phi)^2 +g^{-1} (\partial_x\theta)^2] 
  \nonumber \\
  &- \kappa_1 \cos(2M\phi)- \kappa_2 \cos(2\theta)\bigg{\}}~.\label{Z_2M_bosonized}
\end{align}
Relevant $\kappa_1>0$ and $\kappa_2>0$ couplings respectively generate the ferromagnetic and paramagnetic phases in clock language.  Next we turn to the phases stabilized by relevant couplings of the opposite sign, which are generalizations of the canted and SPT phases explored previously for the $\mathbb{Z}_4$ case.

\subsection{Canted and SPT phases}

The canted-ferromagnet phase discussed in Sec.~\ref{FM_phase2} generalizes to a state with $\langle \sigma_a \rangle \sim e^{i\frac{\pi}{M}(k+\frac{1}{2})}$ for integer $k$, i.e., the clock spins orient `halfway' between adjacent $\sigma_a$ eigenvalues.  We construct trial wavefunctions that exhibit this ordering as
\begin{equation}
  \ket{e^{i\frac{\pi}{M}(k+\frac{1}{2})}} = \prod_a \frac{1+\tau_a}{\sqrt{2}}\ket{\sigma = e^{i\frac{\pi}{M} k},e^{i\frac{\pi}{M} k},\ldots}~.
  \label{CantedStates2M}
\end{equation}
These states contain no $\tau = -1$ components; moreover, all nearest-neighbor bonds involve only configurations with $\sigma^\dagger_a \sigma_{a+1} = 1, e^{i\frac{\pi}{M}},$ or $e^{-i\frac{\pi}{M}}$.  Our trial wavefunctions are therefore exact ground states of the Hamiltonian
\begin{equation}
H_{\rm canted} = -\sum_{a = 1}^{N-1}\mathcal{P}_{\sigma_a^\dagger \sigma_{a+1} = 1, e^{i\frac{\pi}{M}}, e^{-i\frac{\pi}{M}}} + \sum_{a = 1}^N \mathcal{P}_{\tau_a = -1}~,
\label{Hcanted}
\end{equation}
where $\mathcal{P_\kappa}$ projects onto states satisfying property $\kappa$.  In the $\mathbb{Z}_4$ limit $H_{\rm canted}$ is equivalent to the Ashkin-Teller model at $f = 0$ and $\lambda = 1$, which contains many other ground states as well.  Thus we should again add a perturbation akin to $\delta H$ in Eq.~\eqref{deltaH} that leaves the canted states as the only ground states.  (The specific form of the interaction is not important for us.)  

In the absence of $\mathcal{C}$ symmetry, the canted and ferromagnet states can be smoothly connected.    The parafermion counterpart of these clock phases must therefore share exactly the same symmetry-independent topological characteristics---i.e., both phases must support a $2M$-fold robust ground-state degeneracy.  An identical conclusion holds for the fermion realization: Both phases yield Majorana end states whose structure depends on the order parameter, but with a different expectation value $\langle \mathcal{O}_a \rangle = \langle \sigma_a^2 \rangle \sim e^{i\frac{\pi}{M}(2k+1)}$ in the canted state.

The dual of $H_{\rm canted}$ is given by
\begin{equation}
  H_{\rm SPT} = -\sum_{a = 1}^N \mathcal{P}_{\tau_a = 1, e^{i\frac{\pi}{M}}, e^{-i\frac{\pi}{M}}} + \sum_{a = 1}^{N-1}\mathcal{P}_{\sigma^\dagger_a \sigma_{a+1} = -1}~.
\end{equation}
The four wavefunctions $\ket{\eta^z_1\eta^z_2}$ defined for the $\mathbb{Z}_4$ case in Eqs.~\eqref{psiSPT} and \eqref{other_states} are unfrustrated ground states of $H_{\rm SPT}$ for any $M$.  One can always add a perturbation $\widetilde {\delta H}$ to ensure that no other ground states exist; we will assume that such a perturbation has been included.  The resulting four-fold degeneracy again arises from pseudospin-1/2 edge degrees of freedom $\vec \eta_{1,2}$ for the clock chain.  These edge modes can be related to microscopic operators projected into the ground-state manifold: 
\begin{align}
  &\mathcal{P}\dfrac{i(\tau_1-\tau_1^\dagger)}{\sin{\pi/M}}\mathcal{P} =\eta^z_1, &&\mathcal{P}\dfrac{i(\tau_N-\tau_N^\dagger)}{\sin{\pi/M}}\mathcal{P} = \eta^z_2 
  \label{Z2M_etaz} \\
  &\mathcal{P}\sigma_1\mathcal{P} = (\eta^x_1 + i \eta^y_1)/2, &&\mathcal{P}\sigma_N\mathcal{P} = (\eta^x_2 + i \eta^y_2)/2,
\end{align}
which straightforwardly generalize Eqs.~\eqref{etaz} and \eqref{etaxy}.   

The parafermion and fermion realizations exhibit edge zero modes as well, though the statistics of the boundary operators naturally changes compared to the clock case.  (The transcription between representations can be carried out using the same procedure adopted in Sec.~\ref{SPT}.). In particular, for the fermion case the edge degrees of freedom can be described by a pair of Majorana zero modes at each end, precisely as for the TRITOPS phase found in the $\mathbb{Z}_4$ limit.  For any representation, the edge zero modes are robust in the presence of $\mathbb{Z}_{2M}, \mathcal{C}$, and $\mathcal{T}$ but can be eliminated when all breakable symmetries are abandoned---strongly suggesting the onset of an SPT phase for any $M \geq 2$.

\subsection{Anomalous $\mathbb{Z}_{2M}$ pumps}

The parafermion fusion setups in Figs.~\ref{Fusion_fig_PF}(a) and \ref{Fusion_fig_elec_2}(a) extend straightforwardly to the $\mathbb{Z}_{2M}$ case by simply replacing $\cos(4\phi) \rightarrow \cos(2M\phi)$ and $\cos(4\phi-\phi_0) \rightarrow \cos(2M\phi-\phi_0)$ in the appropriate domains.  For the generalized Fig.~\ref{Fusion_fig_PF}(a), coupling of parafermions $\alpha_{1,2}$ is governed by 
\begin{equation}
  H_{1,2} =-t\left[e^{i\frac{\pi-\theta_0}{2M}}\alpha_1^\dagger \alpha_2 + H.c.\right].  
  \label{H12ZM}
\end{equation}
Eigenstates of $H_{1,2}$ have $e^{i \frac{\pi}{2M}} \alpha_1^\dagger \alpha_2 = e^{i\frac{\pi}{M} n}$ with $n = 0,\ldots,2M-1$, yielding energies 
\begin{equation}
  E_n(\theta_0) = -2t\cos\left(\frac{n\pi}{M}-\frac{\theta_0}{2M}\right)
  \label{En}
\end{equation}
that are each $4M\pi$-periodic in $\theta_0$.  Level crossings occur only at $\theta_0 = 0 \mod \pi$; they are all protected by an unbreakable $\mathbb{Z}_{2M}$ symmetry in this realization---implying a $4M\pi$-periodic response to $\theta_0$ sweeps.  Once again, this anomalous periodicity reflects the fact that shifting $\theta_0$ by $2\pi$ cycles the system among the  $2M$ available fusion channels for the corresponding non-Abelian defects.  For the generalized Fig.~\ref{Fusion_fig_elec_2}(a), parafermions $\alpha_{2,3}$ couple via
\begin{equation}
  H_{2,3} =-t\left[e^{i\frac{\pi+\phi_0}{2M}}\alpha_2^\dagger \alpha_3 + H.c.\right].  
  \label{H23ZM}
\end{equation}
Identical logic applied to this setup implies a $4M\pi$-periodic response to $\phi_0$ sweeps as well.

The fermionic setups from Figs.~\ref{Fusion_fig_elec}(a) and \ref{Fusion_fig_elec_2}(b) admit the same $\mathbb{Z}_{2M}$ generalization, though here we must also promote the magnetizations $m_i$ to $\mathbb{Z}_M$ order parameters $\mathcal{O}_i$.  As in our analysis of the $\mathbb{Z}_4$ case, we will allow for additional physical perturbations in this setting, e.g., those that break $\mathbb{Z}_{2M}$ symmetry.  If the level crossings that underlie the anomalous periodicity for the parafermion platform persist, then the $4M\pi$-periodic response survives; otherwise the periodicity will be reduced.  

Consider the generalized Fig.~\ref{Fusion_fig_elec}(a) first.  Suppose for now that the Hamiltonian is given by Eq.~\eqref{H12ZM} re-expressed in terms of fermions, so that the energies are again given by Eq.~\eqref{En}.  At $\theta_0 = 0$, the $n = 0$ and $n = M$ levels are non-degenerate, while all other energy levels form doublets comprised of states with $n = p$ and $n = -p \mod 2M$.  This structure persists even in the presence of additional perturbations provided the Hamiltonian preserves $\mathcal{T}' = \mathbb{Z}_{2M} \mathcal{T}$---which guarantees degeneracy of the doublets via a generalization of Kramer's theorem.  At $\theta_0 = \pi$ the Hamiltonian describes the boundary between the fermionic vacuum and the SPT phase described in the previous subsection.  This interface hosts a single Dirac-fermion zero mode and, accordingly, all energy levels in Eq.~\eqref{En} are doubly degenerate.  The resulting level crossings at $\theta_0 = \pi$ are protected by fermion parity and remain robust to arbitrary local perturbations.  Hence, the fermionic system retains the anomalous $4M\pi$-periodic response to $\theta_0$ sweeps provided $\mathcal{T}'$ symmetry is enforced at $\theta_0 = 0 \mod 2\pi$.  

Finally, consider the generalized Fig.~\ref{Fusion_fig_elec_2}(b).  Just as for the $\mathbb{Z}_4$ limit, locality and fermion-parity constraints alone guarantee an anomalous $4M\pi$-periodic response to $\phi_0$ (no special symmetries required).  The Hamiltonian governing the left junction in the figure can only depend on $i\gamma_2 \gamma_3$ and the order parameters $\mathcal{O}_{1,2}$, all of which are necessarily conserved quantities in the effective low-energy description.  Eliminating the level crossings that underlie the anomalous periodicity would require either transitioning between macroscopically distinct order-parameter configurations, or a source of low-energy fermions to flip $i\gamma_2\gamma_3$.  Neither process is available in our setup.  We can see this result explicitly by rewriting Eq.~\eqref{H23ZM} in the fermionic representation:
\begin{equation}
  H_{2,3} = -2t \cos\left[\frac{\pi}{M}(\hat{q}_1-\hat{q}_2) + \frac{\phi_0}{2M} \right] i\gamma_2\gamma_3,
  \label{Z2MPumpingHamiltonian}
\end{equation}
where we introduced integer-valued operators $\hat{q}_{1,2}$ that specify the order parameters via $\mathcal{O}_{1,2} = e^{i \frac{2\pi}{M} \hat{q}_{1,2}}$.  As deduced on general grounds, Eq.~\eqref{Z2MPumpingHamiltonian} predicts energies that are $4M\pi$ periodic in $\phi_0$, with each branch corresponding to fixed order-parameter and parity configurations.  Transitions between these levels are therefore forbidden.  

As an aside, Eq.~\eqref{Z2MPumpingHamiltonian} in the $M = 2$ limit describes precisely the same setup as Eq.~\eqref{H23m}, though the Hamiltonians look rather different.  In terms of the magnetizations appropriate for the $\mathbb{Z}_4$ case, we have $e^{i \frac{\pi}{2}\hat{q}_{1,2}} = [(1+m_{1,2}) +i(1-m_{1,2})]/2$.  Using this relation and sending $\gamma_2 \rightarrow m_1 \gamma_2$ in Eq.~\eqref{Z2MPumpingHamiltonian} reproduces Eq.~\eqref{H23m}, i.e., they indeed provide equivalent descriptions.  

\begin{figure}
\includegraphics[width=\columnwidth]{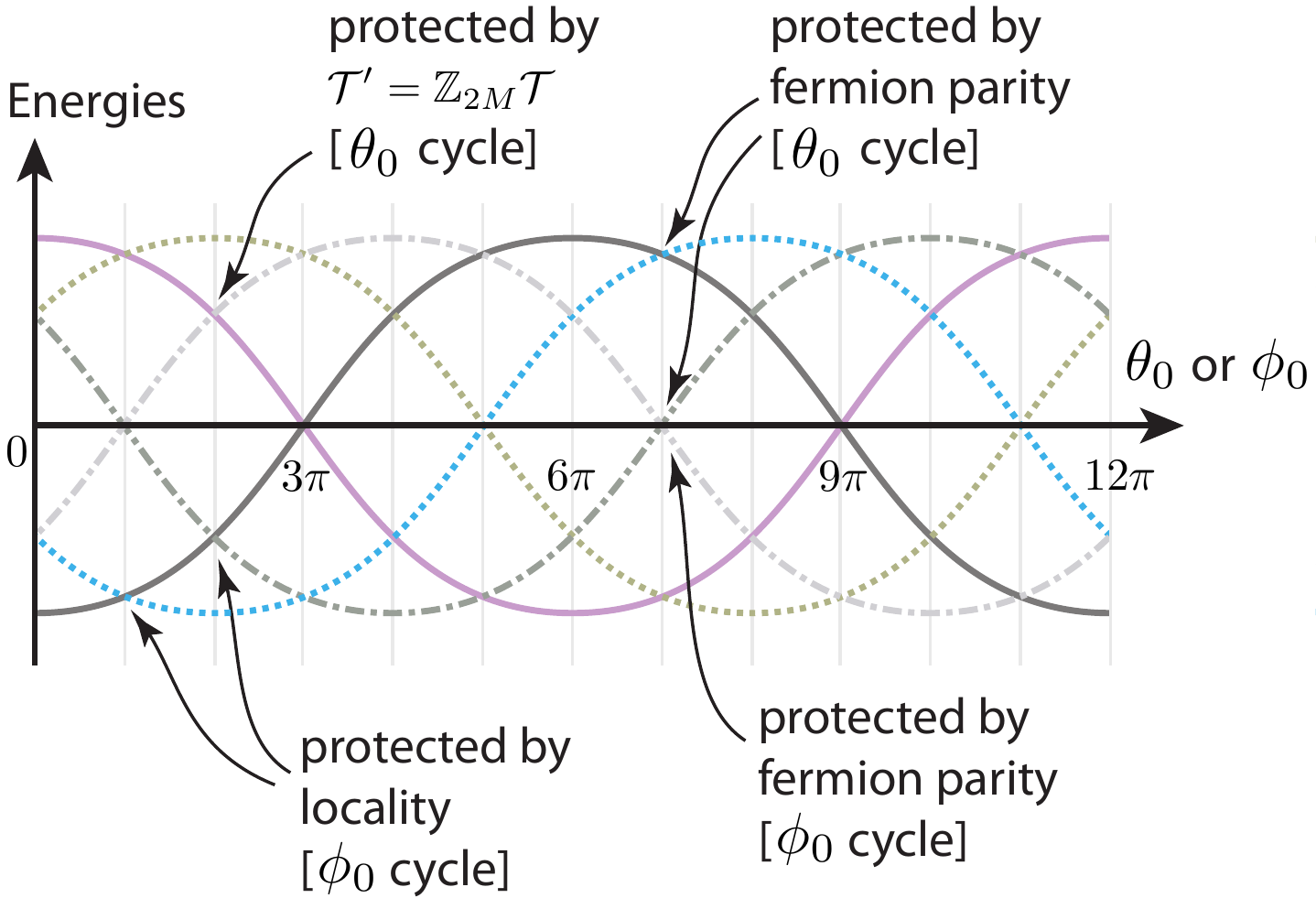}
\caption{Energies versus pumping parameters $\theta_0$ or $\phi_0$ for the fermionic setups in Figs.~\ref{Fusion_fig_elec}(a) and \ref{Fusion_fig_elec_2}(b), generalized to the $\mathbb{Z}_6$ case (i.e., $M = 3$).  For the generalized Fig.~\ref{Fusion_fig_elec}(a), the level crossings at $\theta_0 = 0 \mod 2\pi$ are protected by the antiunitary symmetry $\mathcal{T'} = \mathbb{Z}_{2M}\mathcal{T}$, while level crossings at $\theta_0 = \pi \mod 2\pi$ exhibit fermion-parity protection.  As long as these level crossings are maintained, the system exhibits an anomalous $12\pi$-periodic response to $\theta_0$ sweeps.  For the generalized Fig.~\ref{Fusion_fig_elec_2}(b), the level crossings at zero energy are fermion-parity protected; all others occur between states with different order-parameter configurations and are protected by locality.  The system thus generically exhibits $12\pi$-periodic response to $\phi_0$, with no additional symmetries required.  These enlarged periodicities are an imprint of the nontrivial fusion rules in the corresponding $\mathbb{Z}_6$ parafermion platforms. }
\label{Fusion_fig_Z2M}
\end{figure}

Figure~\ref{Fusion_fig_Z2M} summarizes the structure of the energy levels in both fermionic platforms considered above, specializing to the $M = 3$ case.  

\section{Discussion}
\label{Discussion}

We have established an exact correspondence between $\mathbb{Z}_{\rm even}$ parafermion chains and 1D fermionic systems, using clock spins as an intermediary.  From the clock viewpoint, our formalism extends the familiar fermionization of the Ising model into a much broader class of discrete spin systems.  We were most interested, however, in understanding how the physics of bona fide parafermions, which (to our knowledge) require a fractionalized host, filters into the fermionic realm.  

Most of our effort centered around the $\mathbb{Z}_4$ case.  There we introduced a judicious fermionization algorithm that maps $\mathbb{Z}_4$ parafermions to ordinary spinful electrons, a result foreshadowed by earlier works on anomalous quantum-spin-Hall edge modes~\cite{ZhangKane,Orth}.  Moreover, we saw that symmetries of the parafermion system can be repackaged into familiar operations for fermions---notably electronic time reversal and global spin rotations. Phases for $\mathbb{Z}_4$ parafermions, in turn, translate into physically relevant electronic states as summarized in Fig.~\ref{Phases_fig}: The trivial gapped parafermion phase maps to an electronic insulator; the topological phase with unpaired parafermion zero modes \cite{Fendley:2012} maps to a topological superconductor hosting symmetry-enriched Majorana zero modes whose structure intertwines with a spontaneously chosen magnetization order parameter; and an SPT phase for parafermions maps to the widely studied time-reversal-invariant topological superconductor (TRITOPS) for electrons.  

Interestingly, symmetry-enriched Majorana zero modes may have already been experimentally realized in proximitized Fe chains \cite{Nadj-Perge,Ruby,Pawlak,Feldman,Jeon}.  The Fe-chain Hamiltonian of course differs markedly from the toy models studied in Sec.~\ref{FM_phase}, but shares the all-important feature of spontaneous time-reversal symmetry breaking.  Majorana zero modes appearing in Fe chains must then conform to Eq.~\eqref{gamma_transformation} on very general grounds, indicating symmetry enrichment in the sense defined in this paper.  The precise connection to parafermion physics highlights a surprising new perspective on these experiments.

Our exact mappings further enabled a rigorous comparison between non-Abelian-anyon physics arising from $\mathbb{Z}_4$ parafermion zero modes and from symmetry-enriched Majorana modes.  We showed that their braiding properties differ starkly and pinpointed the origin of this distinction (the parafermion braiding Hamiltonian becomes nonlocal when mapped to fermions).  Symmetry-enriched Majorana modes do, nevertheless, underlie braid transformations that can not arise in conventional Majorana systems, since the order parameter need not return to its original value under an adiabatic closed cycle of the Hamiltonian.  It is unclear whether this additional flexibility offers any advantages for quantum computing, though this question certainly warrants serious consideration.  

Fusion properties arising from $\mathbb{Z}_4$ parafermion zero modes are more directly inherited by electrons in the following sense.  Parafermion platforms admit a pumping cycle that returns the Hamiltonian to its original form but cycles the system among four possible `fusion channels' for the parafermions---yielding an anomalous $8\pi$-periodic response.  Precisely the same $8\pi$ periodicity can be harnessed in the corresponding 1D electronic setting.  We introduced `weak' and `strong' implementations that can both be understood in terms of hybridization of symmetry-enriched Majorana modes.  The `weak' version (summarized in Fig.~\ref{Fusion_fig_elec}) cyclically modulates a wire between TRITOPS and trivial phases; provided time-reversal symmetry is maintained at certain points of the cycle, the magnetization at the ends of the system exhibits $8\pi$ periodicity.  This phenomenon is a cousin of the $8\pi$-periodic Josephson effect that can arise at a quantum-spin-Hall edge \cite{ZhangKane,Orth,Peng,Hui,Vinkler}.  The `strong' version (Fig.~\ref{Fusion_fig_elec_2}) 
realizes an anomalous $8\pi$-periodic pumping cycle that eschews symmetry requirements altogether, but necessitates strong correlation together with tunable interactions.  Implementation in proximitized Fe chains poses a tantalizing possibility worth exploring in detail.

We generalized our $\mathbb{Z}_4$ results by using a modified algorithm that recasts $\mathbb{Z}_{2M}$ parafermions in terms of a single species of fermions coupled to a $\mathbb{Z}_M$ order parameter.  Weak and strong anomalous pumping cycles, now with $4M\pi$ periodicity, were identified also in this broader setting.  Experimental connections are less obvious compared to the $\mathbb{Z}_4$ case, however, and would be useful to develop in future work.  One potentially promising avenue is to explore an array of wires with a $Z_M$ rotational symmetry (similar to the bundles examined in Ref.~\onlinecite{Klinovaja:2014a}) that is spontaneously broken, yielding a nontrivial interplay with Majorana zero modes. It is natural to also ask about $Z_{\rm odd}$ parafermions.  Our fermionization approach does not readily extend to this case due to a `mismatch' in Hilbert-space dimensions.  Nevertheless, it would be interesting to pursue variations of our approach, for instance that map $Z_{\rm odd}$ parafermions to fermions with a constrained Hilbert space.  

The classifications of interacting gapped 1D phases from Refs.~\onlinecite{FidkowskiKitaev1,Turner:2010,ChenGuWen,Schuch} strongly constrain the kinds of non-Abelian-anyon defects that 1D systems can support.  Specifically, these works capture only `Ising' defects that trap Majorana zero modes.  One of the general messages of our work is that the interplay between Majorana modes and local order parameters can nonetheless enrich their properties as summarized above.  In light of this perspective, it would be interesting to revisit earlier works aimed at mimicking parafermion physics in strictly 1D setups \cite{Oreg:2014,Klinovaja:2014a,Klinovaja:2014b,Tsvelik:2014b}: Might such setups provide additional platforms for symmetry-enriched Majorana modes?  Pursuing realizations of symmetry-enhanced non-Abelian defects using cold atoms poses another natural direction---building, e.g., off of Ref.~\onlinecite{Hu}.  Cold-atoms proposals for obtaining genuine parafermions do exist \cite{Maghrebi}, but the requisite topologically ordered host platforms have not yet been demonstrated.  

We conclude by highlighting several other future directions.  A more exhaustive dictionary linking phases for parafermions and fermions would certainly be welcome.  We have focused on a select few examples, and there are likely deeper insights to be gained from other such correspondences.  Majorana zero modes can also of course arise in two-dimensional topological superconductors.  Can one harvest a fruitful interplay with order-parameter physics also in this setting?  On a more formal level, we saw that a duality transformation for clock spins maps fermions onto dual fermions, which (roughly) are related to one another by attaching a parafermion [recall, e.g., Eq.~\eqref{fermion_dualfermion}].  It is interesting to ask whether a similar nontrivial connection between fermions and dual fermions can exist in higher dimensions \cite{diracduality.son,diracduality.wang,diracduality.metlitski1,diracduality.metlitski2,diracduality.mross}.  Explorations along these lines may contribute to the growing `duality web' that has recently been established in $(2+1)$-dimensional field theories \cite{dualityweb.SeibergSenthilWangWitten,dualityweb.KarchTong,dualityweb.murugan,dualityweb.mirror.kachru1,dualityweb.mirror.kachru2,dualityweb.mross,dualityweb.lattice.ChenRaghu,dualityweb.goldman}.

\begin{acknowledgments} 
We are indebted to D.~Aasen, X.~Chen, D.~Clarke, P.~Fendley, and A.~Jermyn for illuminating discussions.  We gratefully acknowledge support from the National Science Foundation through grants DMR-1341822 and DMR-1723367 (A.~C.~and J.~A.); the Army Research Office under Grant Award W911NF-17-1-0323 (A.~C.~and J.~A.); the Israel Science Foundation grant No.~1866/17 (D.~F.~M.); grant No.~2016258 from the United States-Israel Binational Science Foundation (BSF); the Dominic Orr Graduate Fellowship (A.~C.); the Yunni and Maxine Pao Graduate Fellowship (A.~C.); the Caltech Institute for Quantum Information and Matter, an NSF Physics Frontiers Center with support of the Gordon and Betty Moore Foundation through Grant GBMF1250; and the Walter Burke Institute for Theoretical Physics at Caltech.  
\end{acknowledgments}

\appendix

\section{Expressing hard-core bosons in terms of $\mathbb{Z}_4$ clock operators}
\label{Inversion_Appendix}

Inverting Eqs.~\eqref{tau_expansion} and \eqref{sigma_expansion} is nontrivial since the expansion of $\sigma_a$ involves terms that are both linear and cubic in hard-core boson operators.  We perform the inversion by first assembling linear combinations that cancel the cubic components.  Some algebra yields
\begin{align}
  \sigma\left(\frac{1-\tau^2}{2}\right) + H.c. &= b_\uparrow^\dagger + b_\uparrow
  \\
  \left(\frac{1-\tau^2}{2}\right)\sigma + H.c. &= b_\downarrow^\dagger + b_\downarrow
  \\
  \sigma\left(\frac{\tau^\dagger-\tau}{2}\right)+H.c. &= i(b_\uparrow^\dagger-b_\uparrow)
  \\
  \left(\frac{\tau^\dagger-\tau}{2}\right)\sigma + H.c. &= i(b_\downarrow^\dagger-b_\downarrow),
\end{align}
where for notational simplicity we suppressed the site label $a$.  From here it is straightforward to take superpositions that isolate $b_\uparrow$ and $b_\downarrow$, yielding Eqs.~\eqref{b_up} and \eqref{b_down} from the main text.

\section{Symmetry properties of spinful fermions}
\label{symmetries}

In this Appendix we derive the action of $\mathbb{Z}_4$, $\mathcal{T}$, and $\mathcal{C}$ symmetries on spinful fermions.  The string $S_a$ is invariant under each of these operations; thus all the action arises from the bosons and the additional phase factors in Eqs.~\eqref{f_up} and \eqref{f_down}.  

Consider first $\mathbb{Z}_4$.  The following relations, which can be obtained from Eqs.~\eqref{tau_expansion} and \eqref{sigma_expansion}, are helpful for evaluating this symmetry:
\begin{align}
  i\sigma\left(\frac{1-\tau^2}{2}\right) + H.c. &= -i e^{i \pi n_{\downarrow}} (b_\uparrow^\dagger-b_\uparrow)
  \\
  i\left(\frac{1-\tau^2}{2}\right)\sigma + H.c. &= i e^{i \pi n_{\uparrow}} (b_\downarrow^\dagger - b_\downarrow)
  \\
  i\sigma\left(\frac{\tau^\dagger-\tau}{2}\right)+H.c. &= e^{i \pi n_\downarrow}(b_\uparrow^\dagger + b_\uparrow)
  \\
  i\left(\frac{\tau^\dagger-\tau}{2}\right)\sigma + H.c. &= -e^{i \pi n_\uparrow}(b_\downarrow^\dagger + b_\downarrow).
\end{align}
(We continue to suppress site indices for simplicity.)  Using the above together with Eqs.~\eqref{Z_4}, \eqref{b_up}, and \eqref{b_down}, one sees that the hard-core bosons transform under $\mathbb{Z}_4$ as
\begin{align}
  Q b_\uparrow Q^\dagger &= i e^{i \pi n_\downarrow} b_\uparrow,~~~~
  Q b_\downarrow Q^\dagger = -i e^{i \pi n_\uparrow} b_\downarrow.
\end{align}
The fermions transform in an identical fashion:
\begin{equation}
  Q f_\uparrow Q^\dagger = i e^{i \pi n_\downarrow} f_\uparrow,~~~~
  Q f_\downarrow Q^\dagger = -i e^{i \pi n_\uparrow} f_\downarrow.
\end{equation}

The action of time-reversal $\mathcal{T}$ on hard-core bosons follows straightforwardly from Eqs.~\eqref{T}, \eqref{b_up}, and \eqref{b_down}; we find
\begin{equation}
  \mathcal{T} b_\uparrow \mathcal{T} = b_\downarrow,~~~~ \mathcal{T} b_\downarrow \mathcal{T} = b_\uparrow.
\end{equation}
In this case the phase factors in Eqs.~\eqref{f_up} and \eqref{f_down} result in a more nontrivial action on the fermions,
\begin{equation}
  \mathcal{T} f_\uparrow \mathcal{T} = i e^{i \pi n_\uparrow} f_\downarrow,~~~~ \mathcal{T} f_\downarrow \mathcal{T} = i e^{i \pi n_\downarrow} f_\uparrow.
\end{equation}

An analogous situation arises for charge conjugation $\mathcal{C}$.  For the bosons we obtain the simple transformation
\begin{equation}
  \mathcal{C} b_\uparrow \mathcal{C} = b_\downarrow,~~~~ \mathcal{C} b_\downarrow \mathcal{C} = b_\uparrow,
\end{equation}
which yields 
\begin{equation}
  \mathcal{C} f_\uparrow \mathcal{C} = e^{i \pi n_\uparrow} f_\downarrow,~~~~ \mathcal{C} f_\downarrow \mathcal{C} = e^{i \pi n_\downarrow}f_\uparrow
\end{equation}
for the fermions.

\section{Spin-$1/2$ representations and symmetries}
\label{app.spin}

To better understand the structure behind our fermionization, and compare to earlier works, it is instructive to express the clock operators $\sigma_a,\tau_a$ in terms of spin-1/2 degrees of freedom. References~\onlinecite{Kohmoto,Kohmoto2} employed one possible decomposition given by
\begin{align}
&\sigma_a = \frac{1+i}{2}(s^z_{a+\frac{1}{4}}+ i  s^z_{a - \frac{1}{4}})~,\label{clocktospin}\\
&\tau_a=\frac{1}{2} (s^x_{a+\frac{1}{4}} + s^x_{a-\frac{1}{4}})+\frac{1}{2}(s^x_{a+\frac{1}{4}} - s^x_{a-\frac{1}{4}}) s^z_{a+\frac{1}{4}} s^z_{a-\frac{1}{4}}~\nonumber.
\end{align}
The inverse relationship is 
\begin{align}
s^z_{a-\frac{1}{4}} &=- \frac{1+i}{2}(\sigma_a - i \sigma_a^\dagger)~,\\
s^z_{a+\frac{1}{4}} &= \frac{1-i}{2}(\sigma_a + i \sigma_a^\dagger)~,\\
s^x_{a-\frac{1}{4}} &=\frac{1}{2}\left(\tau_a +\tau_a^\dagger + \sigma_a^2 (\tau_a - \tau_a^\dagger)\right)~,\\
s^x_{a+\frac{1}{4}} &=\frac{1}{2}\left(\tau_a +\tau_a^\dagger - \sigma_a^2 (\tau_a - \tau_a^\dagger)\right)~.
\end{align}
In these variables the Ashkin-Teller model, Eq.~\eqref{AshkinTeller}, maps onto two coupled transverse-field Ising models: 
\begin{align}
H =& -J \sum_{a } (s^z_{a+\frac{1}{4}}s^z_{a+1+\frac{1}{4}}+s^z_{a-\frac{1}{4}}s^z_{a+1-\frac{1}{4}} ) \label{ATdoubleIsing}
 \\
  &- f \sum_{a } (s^x_{a+\frac{1}{4}} + s^x_{a-\frac{1}{4}} )\nonumber\\
  &+ \lambda \sum_{a } (J s^z_{a-\frac{1}{4}}s^z_{a+\frac{1}{4}} s^z_{a+1-\frac{1}{4}}s^z_{a+1+\frac{1}{4}}  +f s^x_{a-\frac{1}{4}}  s^x_{a+\frac{1}{4}})~.\nonumber
\end{align}
Next we will show that this spin-1/2 model admits two useful alternative forms: `Spin model A'  exhibits translation symmetry, with duality implemented as a non-symmorphic spin rotation.  `Spin model B' is invariant under a continuous spin-rotation symmetry, with duality instead implemented as a translation. 

\subsection{Spin model A}
\label{spinmodela}
Suppose that we perform the familiar Ising-model duality transformation that trades in $s^{x,y,z}_{a\pm \frac{1}{4}}$ variables for dual spins $t^{x,y,z}$ living on integer as well as half-integer sites:
\begin{align}&t^x_a =s^z_{a-\frac{1}{4}}s^z_{a+\frac{1}{4}} \qquad
&t^z_a =\prod_{a'<a}s^x_{a'}~.\label{Ising.duality}
\end{align}
(In the second expression, the variable $a'$ in the product runs over all integers and half-integers.) The Ashkin-Teller model then takes the form
\begin{align}
H_{J,f} =& - J\sum_{a}(t^x_{a-\frac{1}{2}}t^x_{a}+t^x_{a}t^x_{a+\frac{1}{2}})~\label{eqn.atrot}\\
  & -f \sum_{a}(t^z_{a-\frac{1}{2}}t^z_{a}+t^z_{a}t^z_{a+\frac{1}{2}})~\nonumber\\
  &+ \lambda \sum_{a} (J t^x_{a} t^x_{a+1}  +f t^z_{a-\frac{1}{2}} t^z_{a+\frac{1}{2}} )~.\nonumber
\end{align}
For later convenience, on the left side we have explicitly displayed the $J,f$ couplings as subscripts of $H$.  
The $\lambda$ terms only involve operators on the same sublattice (integer or half-odd-integer sites). Translations $T: a \rightarrow a+1$ and inversions ${\cal I}$ that preserve these sublattices leave $H_{J,f}$ invariant. We also introduce the `half-translation' $T_{\frac{1}{2}}: a \rightarrow a+ \frac{1}{2}$, which interchanges the sublattices, and a $\frac{\pi}{2}$ spin rotation $U = \exp \left[i \frac{\pi}{4} t_y\right]$. The model of Eq.~\eqref{eqn.atrot} satifies
\begin{align}
 H_{J,f}[\vect t ] =  H_{f,J}[T_{\frac{1}{2}} U\vect t  U^{-1}T^{-1}_{\frac{1}{2}}]~\label{eqn.dualrotation},
\end{align}
i.e., duality is realized as a local spin rotation combined with a change of sublattice. This implementation of duality is specific to the Ashkin-Teller Hamiltonian and does not hold for more generic models that are only constrained by $\mathbb{Z}_4$, ${\cal C}$ and ${\cal T}$ symmetries. We already encountered an example of such a term in Eq.~\eqref{deltaH}. Specifically, we find
\begin{align}
 \sigma_a^\dagger \sigma_{a+2} + H.c. =& [t^x_{a+\frac{1}{2}}t^x_{a+\frac{3}{2}}][t^x_{a}t^x_{a+1}+t^x_{a+1}t^x_{a+2}]~,\label{dualnosym1}\\
 \tau_a \tau_{a+1}+ H.c. 
  =& [t^z_{a} t^z_{a+1}-t^y_{a} t^y_{a+1}][t^z_{a-\frac{1}{2}}t^z_{a+\frac{1}{2}}+t^z_{a+\frac{1}{2}}t^z_{a+\frac{3}{2}}]\nonumber \\
 & +[t^z_{a} t^z_{a+1}+t^y_{a} t^y_{a+1}][1+t^z_{a-\frac{1}{2}}t^z_{a+\frac{3}{2}}]~.\nonumber 
\end{align}
Clock-model duality interchanges the expressions on the left side, but the corresponding terms on the right side are clearly not related by $T_{\frac{1}{2}}U$. In contrast, for the last term in Eq.~\eqref{deltaH}, the symmetry operation $T_{\frac{1}{2}}U$ does implement duality, i.e.,
\begin{align*}
 \sigma_a^2 \sigma_{a+2}^2 &= t^x_{a}t^x_{a+2}~,\\
  \tau_a^2 \tau_{a+1}^2 &= t^z_{a-\frac{1}{2}}t^z_{a+\frac{3}{2}}~.
\end{align*}

To connect to the treatment in the main text, it is convenient to bosonize this spin model according to
\begin{align}
&t_y \sim \partial_x \phi + (-1)^x\sin 2 \phi~, \label{spinbos1}\\
&t_z \pm i t_x \sim e^{\mp i \theta}[(-1)^x+\sin 2 \phi]~\label{spinbos2}~.
\end{align}
This expansion results in an effective low-energy Hamiltonian
\begin{align}
  \mathcal{H} = \int_x \bigg{\{}&\frac{v}{2\pi}[g(\partial_x \phi)^2 +g^{-1} (\partial_x\theta)^2] 
  \nonumber \\
  &- \kappa_1 \cos(4\phi)- \kappa_2 \cos(2\theta)\bigg{\}} \label{AT.spin.boson}
\end{align}
that has same form as Eq.~\eqref{eqn.abel.boson}, though the relation between $g,\kappa_1,\kappa_2$ and microscopic parameters of the Ashkin-Teller model is different. Firstly, for $\lambda=0$ and $J=f$, Eq.~\eqref{eqn.atrot} is a pure XY model, which in the convention defined by Eqs.~\eqref{spinbos1} and \eqref{spinbos2} corresponds to $g=1$ and $\kappa_1=\kappa_2=0$. Taking $J\neq f$ but $\lambda=0$ introduces an Ising anisotropy with $\kappa_2 \sim J - f$. When instead $\lambda\neq 0$ but $J=f$, Eq.~\eqref{eqn.atrot} is symmetric under $T_{\frac{1}{2}} U$, which acts as $\phi \rightarrow \phi + \pi/2, \theta\rightarrow \theta + \pi/2$---implying that $\kappa_2=0$. Finally, for generic $J,f,\lambda$ all terms in Eq.~\eqref{AT.spin.boson} are present. The broken translation symmetry would appear to permit the additional term $\sim \cos 2 \phi$, but that is forbidden by ${\cal I}$.

In this formulation of the Ashkin-Teller model, all phases discussed in Sec.~\ref{Phases} are readily identified. The ferromagnetic and paramagnetic phases of the clock model are driven by $\kappa_2$. When $\kappa_2<0$, $\theta$ is pinned to $\pi/2 \mod \pi$ and $\langle t^x\rangle \neq 0$
while for $\kappa_2>0$ it is pinned to $0 \mod \pi$ and consequently $\langle t^z\rangle \neq 0$. The phases driven by $\kappa_{1}$ are characterized by magnetization in the $y$ direction ($\kappa_1<0$) or by valence-bond order ($\kappa_1 >0$). 
Finally, `hybrid order' can be read off from the $\lambda \rightarrow \infty$ limit of Eq.~\eqref{eqn.atrot} and amounts to $\langle t^x\rangle \neq 0$ on integer sites and $\langle t^z\rangle \neq 0$ on half-integer sites. 

\subsection{Spin model B}
\label{spinmodelb}
We now return to Eq.~\eqref{ATdoubleIsing} and perform the Ising-model duality of Eq.~\eqref{Ising.duality} for \textit{half} of the spins, i.e.,
$s^{\prime x}_{a-1/4} =s^z_{a-\frac{1}{4}}s^z_{a-1-\frac{1}{4}}$ and $s^{\prime z}_{a-\frac{1}{4}} =\prod_{a'<a}s^x_{a'-\frac{1}{4}}$ (the product now runs only over integer sites $a'$). The Ashkin-Teller Hamiltonian becomes
\begin{align}
H_{J,f} =& -J \sum_{a } (s^{ z}_{a+\frac{1}{4}}s^{ z}_{a+1+\frac{1}{4}}+ s^{\prime x}_{a-\frac{1}{4}})
 \\
  &- f \sum_{a } (s^{x}_{a+\frac{1}{4}}+s^{z}_{a-\frac{1}{4}}s^{\prime z}_{a+1-\frac{1}{4}}   )\nonumber\\
  &+ \lambda \sum_{a } (J s^{\prime z}_{a+\frac{1}{4}} s^{\prime z}_{a+1+\frac{1}{4}} s^{x}_{a+1-\frac{1}{4}}
  \nonumber \\
  &+f s^{z}_{a-\frac{1}{4}}s^{z}_{a+1-\frac{1}{4}}    s^{\prime x}_{a+\frac{1}{4}})~.\nonumber
\end{align}
A second application of Eq.~\eqref{Ising.duality}, this time for all $s$ and $s'$, yields \cite{Kohmoto,Kohmoto2}
\begin{align}
H_{J,f} =& -J \sum_{a } (t^{\prime x}_{a-\frac{1}{2}}t^{\prime x}_{a}+ t^{\prime z}_{a-\frac{1}{2}}t^{\prime z}_{a}-\lambda t^{\prime y}_{a-\frac{1}{2}}t^{\prime y}_{a})
 \\
  &- f \sum_{a } (t^{\prime z}_{a}t^{\prime z}_{a+\frac{1}{2}}+t^{\prime x}_{a}t^{\prime x}_{a+\frac{1}{2}}-\lambda  t^{\prime y}_{a}t^{\prime y}_{a+\frac{1}{2}}   )~.\nonumber
\end{align}
This formulation is invariant under 
continuous global spin rotations about $t'^y$ and satisfies
\begin{align}
 H_{J,f}[\vect t' ] =  H_{f,J}[T_{\frac{1}{2}} \vect t' T^{-1}_{\frac{1}{2}}]~,
\end{align} 
i.e., duality is implemented as a translation. Bosonizing as before, one finds
\begin{align}
  \mathcal{H} = \int_x \bigg{\{}&\frac{v'}{2\pi}[g'(\partial_x \phi')^2 +g'^{-1} (\partial_x\theta')^2] 
  \nonumber \\
  &- \kappa_1' \cos(4\theta')- \kappa_2' \cos(2\phi')\bigg{\}} 
\end{align}
with $\kappa_2' \sim J-f$. It follows that the low-energy descriptions of spin models A and B are related by interchanging $\phi$ and $\theta$. 

\section{Alternative fermionization schemes}
\label{app.altferm}
The spin-1/2 representations of Appendix~\ref{app.spin} provide an alternative route to fermionizing clock Hamiltonians by using the conventional Jordan-Wigner transformation. The form of Eq.~\eqref{eqn.atrot} suggests introducing spinless Jordan-Wigner fermions as
\begin{align}
\mathtt{c}'_a&= \frac{1}{2} (t^z_a - i t^x_a)\prod_{a'<a} t^y_{a'}~.
\end{align}
When these fermions are bosonized via  $\mathtt{c}'_a \sim e^{i k_F a} e^{i(\theta + \phi)}+e^{-i k_F a} e^{i( \theta - \phi)}$, the Pauli operators take the form given in Eqs.~\eqref{spinbos1} and \eqref{spinbos2}, and the low-energy Hamiltonian is the one of Eq.~\eqref{AT.spin.boson}. Note that the bosonization convention employed above differs from that in Sec.~\ref{long_wavelength_limit}, which is useful since the low-energy descriptions obtained in the two approaches then exactly match up.  

\subsection{Spinful fermions}
To connect to the fermionization performed in the main text, it is instructive to adopt the alternative convention
\begin{align}
\mathtt{c}_{a} &= \frac{1}{2} (t^y_a - i t^z_a)\prod_{a'<a} t^x_{a'}~.
\end{align}
This is related to the one above by a global spin rotation---a highly non-local transformation on the fermions. Using Eqs.~\eqref{clocktospin} and \eqref{Ising.duality} we find for integer $a$
\begin{align}
\mathtt{c}_{a} 
&=  \frac{i}{2} (\sigma_a + \sigma_a^\dagger)\tau_a^\dagger  \prod_{a' <a}\tau_{a'}^2~,\nonumber\\
\mathtt{c}_{a+\frac{1}{2}}^\dagger-\mathtt{c}_{a+\frac{1}{2}} &=\frac{1+i}{2}(\sigma_a + i \sigma_a^\dagger) \tau_a^2\prod_{a' <a}\tau_{a'}^2\nonumber~,\\
\mathtt{c}_{a-\frac{1}{2}}^\dagger+\mathtt{c}_{a-\frac{1}{2}} &= - \frac{1-i}{2}(\sigma_{a} - i \sigma_{a}^\dagger) \tau_a^2\prod_{a' <a}\tau_{a'}^2\nonumber~,
\end{align}
where we omitted a boundary term $s^z_{0}$. Note that the spinful fermions introduced in Sec.~\ref{fermions} have exactly the same structure, i.e., a string of $\tau^2$'s that is terminated by an odd power of $\sigma$ operators.  This similarity implies a local relationship between the two kinds of fermions, which we already provided explicitly in Eqs.~\eqref{JWferm1} and \eqref{JWferm2}.

\subsection{Dual fermions}
To connect to the dual fermions of Sec.~\ref{fermions}, recall that the spin-1/2 representation of Eq.~\eqref{eqn.atrot} implements duality as a $\pi/2$ rotation combined with a half-translation. This suggests that the spinful fermions $\tilde{\mathtt{f}}_{a,\alpha}$ defined by replacing $c_{a}$ in Eqs.~\eqref{JWferm1} and \eqref{JWferm2} by
\begin{align}
\tilde c_{a} &= T_{\frac{1}{2}} U c_{a} U^{-1}T_{\frac{1}{2}}^{-1}=\frac{1}{2} (t^y_{a+\frac{1}{2}} + i t^x_{a+\frac{1}{2}})\prod_{a'<a+\frac{1}{2}} t^z_{a'}
\end{align}
correspond to the dual fermions introduced in the main text. Indeed, for the Ashkin-Teller model one finds
\begin{align}
H_{J,f}[\tilde f_\alpha] =&  H_{f,J}[f_\alpha] =  H_{f,J}[\mathtt{f}_\alpha]\\
= 
&H_{J,f}[T_{\frac{1}{2}} U  \mathtt{f}_\alpha U^{-1}T_{\frac{1}{2}}^{-1} ]=H_{J,f}[\tilde { \mathtt{f}}_\alpha]~\nonumber, 
\end{align}
where the third equality holds due to Eq.~\eqref{eqn.dualrotation}. This relationship breaks down, e.g., in the presence of the perturbation described by Eq.~\eqref{dualnosym1}. Unlike $\mathtt{f}_\alpha$ and $f_\alpha$, the single-particle operators $\tilde{\mathtt{f}}_\alpha$ and $\tilde f_\alpha$ are related non-locally as noted in the main text.

\section{Explicit map between $\mathbb{Z}_4$ parafermions and fermions}
\label{explicit_map}

Here we furnish explicit maps that fermionize the $\mathbb{Z}_4$ parafermion operators defined in Eq.~\eqref{alpha}.  We first use Eqs.~\eqref{tau_expansion} through \eqref{string} to write $\sigma_a$ and $\sigma_a\tau_a$ in terms of fermions:
\begin{align}
  \sigma_a &= S_a[(\bar w f_{a,\downarrow}^\dagger + w f_{a,\uparrow}) + (\bar w f_{a,\uparrow}^\dagger - w f_{a,\uparrow}) n_{a,\downarrow}
  \nonumber \\
  &- (w f_{a,\downarrow} + \bar w f_{a,\downarrow}^\dagger) n_{a,\uparrow}]
  \label{sigma_fermionized}
  \\
  \sigma_a \tau_a &= S_a[\bar w(f_{a,\downarrow}^\dagger - f_{a,\uparrow}) + (\bar w f_{a,\uparrow} - w f_{a,\uparrow}^\dagger) n_{a,\downarrow}
  \nonumber \\
  &+ (w f_{a,\downarrow} - \bar w f_{a,\downarrow}^\dagger) n_{a,\uparrow}],
  \label{sigmatau_fermionized}
\end{align}
where $w = e^{i \frac{\pi}{4}}$.  These expressions simplify considerably upon introducing Majorana operators and projectors as follows,
\begin{align}
  f_{a,\alpha} &= \bar w(\gamma_{a,1\alpha} + i \gamma_{a,2\alpha})/2 
  \label{fMajorana} \\
  \mathcal{P}_{a,1\pm} &= \frac{1}{2} (1 \pm i\gamma_{a,1\downarrow}\gamma_{a,2\uparrow})\\
  \mathcal{P}_{a,2\pm} &= \frac{1}{2} (1 \pm i\gamma_{a,1\uparrow}\gamma_{a,2\downarrow}).
  \label{projectors}
\end{align}
We then obtain
\begin{align}
  \sigma_a &= S_a[\mathcal{P}_{a,1+} \gamma_{a,1\uparrow} -i \mathcal{P}_{a,1-} \gamma_{a,2\downarrow}]  \label{sigma_fermionized2} \\
  \bar w \sigma_a \tau_a &= \frac{S_a}{\sqrt{2}}e^{\frac{\pi}{2}i(\mathcal{P}_{a,2+} -1)}(\gamma_{a,1\downarrow}-\gamma_{a,2\uparrow}) \label{sigmatau_fermionized2}.
\end{align}  As an aside, Eq.~\eqref{sigma_fermionized2} provides an alternative means of recovering Eq.~\eqref{Hf} directly from the clock model.  The fermionic operators $c_a$ and $d_a$ are respectively given by $c_a = \frac{1}{2}(\gamma_{a,1\uparrow} + i\gamma_{a,2\downarrow})$ and $d_a = \frac{1}{2} (\gamma_{a, 1\downarrow} + i\gamma_{a, 2\uparrow})$.  Moreover, $P_{a, 1+}$ and $P_{a, 1-}$ project onto the magnetization sectors $m_a = -1$ and $m_a = +1$, respectively, while the strings combine to yield a simple multiplicative factor of $m_a (i\gamma_{a,1\uparrow} \gamma_{a,2\downarrow})$.

The parafermion operators $\alpha_{2a-1},\alpha_{2a}$ arise from Eq.~\eqref{sigma_fermionized2},~\eqref{sigmatau_fermionized2} multipled by the disorder operator $\mu_{a-\frac{1}{2}}$, respectively.  Both cases contain a factor 
\begin{align}
  S_a \mu_{a-\frac{1}{2}} &= \mu_{a-\frac{1}{2}}^\dagger = e^{-i \frac{\pi}{2}\sum_{b<a}(n_{b,\uparrow}-n_{b,\downarrow}+2n_{b,\uparrow}n_{b,\downarrow})}
  \nonumber \\
  &= e^{-i \frac{\pi}{4}\sum_{b<a}[1 + i \gamma_{b,1\uparrow}\gamma_{b,2\uparrow}(2+i \gamma_{b,1\downarrow}\gamma_{b,2\downarrow})]}.
\end{align}
Putting everything together yields
\begin{align}
  \alpha_{2a-1} &= \frac{1}{2}e^{-i \frac{\pi}{4}\sum_{b<a}[1 + i \gamma_{b,1\uparrow}\gamma_{b,2\uparrow}(2+i \gamma_{b,1\downarrow}\gamma_{b,2\downarrow})]}
  \nonumber \\
  &\times [\mathcal{P}_{a,1+} \gamma_{a,1\uparrow} -i \mathcal{P}_{a,1-} \gamma_{a,2\downarrow}] 
  \nonumber \\
  \label{alpha_odd_fermionized}
  \\
  \alpha_{2a} &= \frac{1}{\sqrt{2}}e^{-i \frac{\pi}{4}\sum_{b<a}[1 + i \gamma_{b,1\uparrow}\gamma_{b,2\uparrow}(2+i \gamma_{b,1\downarrow}\gamma_{b,2\downarrow})]}
  \nonumber \\
  &\times e^{\frac{\pi}{2}i(\mathcal{P}_{a,2+} -1)}(\gamma_{a,1\downarrow}-\gamma_{a,2\uparrow}).
  \label{alpha_even_fermionized}
\end{align}

Equations~\eqref{alpha_odd_fermionized} and \eqref{alpha_even_fermionized} explicitly relate parafermions to non-local products of fermions.  We will now derive an alternative decomposition that involves local combinations of fermions and dual fermions.  To this end define the dual analogue of Eqs.~\eqref{fMajorana} through \eqref{projectors},
\begin{align}
  \tilde f_{\tilde a,\alpha} &= \bar w(\tilde \gamma_{\tilde a,1\alpha} + i \tilde \gamma_{\tilde a,2\alpha})/2 \\
  \tilde{\mathcal{P}}_{\tilde a,1\pm} &= \frac{1}{2} (1 \pm i\tilde \gamma_{\tilde a,1\downarrow}\tilde \gamma_{\tilde a,2\uparrow})\\
  \tilde{\mathcal{P}}_{\tilde a,2\pm} &= \frac{1}{2} (1 \pm i\tilde \gamma_{\tilde a,1\uparrow} \tilde \gamma_{\tilde a,2\downarrow}),
\end{align}
as well as the dual analogue of Eq.~\eqref{sigma_fermionized2},
\begin{align}
  \mu_{\tilde a} &= \tilde S_{\tilde a}[\tilde{\mathcal{P}}_{\tilde a,1+} \tilde \gamma_{\tilde a,1\uparrow} - i \tilde{\mathcal{P}}_{\tilde a,1-} \tilde \gamma_{\tilde a,2\downarrow}],  \label{mu_fermionized}
\end{align}
where $\tilde a = a + \frac{1}{2}$.  
The string in the above equation reads
\begin{equation}
  \tilde S_{\tilde a} = \prod_{\tilde b < \tilde a} \nu_{\tilde b}^2 = \sigma_{a}^2 = i \gamma_{a,1\downarrow}\gamma_{a,2\uparrow}.
  \label{sigmasquared}
\end{equation}
Here we neglected the termination of the $\nu^2$ string; that is, we discarded a $\sigma_{-\infty}^2$ factor. To obtain the right-hand side, we used Eq.~\eqref{sigma_fermionized2} to express $\tilde S_{\tilde a}$ as a purely local product of the original fermions.  One can similarly express the string in Eqs.~\eqref{sigma_fermionized2} and \eqref{sigmatau_fermionized2} as a local product of dual fermions:
\begin{equation}
  S_a = \mu^2_{\tilde a - 1} = i \tilde \gamma_{\tilde a - 1,1\downarrow}\tilde \gamma_{\tilde a - 1,2\uparrow}.
  \label{musquared}
\end{equation}
We can now obtain the desired form of the parafermion operators, 
\begin{align}
  \alpha_{2a-1} &= \sigma_a \mu_{\tilde a - 1}  \nonumber \\
  &= [\mathcal{P}_{a,1+} \gamma_{a,1\uparrow} + i \mathcal{P}_{a,1-} \gamma_{a,2\downarrow}] (\tilde{\mathcal{P}}_{\tilde a-1,2+}-\tilde{\mathcal{P}}_{\tilde a-1,2-}) \nonumber \\
  &\times [\tilde {\mathcal{P}}_{\tilde a-1,1+} \tilde \gamma_{\tilde a-1,1\uparrow} - i \tilde{\mathcal{P}}_{\tilde a-1,1-} \tilde \gamma_{\tilde a-1,2\downarrow}]  \label{parafermion_1} \\
  \alpha_{2a} &= \bar{w} \sigma_a \mu_{\tilde a} \nonumber \\
  &= \bar w [\mathcal{P}_{a,1+} \gamma_{a,1\uparrow} - i \mathcal{P}_{a,1-} \gamma_{a,2\downarrow}] (\mathcal{P}_{a,2+}-\mathcal{P}_{a,2-})\nonumber \\
  &\times [\tilde{\mathcal{P}}_{\tilde a,1+} \tilde \gamma_{\tilde a,1\uparrow} - i \tilde{\mathcal{P}}_{\tilde a,1-} \tilde \gamma_{\tilde a,2\downarrow}].  
  \label{parafermion_2}
\end{align}
The factor $\mu_{\tilde a-1}$ above involves a string $\tilde S_{\tilde a - 1} = \sigma_{a-1}^2$.  To derive Eq.~\eqref{parafermion_1} we equivalently wrote this string as $\tilde S_{\tilde a - 1} = \sigma_a^2 \nu_{\tilde a - 1}^2$, expressed $\sigma_a^2$ in terms of a local product of fermions using Eq.~\eqref{sigmasquared}, and expressed $\nu_{\tilde a-1}^2$ in terms of dual fermions.  Similarly, the $\sigma_a$ in Eq.~\eqref{parafermion_2} involves a string $S_a = \mu_{\tilde a - 1}^2$ which we can rewrite as $\tau_a^2 \mu_{\tilde a}^2$.  Here we expressed $\mu_{\tilde a}^2$ as a local product of dual fermions using Eq.~\eqref{musquared} and wrote $\tau_a^2$ in terms of fermions.  We adopted this approach to express the parafermions as products of fermion operators living on a single site and dual fermions living on another.  

\section{Self-duality of the hybrid-order ground states}
\label{SelfDualityHybridOrder}

As discussed in Sec.~\ref{HybridOrder}, ground states of the Hamiltonian $H_{\rm hybrid~order} = -J_2 \sum_a \sigma_a^2 \sigma_{a+1}^2 - f_2 \sum_a \tau^2_a$ can be expressed as
\begin{align}
  \ket{+} &= \prod_a \dfrac{1+\tau_a^2}{\sqrt{2}} \ket{\sigma = 1,\ldots,1} \\
\ket{-} &= \prod_a \dfrac{1+\tau_a^2}{\sqrt{2}} \ket{\sigma = i,\ldots,i}
\end{align}
for any positive couplings $f_2, J_2$.  Our goal here is to show that these states take essentially the same form after a duality transformation.  

For this purpose, one can profitably view $|\pm\rangle$ as follows: Start from `root states' $\ket{\sigma = 1,\ldots,1}$ and $\ket{\sigma = i,\ldots,i}$ that satisfy the $J_2$ term, and then apply $(1+\tau_a^2)$ factors that project away $\tau_a^2 = -1$ configurations to satisfy the $f_2$ term.  (Choosing root states $\ket{\sigma = -1,\ldots,-1}$ and $\ket{\sigma = -i,\ldots,-i}$ produces the same end result.)  From a dual viewpoint, we can construct one ground state by taking the root state $|\tau = 1,\ldots, 1\rangle$ that satisfies the $f_2$ term, and then applying $(1+\sigma_a^2\sigma_{a+1}^2)$ factors to satisfy $J_2$.  The corresponding wavefunction reads
\begin{equation}
  |\tilde+\rangle = \prod_a \frac{1+\sigma_a^2\sigma_{a+1}^2}{\sqrt{2}} \ket{\tau = 1,\ldots,1}
\end{equation}
and obeys $Q|\tilde+\rangle = |\tilde +\rangle$ (as usual $Q$ is the $\mathbb{Z}_4$ generator).  Taking the root state $\ket{\tau = -1,1,\ldots,1}$---which also satisfies $f_2$---yields a second ground state
\begin{equation}
  |\tilde-\rangle = \prod_a \frac{1+\sigma_a^2\sigma_{a+1}^2}{\sqrt{2}} \ket{\tau = -1,1,\ldots,1} = \sigma_1^2|\tilde +\rangle
\end{equation}
with $Q|\tilde-\rangle = -|\tilde -\rangle$.  Despite appearances, $|\tilde\pm\rangle$ represent product states for the clock chain.  Applying the basis change $|\tau = 1\rangle = \frac{1}{2}\sum_\sigma|\sigma\rangle$ allows us to write
\begin{eqnarray}
  \frac{|\tilde+\rangle + |\tilde - \rangle}{\sqrt{2}} &=& \frac{1}{2^N}\sum_{\sigma_1,\ldots,\sigma_N}\frac{1+\sigma_1^2}{\sqrt{2}}
  \nonumber \\
  &\times& \prod_a\frac{1+\sigma_a^2\sigma_{a+1}^2}{\sqrt{2}}|\sigma_1,\ldots,\sigma_N\rangle.
\end{eqnarray}
The $(1+\sigma_1^2)$ factor restricts the $\sigma_1$ sum to $\pm 1$; the product $(1+\sigma_a^2\sigma_{a+1}^2)$ then propagates this same restriction to all other sites.   We therefore obtain the relation
\begin{equation}
  \frac{|\tilde+\rangle + |\tilde - \rangle}{\sqrt{2}} = |+\rangle,
\end{equation}
while analogous logic yields
\begin{equation}
  \frac{|\tilde+\rangle - |\tilde - \rangle}{\sqrt{2}} = |-\rangle.
\end{equation}
Duality indeed merely introduces a basis change.  The situation should be contrast to the broken-symmetry canted-ferromagnet states defined in Eq.~\eqref{CantedStates}, which dualize into ground states of an SPT phase [Eqs.~\eqref{psiSPT} and \eqref{other_states}].

\section{Zero-mode anomalies in the SPT phases}
\label{ProjectiveRepsAppendix}

This Appendix rigorously shows that the $\kappa_2 < 0$ states discussed in Sec.~\ref{SPT} correspond to SPT phases.  To do so, we will appeal to the theory of projective representations for SPT's put forward by Refs.~\onlinecite{ChenSPT1,ChenGuWen,ChenSPT2}.  
The relevant symmetries are $\mathbb{Z}_4$, $\mathcal{C}$, and $\mathcal{T}$.  Generators of these symmetries---which we respectively denote by $Q, c$, and $t$---form a linear representation of the symmetry group when acting on physical degrees of freedom.  For example, take $\mathbb{Z}_4$.  In the clock representation, we can choose $\sigma$ eigenstates as physical kets; $Q$ `winds' these states according to
\begin{align}
\ket{\sigma=1} &\rightarrow \ket{\sigma=-i} \rightarrow \nonumber \ket{\sigma=-1} \\ 
&\rightarrow \ket{\sigma=i} \rightarrow \ket{\sigma=1}.
\end{align}
This action leads to an example of a linear representation: the matrix representation of the symmetry generator $Q$,
\begin{equation}
  N(Q) = \left[ {\begin{array}{cccc}
   0 & 0 & 0 & 1 \\
   1 & 0 & 0 & 0 \\
   0 & 1 & 0 & 0 \\
   0 & 0 & 1 & 0 
  \end{array} } \right],
\end{equation}  
obeys the same multiplication rules as the symmetry generators themselves. That is,
\begin{equation}
  N(g)N(h) = N(gh)
  \label{linear_rep}
\end{equation}
where $g,h$ are symmetry-group elements and $N$ is the corresponding matrix representation.  When considering $\mathbb{Z}_4$ alone, one has $g = Q^a$ and $h = Q^b$ for integers $a,b$, though Eq.~\eqref{linear_rep} defines a linear representation for general symmetry groups as well.  

For an SPT, an interesting loophole arises---the edge modes are anomalous and carry a \emph{projective} representation of the symmetry group.  Specifically, if $M(g)$ is the matrix representation that specifies how the edge modes transform under a symmetry element $g$, then
\begin{equation}
  M(g)M(h) = \omega(g,h)M(gh).
\end{equation}
Here $\omega$ is a phase factor that cannot be gauged to $1$ by a phase redefinition of the form $M(g) \rightarrow  M'(g) = e^{i\theta_g}M(g)$.  In what follows we will show that the edge zero modes in the $\kappa_2<0$ phases indeed transform projectively under appropriate symmetries, indicating that the bulk forms an SPT. 

We will first address the clock representation (see below for an extension to the fermion and parafermion cases).
Let us focus on the left zero mode, which as discussed in Section~\ref{SPT} encodes a twofold degeneracy corresponding to pseudospin-1/2 states with $\eta^z_1 = \pm 1$.  According to Eq.~\eqref{Q_proj}, the action of $Q$ on the zero mode is given by the operator $e^{i\frac{\pi}{4} \eta^z_1}$, which yields the matrix representation 
\begin{equation}
M(Q)=
\begin{bmatrix}
e^{i\frac{\pi}{4}} & 0 \\
0 & e^{-i\frac{\pi}{4}}
\end{bmatrix}.
\label{MQ}
\end{equation}
Although $Q^4 = 1$ by definition, the matrix above satisfies $[M(Q)]^4 = -1$.  In this case the $-1$ on the right side can be gauged away by defining $M'(Q) = e^{i\frac{\pi}{4}} M(Q)$. Then $M'(Q)^4 = M'(Q^4) = 1$, yielding a linear representation.  Hence the clock chain does \emph{not} form an SPT if $\mathbb{Z}_4$ alone is present.

Suppose that we instead enforce the combination $\mathbb{Z}_4 \mathcal{T}$.  The symmetry properties from Table~\ref{SPT_symmetry_table} imply that $\mathcal{T}$ transforms the zero mode according to the matrix
\begin{equation}
M(t)=
\begin{bmatrix}
0 & 1 \\
1 & 0
\end{bmatrix} \mathcal{K},
\end{equation}
where $\mathcal{K}$ denotes complex conjugation, reflecting antiunitary of $\mathcal{T}$.  The matrix representation of the generator $Qt$ follows as
\begin{equation}
M(Qt)=
\begin{bmatrix}
0 & e^{i\frac{\pi}{4}} \\
e^{-i\frac{\pi}{4}} & 0
\end{bmatrix} \mathcal{K}.
\end{equation}
Similar to the case of $\mathbb{Z}_4$ by itself, we see that $[M(Qt)]^4 = -1$ even though $(Qt)^4 = 1$.  Crucially, however, here there is no phase factor that we can append to remove the $-1$.  Thus the zero mode transforms projectively under $\mathbb{Z}_4\mathcal{T}$, and the clock chain forms an SPT in the presence of this composite symmetry.  

Alternatively, the clock chain can form an SPT protected by $\mathbb{Z}_4$ and $\mathcal{C}$.  
Under $\mathcal{C}$ symmetry $\sigma$ eigenstates transform as
\begin{align}
\ket{\sigma=1} &\rightarrow \ket{\sigma=1}, ~~ \ket{\sigma=-1} \rightarrow \ket{\sigma=-1} \\
\ket{\sigma=i} &\rightarrow \ket{\sigma=-i}, ~~ \ket{\sigma=-i} \rightarrow \ket{\sigma=i}.
\end{align}
Furthermore, $\mathcal{C}$ transforms the zero mode according to
\begin{equation}
M(c)=
\begin{bmatrix}
0 & 1 \\
1 & 0
\end{bmatrix},
\label{Mc}
\end{equation}
where we again used the symmetry properties from Table~\ref{SPT_symmetry_table}.  It is useful to now associate $\sigma$ eigenstates with the four compass directions; from this viewpoint $\mathbb{Z}_4$ effects a rotation while $\mathcal{C}$ yields a reflection.  The corresponding symmetry group is $D_8$, the dihedral group on $4$ elements.  
In order for the zero modes to transform as a linear representation with respect to $\mathbb{Z}_4$ and $\mathcal{C}$ symmetries, 
we must be able to deform the matrices in Eqs.~\eqref{MQ} and \eqref{Mc} so that
\begin{align}
[M(Q)]^4 &= M(Q^4) = 1 \\
{[}M(c)]^2 &= M(c^2) = 1 \\
M(c)M(Q)M(c) &= M(cQc = Q^{-1}) = M(Q)^{-1}.
\end{align}
In the last line we invoked properties of the dihedral group.  Such a deformation is impossible---no matter what phases we append to $M(Q)$ and $M(c)$, we can not simultaneously satisfy all three conditions above.  So the zero modes indeed transform projectively in the presence of $\mathbb{Z}_4$ and $\mathcal{C}$, again guaranteeing an SPT for the clock chain.  

Note that \emph{both} $\mathbb{Z}_4$ and $\mathcal{C}$ symmetries must be enforced for the conclusion above to apply, as similar logic shows that the combination $\mathbb{Z}_4\mathcal{C}$ by itself does not protect the SPT.  However, an SPT does emerge upon supplementing $\mathbb{Z}_4\mathcal{C}$ with $\mathbb{Z}_4^2$, which together form the group $\mathbb{Z}_2 \times \mathbb{Z}_2$.  The associated matrix representations are
\begin{equation}
M(Qc)=
\begin{bmatrix}
0 & e^{i\frac{\pi}{4}} \\
e^{-i\frac{\pi}{4}} & 0
\end{bmatrix},~~~
M(Q^2)=
\begin{bmatrix}
i & 0 \\
0 & -i
\end{bmatrix}.
\end{equation}
A linear representation arises if we can redefine the matrices such that
\begin{align}
[M(Q^2)]^2 &= M(Q^4) = 1 \\
{[}M(Qc)]^2 &= M((Qc)^2) = 1 \\
M(Q^2)M(Qc) &= M(Qc)M(Q^2),
\end{align}
which again is impossible.

We can readily extend these results to the parafermion and fermion realizations.  Above we saw that the clock-chain SPT can be protected by $(i)$ $\mathbb{Z}_4 \mathcal{T}$, $(ii)$, $\mathbb{Z}_4$ and $\mathcal{C}$, or $(iii)$ $\mathbb{Z}_4\mathcal{C}$ and $\mathbb{Z}_4^2$.  For parafermions and fermions, some of these symmetries are enforced automatically---thus weakening the symmetry requirements for obtaining an SPT in these realizations.  Parafermions realize $\mathbb{Z}_4$ automatically, so that we need only impose $\mathcal{T}$ or $\mathcal{C}$.  Fermions realize $\mathbb{Z}_4^2$ automatically---which corresponds to global fermion parity---though $\mathbb{Z}_4$ itself can be broken.  Thus electronic time reversal $\mathcal{T}_{\rm elec} = \mathbb{Z}_4\mathcal{T}$ or spin rotation symmetry $U_{\rm spin} = \mathbb{Z}_4 \mathcal{C}$ protect the fermionic SPT.  Incidentally, the existence of an SPT in the latter context is clear even without the analysis in this Appendix, since the fermions realize the well-studied TRITOPS phase.

\section{Parafermion braid matrices in fermion language}
\label{BraidingAppendix}

As noted in Sec.~\ref{nonAbelian1D}, rewriting parafermion braid matrices in terms of Majorana operators enables a direct comparison with braid matrices that arise in the spinful-fermion realization.  Adapting the machinery from Sec.~\ref{FM_phase} yields the following dictionary:
\begin{align}
\alpha_1 &= -e^{i\frac{\pi}{4}(m_L-1)}\gamma_1 \\
\alpha_2 &= -e^{-i\frac{\pi}{4}[p_L(m_L+1)+1]}\Gamma_2 \\
\alpha_3 &= -e^{i\frac{\pi}{4}[m_R-m_L+p_L(m_L+1)]}\gamma_3\gamma_2\Gamma_1 \\
\alpha_4 &= -e^{-i\frac{\pi}{4}[p_R(m_R+1)-p_L(m_L+1)+m_L]}\Gamma_1\Gamma_4\gamma_2.
\end{align}
Here $\alpha_j$ and $\gamma_j$ denote zero-mode operators in Fig.~\ref{Braiding_fig}; $p_L = i \gamma_1\gamma_2$ and $p_R = i \gamma_3\gamma_4$; and $m_L = i \Gamma_1\Gamma_2$ and $m_R = i \Gamma_3\Gamma_4$ denote the magnetizations in Fig.~\ref{Braiding_fig}(b).  The total fermion parities in the left and right topological segments are $P_{{\rm tot},L} = m_L p_L$ and $P_{{\rm tot},R} = m_R p_R$, respectively.  Inserting the decomposition above into Eq.~\eqref{PFbraid} yields the braid matrices given in Eqs.~\eqref{U12fermionic} and \eqref{U23fermionic} from Sec.~\ref{nonAbelian1D}.

To see that the parafermionic braid matrix $U_{1,2}$ generates cat states when acting on physical fermion wavefunctions, consider its action on states $\ket{m_L,P_{{\rm tot},L}}$ for the left topological segment:  
\begin{align}
\ket{m_L=1,P_{{\rm tot},L}=1} &\rightarrow e^{i\frac{3 \pi}{8}} \ket{m_L=-1,P_{{\rm tot},L}=1} \nonumber \\
\ket{m_L=1,P_{{\rm tot},L}=-1} &\rightarrow e^{i\frac{\pi}{8}} \ket{m_L=1,P_{{\rm tot},L}=-1} \nonumber \\
\ket{m_L=-1,P_{{\rm tot},L}=1} &\rightarrow e^{i\frac{3 \pi}{8}} \ket{m_L=1,P_{{\rm tot},L}=1} \nonumber \\
\ket{m_L=-1,P_{{\rm tot},L}=-1} &\rightarrow e^{i\frac{\pi}{8}} \ket{m_L=-1,P_{{\rm tot},L}=-1}. \nonumber
\end{align}
The total parity is preserved under $U_{1,2}$ as expected, though the magnetization flips in the $P_{{\rm tot},L} = +1$ sector.  Applying $U_{1,2}$ to a physical fermion state
\begin{align}
  \ket{\psi} &= a \ket{m_L,P_{{\rm tot},L};m_R,P_{{\rm tot},R}} 
  \nonumber \\
  &+ b \ket{m_L,-P_{{\rm tot},L};m_R,-P_{{\rm tot},R}}
  \nonumber
\end{align}
then yields a cat state whenever $a$ and $b$ are both nonzero.

\section{Derivation of parafermion fusion Hamiltonians}
\label{FusionHamiltonian}

We will now analyze Fig.~\ref{Fusion_fig_PF}(a) and derive the minimal Hamiltonian coupling parafermions $\alpha_1$ and $\alpha_2$.  Following Sec.~\ref{FM_phase}, we parametrize the pinned bosonized fields in the adjacent domains as follows: $\theta = 0$ on the left, $\phi = \pi \hat a/2$ between $\alpha_{1,2}$, and $\theta = \pi\hat b +\theta_0/2$ in the central region.  With these definitions (in particular, including the $\theta_0/2$ shift) $\hat a, \hat b$ are once again integer-valued operators that define parafermions via $\alpha_1 = e^{i \frac{\pi}{2} \hat a}$ and $\alpha_2 = e^{i \frac{\pi}{2}(\hat a-\hat b)}$, precisely as in Eq.~\eqref{alphas_bosonized}.  

Now consider the bosonized perturbation 
\begin{equation}
  H_{1,2} = -2t\cos\left[\frac{\theta(x_2)-\theta(x_1)}{2}\right],
\end{equation}
where $x_1$ sits just to the left of $\alpha_1$ while $x_2$ sits just to the right of $\alpha_2$.  This coupling cycles $\phi$ in the intervening region among adjacent pinned values and is physical provided $\alpha_{1,2}$ are close to one another.  Note also that $H_{1,2}$ preserves $\mathbb{Z}_4, \mathcal{C}$, and $\mathcal{T}$---which are present at least when $\theta_0 = 0\mod \pi$.   Away from these special $\theta_0$ values we can in principle introduce a non-universal shift inside of the cosine in $H_{1,2}$, though such a shift is benign for our purposes.  We will also ignore higher harmonics, i.e., terms like $\cos[\theta(x_2)-\theta(x_1)]$, since they also do not affect our conclusions.  Projecting $H_{1,2}$ into the low-energy subspace yields
\begin{align}
  H_{1,2} &\rightarrow - 2t \cos\left(\frac{\pi}{2}\hat b + \frac{\theta_0}{4}\right) 
  \nonumber \\
  &= -t\left[ e^{i\frac{\pi-\theta_0}{4}} \alpha_1^\dagger \alpha_2 + H.c.\right],
\end{align}
corresponding to Eq.~\eqref{FusionH} from the main text.  

One can similarly examine the parafermion setup from Fig.~\ref{Fusion_fig_elec_2}(a).  Here we parametrize the pinned bosonized fields as $\phi = \pi \hat a/2$ on the left, $\theta = \pi \hat b$ between $\alpha_{2,3}$, and $\phi = \pi \hat c/2 + \phi_0/4$ in the middle domain ($\hat a,\hat b,\hat c$ are integer-valued operators).  In this case the parafermion operators are given by $\alpha_2 = e^{i \frac{\pi}{2}(\hat a-\hat b)}$ and $\alpha_3 = e^{i \frac{\pi}{2}(\hat c-\hat b)}$.  Define a bosonized perturbation that cycles $\theta$ between adjacent pinned values:
\begin{equation}
  H_{2,3} = -2t \cos[\phi(x_3)-\phi(x_2)]
\end{equation}
with $x_2$ now taken just to the left of $\alpha_2$ and $x_3$ taken just to the right of $\alpha_3$.  This term projects to
\begin{align}
  H_{2,3} &\rightarrow - 2t \cos\left[\frac{\pi}{2}(\hat c-\hat a) + \frac{\phi_0}{4}\right] 
  \nonumber \\
  &= -t \left[ e^{i \frac{\pi + \phi_0}{4}} \alpha_2^\dagger \alpha_3 + H.c.\right].
\end{align}

\section{Dictionary for higher parafermions}
\label{HigherDictionaryAppendix}

In this Appendix we will invert Eqs.~\eqref{sigma2M} and \eqref{tau2M} so that we can express fermions and order-parameter operators in terms of clock variables.  This exercise will enable us to relate the fermions in the $M = 2$ limit to the alternate set of fermions that we obtained for the $\mathbb{Z}_4$ case in Sec.~\ref{FM_phase}.  

As we already observed, the order parameter $\mathcal{O}_a$ is easily related to clock operators by squaring Eq.~\eqref{sigma2M}, which yields 
\begin{equation}
  \mathcal{O}_a = \sigma_a^2.
\end{equation}
Next we will solve for the hard-core bosons $B_a$.  It is useful to observe that
\begin{align}
  \tau_a^M = e^{i\pi B_a^\dagger B_a},
  \label{tauM}
\end{align}
which follows from Eqs.~\eqref{tau2M} and \eqref{ODproperties}.  Using this relation in conjunction with Eq.~\eqref{sigma2M}, we have
\begin{align}
\label{HigherParafermionDictionary}
  \sigma_a &= B_a + \mathcal{O}_a B_a^\dagger \\
  \sigma_a \tau_a^M &= -B_a + \mathcal{O}_a B_a^\dagger
\end{align}
and hence
\begin{equation}
  B_a = \frac{1}{2}\sigma_a(1-\tau_a^M).
  \label{Bsolution}
\end{equation} 
Substituting our expression for $B_a$ into Eq.~\eqref{tau2M} yields
\begin{align}
  \mathcal{D}_a &= \tau\left[\left(\frac{1+\tau^{M}}{2}\right) + e^{-i\frac{\pi}{M}}\left(\frac{1-\tau^{M}}{2}\right)\right].
\end{align}
One can readily verify that $\mathcal{D}_a$ and $\mathcal{O}_a$ commute with $B_a$, as assumed in our decomposition.  Finally, combining Eqs.~\eqref{tauM} and \eqref{Bsolution} allows us to write $C_a$ fermions defined in Eq.~\eqref{Cfermions} as
\begin{equation}
  C_a = \frac{1}{2}\sigma_a(1-\tau_a^M) \prod_{a'<a} \tau_{a'}^M.
\end{equation}

We now specialize to $\mathbb{Z}_4$, i.e., $M = 2$, with the intention of relating operators $\mathcal{O}_a,\mathcal{D}_a,C_a$ to the fermions $c_a,d_a$ defined in Eqs.~\eqref{f_up_transformation} and \eqref{f_down_transformation}.  The order parameter part is trivial, since $\mathcal{O}_a \rightarrow m_a = e^{i \pi d_a^\dagger d_a}$ [recall Eq.~\eqref{m_lattice}].  As an intermediate step for the other pieces, we use Eqs.~\eqref{tau_expansion} and \eqref{sigma_expansion} to express $\mathcal{D}_a$ and $B_a$ in terms of hard-core spinful bosons: 
\begin{align}
\label{Z4Identification}
\mathcal{D}_a &= e ^{i \pi n_{a,\downarrow}} = e^{i \pi f_{a,\downarrow}^\dagger f_{a,\downarrow}} \\
B_a  &= n_{a,\downarrow} b_{a,\uparrow}^\dagger + (1-n_{a,\downarrow}) b_{a,\uparrow}. \label{Bb}
\end{align}
Using Eq.~\eqref{f_down_transformation} in the first equation immediately gives
\begin{equation}
  \mathcal{D}_a = (d_a+d_a^\dagger)(c_a^\dagger - c_a).
\end{equation}
The string that relates $C_a$ fermions to $B_a$ bosons [see Eq.~\eqref{Cfermions}] is built from
\begin{equation}
e^{i\pi B_a^\dagger B_a} = e^{i\pi(n_{a,\downarrow} + n_{a,\uparrow})},
\end{equation}
and thus has exactly the same form as the string in Eq.~\eqref{string} that relates spinful fermions $f_{a,\alpha}$ to $b_{a,\alpha}$.  Thus, $C_a$ should be locally related to $f_{a,\alpha}$ fermions, and in turn $c_a,d_a$ fermions.  Equation~\eqref{Bb} together with Eqs.~\eqref{f_up}, \eqref{f_down}, \eqref{f_up_transformation}, and \eqref{f_down_transformation} specifically yield
\begin{align}
  C_a = \frac{1-m_a}{2} c_a + \frac{1+m_a}{2} c_a^\dagger. 
\end{align}

\bibliography{Fermionization_references}
\end{document}